\documentclass[aps,prd,floatfix,nofootinbib,preprintnumbers]{revtex4}

\usepackage[english]{babel}
\usepackage{amsmath,amssymb}
\usepackage[dvips]{graphicx}

\newcommand{\AddrAHEP}{%
AHEP Group, Institut de F\'{\i}sica Corpuscular --
C.S.I.C. \& Universitat de Val{\`e}ncia \\
Edificio Institutos de Paterna, Apt 22085, E--46071 Valencia, Spain}

\newcommand{\AddrWur}{%
Institut f\"ur Theoretische Physik und Astronomie,
Universit\"at W\"urzburg\\
Am Hubland,
D--97074 W\"urzburg, Germany}

\newcommand{\AddrVienna}{%
Institut f\"ur Theoretische Physik, Universit\"at Wien, \\
A--1090 Vienna, Austria}

\def\rpv{$R_p \hspace{-1em}/\;\:\hspace{0.2em}$}

\begin{document}

\preprint{IFIC/09-09}
\title{LHC phenomenology of the $\mu\nu$SSM}

\author{A.~Bartl} \email{alfred.bartl@univie.ac.at} \affiliation{\AddrVienna}

\author{M.~Hirsch} \email{mahirsch@ific.uv.es} \affiliation{\AddrAHEP}

\author{S.~Liebler}\email{sliebler@physik.uni-wuerzburg.de}\affiliation{\AddrWur}

\author{W.~Porod} \email{porod@physik.uni-wuerzburg.de}\affiliation{\AddrWur}

\author{A.~Vicente} \email{Avelino.Vicente@ific.uv.es} \affiliation{\AddrAHEP}


\begin{abstract}
\vspace*{1cm}
The $\mu\nu$SSM has been proposed to solve simultaneously the 
$\mu$-problem of the MSSM and explain current neutrino data. The 
model breaks lepton number as well as R-parity. In this paper we study 
the phenomenology of this proposal concentrating on neutrino masses 
and the decay of the lightest supersymmetric particle (LSP). At first we
investigate in detail the $\mu\nu$SSM with one generation
of singlets, which can explain all neutrino data, once 1-loop corrections 
are taken into account. Then we study variations of the model with more  
singlets, which can generate all neutrino masses and mixings at tree-level. 
We calculate the decay properties of the lightest supersymmetric 
particle, assumed to be the lightest neutralino, taking into account 
all possible final states. The parameter regions where the LSP 
decays within the LHC detectors but with a length large enough to 
show a displaced vertex are identified. Decay branching ratios of 
certain final states show characteristic correlations with the 
measured neutrino angles, allowing to test the model at the LHC. 
Finally we briefly discuss possible signatures, which allow to 
distinguish between different R-parity breaking models. 

\end{abstract}

\maketitle

\section{Introduction}
\label{sec:Intro}

The Minimal Supersymmetric extension of the Standard Model (MSSM) 
\cite{Martin:1997ns} assumes that R-parity is conserved. R-parity 
($R_p$) \cite{Farrar:1978xj}, defined as $R_p = (-1)^{3 B + L + 2 S}$, 
was originally introduced to guarantee the stability of the proton 
in supersymmetric models \cite{Weinberg:1981wj,Sakai:1981pk}. 
It has two immediate consequences: First, the lightest supersymmetric 
particle (LSP) is stable. For cosmological reasons a stable LSP has 
to be electrically neutral, thus leading to the ``standard'' missing 
momentum signature of SUSY. Second, the MSSM with $R_p$, for the same 
reasons as the SM, predicts zero neutrino masses. 

Neutrino oscillation experiments have demonstrated that at least two 
neutrinos have non-zero mass \cite{Fukuda:1998mi,Ahmad:2002jz,Eguchi:2002dm}. 
Especially remarkable is that data from both atmospheric neutrino 
\cite{Ashie:2004mr} and from reactor neutrino measurements \cite{kl:2008ee} 
now show the characteristic $L/E$ dependence expected from oscillations, 
ruling out or seriously disfavouring other explanations of the observed 
neutrino deficits. It is fair to say that with the most recent data by 
the KamLAND \cite{kl:2008ee}, Super-K \cite{Hosaka:2006zd} and MINOS 
collaborations \cite{Adamson:2008zt} neutrino physics has finally entered 
the precision era. (For the latest evaluation of allowed neutrino parameter 
regions, see for example the updated fits in \cite{Schwetz:2008er}.) 

Non-zero neutrino masses can be easily included into the standard model 
by simply adding right-handed neutrinos, postulating the existence of 
a ($\Delta L=2$) dimension-5 operator \cite{Weinberg:1979sa} of 
unspecified origin or by introducing the seesaw mechanism with either 
fermionic singlets \cite{Minkowski:1977sc,seesaw,MohSen}, a scalar 
triplet \cite{Schechter:1980gr,Cheng:1980qt} or fermionic triplets 
\cite{Foot:1988aq}. Neutrino masses could be induced also at 1-loop-level
\cite{Ma:1998dn} or even at 2-loop order \cite{Cheng:1980qt,%
Zee:1985id,Babu:1988ki}.

While all of the neutrino mass models mentioned above can be 
easily supersymmetrized, there is also an entirely supersymmetric 
possibility to generate Majorana neutrino masses: R-parity violation 
\cite{Aulakh:1982yn,Hall:1983id,Ross:1984yg}. Different models of 
(lepton number violating) R-parity violation have been discussed in 
the literature. Within the MSSM particle content R-parity can be broken 
explicitly either by bilinear or by trilinear terms \cite{Hall:1983id}. 
The huge number of free parameters in the trilinear model, however,
makes such a general ansatz rather arbitrary. Attempts to reduce the
number of free parameters based on discrete symmetries have been discussed
in the literature \cite{Drees:1997id,Dreiner:2007uj,Dreiner:2006xw,Dreiner:2005rd}. 
One could also postulate that lepton number is conserved at the 
superpotential level, broken only by the vacuum expectation value (vev) of some
singlet field \cite{Masiero:1990uj}. This is called spontaneous 
R-parity violation (s-\rpv). \footnote{The first model to propose 
s-\rpv \cite{Aulakh:1982yn} used the left-sneutrinos to break $R_p$. 
This leads to a doublet Majoron, now ruled out by LEP data 
\cite{Amsler:2008zz}.} Bilinear \rpv (b-\rpv) can be understood as the low-energy 
limit of some s-\rpv model, where the new singlet fields are all 
decoupled. Such a bilinear model has only six new \rpv parameters 
and is thus more predictive than the general case with all possible 
bilinear and trilinear couplings.   
\footnote{In \cite{Chun:1998gp} it has been proposed that the trilinear 
parameters follow the hierarchies of the standard model Yukawa couplings. This 
is very similar to the pure bilinear model, which in the mass 
eigenstate basis has effective trilinear parameters given by 
products of bilinear parameters and down quark/charged lepton 
Yukawa couplings.} 

The phenomenology of \rpv SUSY has been studied extensively in 
the past, for reviews see \cite{Barbier:2004ez,Hirsch:2004he}. 
Neutrino masses have been calculated with trilinear couplings 
\cite{Hall:1983id} and for pure bilinear models 
\cite{Hempfling,Hirsch:2000ef,Diaz:2003as}. Neutrino angles are 
not predicted in either schemes, but can be easily fitted to 
experimental data. In bilinear schemes the requirement to correctly 
explain neutrino data fixes all \rpv couplings in sufficiently small 
intervals such that in some specific final states of the decays of 
the LSP correlations with neutrino angles appear. This has been 
shown for a (bino-dominated) neutralino LSP in \cite{Porod:2000hv}, 
for charged scalar LSPs in \cite{Hirsch:2002ys} and for sneutrino, 
chargino, gluino and squark LSPs in \cite{Hirsch:2003fe}. Such 
a tight connection between neutrino physics and LSP decays is 
lost, however, in the general trilinear-plus-bilinear case. 
(For some recent work on collider phenomenology in trilinear \rpv, 
see for example \cite{Dreiner:2008ca,Bernhardt:2008jz,Dreiner:2008rv,%
Allanach:2006st} and references in \cite{Barbier:2004ez}.)

The superpotential of the MSSM contains a mass term for the Higgs 
superfields, $\mu {\widehat H_d}{\widehat H_u}$. For phenomenological 
reasons this parameter $\mu$ must be of the order of the electro-weak 
scale. However, if there is a larger scale in the theory, like the 
grand unification scale, the natural value of $\mu$ lies at this large 
scale. This is, in short, the $\mu$-problem of the MSSM \cite{Kim:1983dt}. 
The Next-to-Minimal SSM (NMSSM) provides a solution to this problem 
\cite{Barbieri:1982eh,Nilles:1982dy}, at the cost of introducing a 
new singlet field. The vev of the singlet produces the $\mu$ term, once 
electro-weak symmetry is broken. (For some recent papers on the 
phenomenology of the NMSSM, see for example 
\cite{Djouadi:2008uw,Ellwanger:2005uu,Ellwanger:2005dv} 
and references therein.)

The $\mu\nu$SSM \cite{LopezFogliani:2005yw} proposes to use the same 
singlet superfield(s) which generate the $\mu$ term to also generate 
Dirac mass terms for the observed left-handed neutrinos. Lepton number 
in this approach is broken explicitly by cubic terms coupling only 
singlets. $R_p$ is broken also and Majorana neutrino masses are generated 
once electro-weak symmetry is broken. Two recent papers have studied the 
$\mu\nu$SSM in more detail. In \cite{Escudero:2008jg} the authors analyze 
the parameter space of the $\mu\nu$SSM, putting special emphasis on 
constraints arising from correct electro-weak symmetry breaking, avoiding 
tachyonic states and Landau poles in the parameters. The phenomenology 
of the $\mu\nu$SSM has been studied also in \cite{Ghosh:2008yh}. In this 
paper formulas for tree-level neutrino masses are given and decays of 
a neutralino LSP to two-body ($W$-lepton) final states have been 
calculated \cite{Ghosh:2008yh}.

We note that similar proposals have been discussed in the literature. 
\cite{Kitano:1999qb} studied a model in which the NMSSM singlet is 
coupled to (right-handed) singlet neutrino superfields. Effectively 
this leads to a model which is very similar to the NMSSM with explicit 
bilinear terms, as studied for example also in \cite{Abada:2006qn}. In 
\cite{Chemtob:2006ur} the authors propose a model similar to the 
$\mu\nu$SSM, but with only one singlet. 

In the present paper, we study the phenomenology of the $\mu\nu$SSM, 
extending previous work 
\cite{LopezFogliani:2005yw,Escudero:2008jg,Ghosh:2008yh}. We consider 
two different variations of the model. In its simplest form the $\mu\nu$SSM 
contains only one new singlet. This version produces one neutrino mass 
at tree-level, while the remaining two neutrinos receive mass at 
the loop-level. This feature is very similar to bilinear R-parity 
breaking, although as discussed below, the relative importance of 
the various loops is different for the explicit bilinear model and 
the $\mu\nu$SSM. As in the explicit bilinear model neutrino angles 
restrict the allowed range of \rpv parameters and correlations between 
certain ratios of decay branching ratios of the LSP and neutrino 
angles appear. In the second version we allow for $n$ singlets. 
Neutrino masses can then be fitted with tree-level physics only. 
However, many of the features of the one generation model remain at 
least qualitatively also in the $n$ singlet variants. LSP decays 
(for a bino or a singlino LSP) can be correlated with either the 
solar or atmospheric angle, thus allowing to construct explicit 
tests of the model for the LHC. 
In contrast to \cite{Ghosh:2008yh} we consider all kinematically
allowed final states. This does not only cover scenarios where 
two-body decays are important, but also those where three-body decays
are dominant. In addition we show that even in the scenarios where
two-body decay modes in singlet Higgs bosons dominante, the lifetime
can be such that the LSP decays outside the detector.

This paper is organized as follows. In the next section we outline the 
model, give the soft breaking terms, discuss the mass matrices and 
calculate approximate formulas for neutrino masses. We will not use 
the approximate formulas in our numerical analysis, but give them 
explicitly because they allow to understand in an easy way our 
numerical results qualitatively. In Section III we discuss existing 
constraints on the model space, apart from neutrino physics, and 
outline the properties of the ``standard'' points, which we will 
use in our numerical analysis. We then turn to the collider phenomenology 
of the model. In Section IV we study the one generation variant of 
the $\mu\nu$SSM. Decays of scalars are briefly discussed, before 
calculating decay properties of the neutralino LSP. Section V 
gives a discussion of the LSP phenomenology for the $n$ generation 
variant, although we will mainly focus on two generations.
Similarities and differences to the one generation model 
are discussed. In Section VI we then give a short, mostly qualitative 
discussion of possible signals which might give some hints which 
R-parity breaking model is indeed realized in nature, before 
closing with a short summary. Mass matrices and couplings are given 
in various appendices.

\section{Model basics}
\label{sec:ModBas}

In this section we introduce the model, work out its most important
properties related to phenomenology and neutrino masses and mixings. 
As explained in the introduction, we will consider the $n$ generations 
case in this section. Approximate formulas are then given for scalar 
masses for the one ($1$) $\hat\nu^c$-model and for neutrino masses 
for the $1$ and $2$ $\hat\nu^c$-model.

\subsection{Superpotential}
\label{subsec:superpot}

The model contains $n$ generations of right-handed neutrino
singlets. The superpotential can be written as
\begin{eqnarray} %
{\cal W} &=& h_U^{ij}\widehat Q_i \widehat U_j\widehat H_u
          +  h_D^{ij}\widehat Q_i\widehat D_j\widehat H_d
          +  h_E^{ij}\widehat L_i\widehat E_j\widehat H_d \nonumber
\\
        & + & h_{\nu}^{is} \widehat L_i \widehat \nu^c_s \widehat H_u
          - \lambda_s \widehat \nu^c_s \widehat H_d \widehat H_u
 +\frac{1}{3!}\kappa_{stu} \widehat \nu^c_s \widehat \nu^c_t \widehat \nu^c_u\quad.
\label{eq:Wsuppot}
\end{eqnarray}
The last three terms include the right-handed neutrino superfields,
which additionally play the role of the $\widehat \Phi$ superfield in
the NMSSM \cite{Barbieri:1982eh}, a gauge singlet with respect to the
SM gauge group. The model does not contain any terms with dimensions
of mass, providing a natural solution to the $\mu$-problem of the
MSSM. Please note, that as the number of right-handed neutrino
superfields can be different from $3$ we use the letters $s$, $t$
and $u$ as generation indices for the $\widehat{\nu}^c$ superfields and
reserve the letter $i$, $j$ and $k$ as generation indices for the
usual MSSM matter fields.

The last two terms in \eqref{eq:Wsuppot} explicitly
break lepton number and thus R-parity giving rise to neutrino
masses. Note that $\kappa_{stu}$ is completely symmetric in all its
indices.  In contrast to other models with R-parity violation, this
model does not need the presence of unnaturally small parameters with
dimensions of mass, like in bilinear R-parity breaking models
\cite{Hirsch:2004he}, and there is no Goldstone boson associated with
the breaking of lepton number
\cite{Chikashige:1980ui,Gelmini:1980re,Aulakh:1982yn}, since 
breaking of $R_p$ is done explicitly.

For practical purposes, it is useful to write the superpotential in
the basis where the right-handed neutrinos have a diagonal mass
matrix. Since their masses are induced by the $\kappa$ term in
\eqref{eq:Wsuppot}, this is equivalent to writing this term including
only diagonal couplings: 
\begin{equation}\label{eq:kappaconv}
\kappa_{stu} \widehat \nu^c_s \widehat \nu^c_t \widehat \nu^c_u 
\qquad\Longrightarrow\qquad \sum_{s=1}^n \kappa_s
(\widehat \nu^c_s)^3
\end{equation}

\subsection{Soft terms}
\label{subsec:soft}

The soft SUSY breaking terms of the model are
\begin{equation}
V_{soft} = V_{soft}^{MSSM - B_\mu} + V_{soft}^{singlets}\quad.
\end{equation}
$V_{soft}^{MSSM - B_\mu}$ contains all the usual soft terms of the MSSM
but the $B_\mu$-term
\begin{eqnarray}
 V_{soft}^{MSSM - B_\mu} & = & {m_Q^{ij}}^2 \tilde{Q}_i^{a \ast}
\tilde{Q}_j^a + {m_U^{ij}}^2 \tilde{U}_i \tilde{U}_j^{\ast} +
{m_D^{ij}}^2 \tilde{D}_i \tilde{D}_j^{\ast} + {m_L^{ij}}^2
\tilde{L}_i^{a \ast} \tilde{L}_j^a + {m_E^{ij}}^2 \tilde{E}_i
\tilde{E}_j^{\ast} \nonumber \\ & + & m_{H_d}^2 H_d^{a \ast} H_d^a +
m_{H_u}^2 H_u^{a \ast} H_u^a - \frac{1}{2} \big[ M_1 \tilde{B}^0
\tilde{B}^0 + M_2 \tilde{W}^c \tilde{W}^c + M_3 \tilde{g}^d
\tilde{g}^d + h.c. \big] \label{eq:softmssm} \\ & + & \epsilon_{ab}
\big[ T_U^{ij} \tilde{Q}_i^a \tilde{U}_j^\ast H_u^b + T_D^{ij}
\tilde{Q}_i^b \tilde{D}_j^\ast H_d^a + T_E^{ij} \tilde{L}_i^b
\tilde{E}_j^\ast H_d^a + h.c. \big] \nonumber
\end{eqnarray}
and $V_{soft}^{singlets}$ includes the new terms with singlets:
\begin{equation}\label{eq:softsing}
V_{soft}^{singlets} = {m_{\tilde{\nu}^c}^{st}}^2 \tilde{\nu}_s^c
 \tilde{\nu}_t^{c \ast} + \epsilon_{ab} \big[ T_{h_\nu}^{st}
 \tilde{L}_s^a \tilde{\nu}_t^c H_u^b - T_\lambda^s \tilde{\nu}_s^c
 H_d^a H_u^b+ h.c. \big] + \big[\frac{1}{3!}T_{\kappa}^{stu} \tilde{\nu}_s^c \tilde{\nu}_t^c
 \tilde{\nu}_u^c +h.c.\big]
\end{equation}
In these expressions the notation for the soft trilinear couplings
introduced in \cite{Skands:2003cj,Allanach:2008qq} is used.
Note that the rotation made in the superpotential does not necessarily
diagonalize the soft trilinear terms $T_{\kappa}^{stu}$ implying 
in general additional mixing between the right-handed sneutrinos.

\subsection{Scalar potential and its minimization}
\label{subsec:scalartadpole}

Summing up the different contributions, the scalar potential
considering only neutral fields reads
\begin{equation}
V = V_D + V_F + V_{soft}
\end{equation}
with
\begin{eqnarray}\label{eq:dterms}
 V_D & = &  \frac{1}{8} (g^2 + g'^2) \big( |H_u^0|^2 - |H_d^0|^2 -
\sum_{i=1}^3 |\tilde{\nu}_i|^2 \big)^2 \\ V_F &=& | h_\nu^{is}
\tilde{\nu}_i \tilde{\nu}_s^c - \lambda_s \tilde{\nu}_s^c H_d^0 |^2 +
| \lambda_s \tilde{\nu}_s^c H_u^0 |^2 + \sum_{i=1}^3 |h_\nu^{is}
\tilde{\nu}_s^c H_u^0 |^2 
+ \sum_{s=1}^n |h_\nu^{is}
\tilde{\nu}_i H_u^0 - \lambda_s H_u^0 H_d^0 + \frac{1}{2} \kappa_s
(\tilde{\nu}_s^c)^2 |^2\quad, \label{eq:fterms}
\end{eqnarray}
where summation over repeated indices is implied. 

This scalar potential determines the structure of the vacuum, inducing
vevs: 
\begin{equation}
\langle H_d^0 \rangle = \frac{v_d}{\sqrt{2}}, \hskip5mm
\langle H_u^0 \rangle = \frac{v_u}{\sqrt{2}}, \hskip5mm
\langle \tilde{\nu}_s^c \rangle = \frac{v_{Rs}}{\sqrt{2}}, \hskip5mm
\langle \tilde{\nu}_i \rangle = \frac{v_i}{\sqrt{2}}
\label{eq:ewsb}
\end{equation}
In particular, the vevs for the right-handed sneutrinos generate
effective bilinear couplings:
\begin{eqnarray}
 h_{\nu}^{is} \widehat L_i \widehat \nu^c_s \widehat H_u - \lambda_s
 \widehat \nu^c_s \widehat H_d \widehat H_u \qquad&\Longrightarrow&
 \qquad h_{\nu}^{is} \widehat L_i \frac{v_{Rs}}{\sqrt{2}} \widehat H_u
 - \lambda_s \frac{v_{Rs}}{\sqrt{2}} \widehat H_d \widehat H_u \equiv
 \epsilon_i \widehat L_i \widehat H_u - \mu \widehat H_d \widehat
 H_u
\end{eqnarray}
Since by electroweak symmetry breaking an effective $\mu$ term is generated, 
it is at the electroweak scale.
Minimizing the scalar potential gives the following tadpole equations at tree-level
\begin{eqnarray}
\frac{\partial V}{\partial v_d} &=& \frac{1}{8}(g^2 + g'^2)u^2 v_d +
m_{H_d}^2 v_d + \frac{1}{2} v_d \lambda_s \lambda_t^* v_{Rs} v_{Rt} +
\frac{1}{2}v_d v_u^2 \lambda_s \lambda_s^* \nonumber \\ && -
\frac{1}{8} v_{Rs}^2 v_u (\kappa_s \lambda_s^* + h.c.) -\frac{1}{4}
v_i (\lambda_s^* h_\nu^{it} v_{Rs} v_{Rt} + h.c.) -\frac{1}{4} v_u^2
v_i (\lambda_s^* h_\nu^{is} + h.c.) \nonumber \\ && - \frac{1}{2
\sqrt{2}} v_u v_{Rs}(T_\lambda^s + h.c.) = 0 \label{eq:tadpolevd} \\
\frac{\partial V}{\partial v_u} &=& -\frac{1}{8}(g^2 + g'^2)u^2 v_u +
m_{H_u}^2 v_u + \frac{1}{2} v_u \lambda_s \lambda_t^* v_{Rs} v_{Rt} +
\frac{1}{2}v_d^2 v_u \lambda_s \lambda_s^* - \frac{1}{8} v_{Ri}^2 v_d
(\kappa_s \lambda_s^* + h.c.) \nonumber \\ && + \frac{1}{8} v_i
v_{Rs}^2 (\kappa_s^* h_\nu^{is} + h.c.) -\frac{1}{2} v_d v_u v_i
(\lambda_s^* h_\nu^{is} + h.c.) + \frac{1}{2} v_u v_i v_j h_\nu^{is}
(h_\nu^{js})^* \nonumber \\ && + \frac{1}{2} v_u h_\nu^{is}
(h_\nu^{it})^* v_{Rs} v_{Rt} - \frac{1}{2 \sqrt{2}} v_d
v_{Rs}(T_\lambda^s + h.c.) + \frac{1}{2 \sqrt{2}} v_i v_{Rs}
(T_{h_\nu}^{is} + h.c.) = 0 \label{eq:tadpolevu} \\ \frac{\partial
V}{\partial v_i} &=& \frac{1}{8}(g^2 + g'^2)u^2 v_i + \frac{1}{2}
({m_L^2}_{ij} + {m_L^2}_{ji}) v_j - \frac{1}{4} v_d v_u^2 (
\lambda_s^* h_\nu^{is} + h.c. ) \nonumber \\ && + \frac{1}{8} v_{Rs}^2
v_u ( \kappa_s^* h_\nu^{is} + h.c. ) - \frac{1}{4} v_d ( \lambda_s^*
v_{Rs} v_{Rt} h_\nu^{it} + h.c. ) + \frac{1}{4} v_j ( v_{Rs} v_{Rt}
h_\nu^{is} (h_\nu^{jt})^* + h.c. ) \nonumber \\ && + \frac{1}{4} v_u^2
v_j ( h_\nu^{is} (h_\nu^{js})^* + h.c. ) + \frac{1}{2 \sqrt{2}} v_u
v_{Rs} (T_{h_\nu}^{is} + h.c. ) = 0 \label{eq:tadpoleneu} \\
\frac{\partial V}{\partial v_{Rs}} &=& {m_{\tilde{\nu}^c}^2}_{ss}
v_{Rs} - \frac{1}{4} v_d v_u v_{Rs} (\kappa_s \lambda_s^* + h.c.) +
\frac{1}{4} \kappa_s \kappa_s^* v_{Rs}^3 \nonumber \\ && + \frac{1}{4}
v_u v_{Rs} v_j ( \kappa_s^* h_\nu^{js} + h.c. ) + \frac{1}{4} (v_u^2 +
v_d^2) ( \lambda_s \lambda_t^* v_{Rt} + h.c. ) + \frac{1}{4} v_u^2 [
h_\nu^{js} (h_\nu^{jt})^* v_{Rt} + h.c. ] \nonumber \\ && +
\frac{1}{4} v_m v_n [ (h_\nu^{ms})^* h_\nu^{nt} v_{Rt} + h.c. ] -
\frac{1}{4} v_d v_j ( \lambda_t^* h_\nu^{js} v_{Rt} + \lambda_s^*
v_{Rt} h_\nu^{jt} + h.c. ) \nonumber \\ && - \frac{1}{2 \sqrt{2}} v_d
v_u (T_\lambda^s + h.c.) + \frac{1}{2 \sqrt{2}} v_u v_j
(T_{h_\nu}^{js} + h.c.) + \frac{1}{4\sqrt{2}} v_{Rt} v_{Ru} \big( T_\kappa^{stu} + h.c. \big)=0 \label{eq:tadpoleright}
\end{eqnarray}
with
\begin{eqnarray}
u^2 &=& v_d^2 - v_u^2 + v_1^2 + v_2^2 + v_3^2  \label{eq:deftkappa}
\end{eqnarray}
and there is no sum over the index $s$ in Equation \eqref{eq:tadpoleright}. 

As usual in R-parity breaking models with right-handed neutrinos, see
for example the model proposed in \cite{Masiero:1990uj}, it is
possible to explain the smallness of the $v_i$ in terms of the
smallness of the Yukawa couplings $h_\nu$, that generate Dirac masses
for the neutrinos. This can be easily seen from Equation \eqref{eq:tadpoleneu}, 
where both quantities are proportional. Moreover, as shown in
\cite{LopezFogliani:2005yw}, taking the limit $h_\nu \to 0$ and,
consequently, $v_i \to 0$, one recovers the tadpole equations of the
NMSSM, ensuring the existence of solutions to this set of equations.

\subsection{Masses of the neutral scalars and pseudoscalars}
\label{sec:scalarmass}

In this subsection we work out the main features of the neutral scalar sector 
mainly focusing on singlets. The complete mass matrices are given
in Appendix \ref{sec:MassMat}. We start with the one generation case
which closely resembles the NMSSM, considered, for example, in
\cite{Franke:1995xn,Miller:2003ay}. This already implies an upper
bound on the lightest doublet Higgs mass $m(h^0)$, where we will focus
on at the end of this subsection.
A correct description of neutrino physics
implies small values for the vevs $v_i$ of the left sneutrinos and
small Yukawa couplings $h_\nu$ as we will see later. Neglecting mixing
terms proportional to these quantities, the $(6\times 6)$ mass matrix of
the pseudoscalars in the basis $Im (H_d^0, H_u^0, \tilde{\nu}^c, \tilde{\nu}_i)$
given in Appendix \ref{sec:MassMat}, Equation \eqref{eq:masspseudoscalars}, can be
decomposed in two $\left(3\times 3\right)$ blocks. By using the
tadpole equations we obtain
\begin{equation}
M_{P^0}^2=\begin{pmatrix}M_{HH}^2&M_{HS}^2&0\\\left(M_{HS}^{2}\right)^T&M_{SS}^2&0\\0&0&M_{\tilde{L}\tilde{L}}^2\end{pmatrix}
\end{equation}
with
{\allowdisplaybreaks
\begin{align}\nonumber
M_{HH}^2&=\begin{pmatrix}\left(\Omega_1+\Omega_2\right)\frac{v_u}{v_d}&\Omega_1+\Omega_2\\\Omega_1+\Omega_2&\left(\Omega_1+\Omega_2\right)\end{pmatrix},\qquad
M_{HS}^2=\begin{pmatrix}\left(-2\Omega_1+\Omega_2\right)\frac{v_u}{v_R}\\\left(-2\Omega_1+\Omega_2\right)\frac{v_d}{v_R}\end{pmatrix}\\
M_{SS}^2&=\left(4\Omega_1+\Omega_2\right)\frac{v_dv_u}{v_R^2}-3\Omega_3,\qquad
\left(M_{\tilde{L}\tilde{L}}^2\right)_{ij}=\tfrac{1}{2}\left(m_{\tilde{L}}^2\right)_{ij}+\tfrac{1}{2}\left(m_{\tilde{L}}^2\right)_{ji}+\delta_{ij}\left[\tfrac{1}{8}\left(g^2+g'^2\right)u^2\right]\quad,
\end{align}
}%
where  $u^2$ is defined in Equation \eqref{eq:deftkappa}. The parameters $\Omega_i$ 
are defined as:
{\allowdisplaybreaks
\begin{align}
\Omega_1 &=\frac{1}{8}\left(\lambda\kappa^*+\lambda^*\kappa\right)v_R^2,\qquad \Omega_2 =\frac{1}{2\sqrt{2}}\left(T_\lambda+T_\lambda^*\right)v_R,\qquad
\Omega_3 =\frac{1}{4\sqrt{2}}\left(T_\kappa+T_\kappa^*\right)v_R
\end{align}
}%
The upper $\left(3\times 3\right)$ block contains the mass terms for
$Im(H_d)$, $Im(H_u)$ and $Im(\tilde{\nu}^c)$ and we get analytic
expressions for the eigenvalues:
{\allowdisplaybreaks
\begin{align}
\nonumber m^2(P^0_1)&=0\\
\nonumber m^2(P^0_2)&=\frac{1}{2}\left(\Omega_1+\Omega_2\right)\left(\frac{v_d}{v_u}+\frac{v_u}{v_d}+\frac{v_dv_u}{v_R^2}\right)-\frac{3}{2}\Omega_3-\sqrt{\Gamma}\\
\nonumber
m^2(P^0_3)&=\frac{1}{2}\left(\Omega_1+\Omega_2\right)\left(\frac{v_d}{v_u}+\frac{v_u}{v_d}+\frac{v_dv_u}{v_R^2}\right)-\frac{3}{2}\Omega_3+\sqrt{\Gamma}\\
\nonumber
&\text{with} \quad \Gamma=\left(\frac{1}{2}\left(\Omega_1+\Omega_2\right)\left(\frac{v_d}{v_u}+\frac{v_u}{v_d}+\frac{v_dv_u}{v_R^2}\right)-\frac{3}{2}\Omega_3\right)^2\\
&\qquad \qquad +3\left(\Omega_1+\Omega_2\right)\Omega_3\left(\frac{v_d}{v_u}+\frac{v_u}{v_d}\right)-9\Omega_1\Omega_2\left(\frac{v_R^2}{v_d^2}+\frac{v_R^2}{v_u^2}\right)
\end{align}
}%
The first eigenvalue corresponds to the Goldstone boson due to spontaneous 
symmetry breaking. To get only positive eigenvalues for the physical 
states, the condition
\begin{equation}
\Omega_3<\frac{v_dv_u}{v_R^2}\frac{3\Omega_1\Omega_2}{\Omega_1+\Omega_2}=:f_1\left(\Omega_2\right) \end{equation}
has to be fulfilled, implying that $T_\kappa$ has in general the
opposite sign of $v_R$. Additional constraints on the parameters are
obtained from the positiveness of the squared masses of the neutral scalars.
Taking the scalar mass matrix from Appendix \ref{sec:MassMat}, Equation \eqref{eq:massscalars}, 
in the basis $Re (H_d^0, H_u^0,
\tilde{\nu}^c, \tilde{\nu}_i)$ in the same limit as above we obtain
\begin{equation}
M_{S^0}^2=\begin{pmatrix}M_{HH}^2&M_{HS}^2&0\\\left(M_{HS}^{2}\right)^T&M_{SS}^2&0\\0&0&M_{\tilde{L}\tilde{L}}^2\end{pmatrix}
\end{equation}
with
{\allowdisplaybreaks
\begin{align}\nonumber
 M_{HH}^2&=\begin{pmatrix}\left(\Omega_1+\Omega_2\right)\frac{v_u}{v_d}+\Omega_6\frac{v_d}{v_u}&-\Omega_1-\Omega_2-\Omega_6+\Omega_4\\-\Omega_1-\Omega_2-\Omega_6+\Omega_4&\left(\Omega_1+\Omega_2\right)\frac{v_d}{v_u}+\Omega_6\frac{v_u}{v_d}\end{pmatrix},\qquad
 M_{HS}^2=\begin{pmatrix}\left(-2\Omega_1-\Omega_2\right)\frac{v_u}{v_R}+\Omega_4\frac{v_R}{v_u}\\\left(-2\Omega_1-\Omega_2\right)\frac{v_d}{v_R}+\Omega_4\frac{v_R}{v_d}\end{pmatrix}\\
 M_{SS}^2&=\Omega_2\frac{v_dv_u}{v_R^2}+\Omega_3+\Omega_5,\quad
 \left(M_{\tilde{L}\tilde{L}}^2\right)_{ij}=\tfrac{1}{4}\left(g^2+g'^2\right)v_iv_j+\tfrac{1}{2}\left(m_{\tilde{L}}^2\right)_{ij}+\tfrac{1}{2}\left(m_{\tilde{L}}^2\right)_{ji}+\delta_{ij}\left[\tfrac{1}{8}\left(g^2+g'^2\right)u^2\right]
\end{align}
}%
using the additional parameters
{\allowdisplaybreaks
\begin{align}
\Omega_4 &=\lambda\lambda^*v_dv_u>0,\qquad \Omega_5 =\frac{1}{2}\kappa\kappa^*v_R^2>0,\qquad
\Omega_6 =\frac{1}{4}\left(g^2+g'^2\right)v_dv_u>0\quad.
\end{align}
}%
An analytic determination of the eigenvalues is possible but not very
illuminating.  However, one can use the following theorem: A symmetric
matrix is positive definite, if all eigenvalues are positive and this
is equal to the positiveness of all principal minors (Sylvester
criterion). This results in the following three conditions
\begin{eqnarray}
\nonumber 
0 &<&
\left(\Omega_1+\Omega_2\right)\frac{v_u}{v_d}+\Omega_6\frac{v_d}{v_u}\\\nonumber
0 &<& \left(\Omega_1+\Omega_2\right)\left(\Omega_6\left(\frac{v_d^2}{v_u^2}+\frac{v_u^2}{v_d^2}\right)-2\Omega_6+2\Omega_4\right)+2\Omega_4\Omega_6-\Omega_4^2\\
0 &<& \Omega_3-f_2\left(\Omega_2\right)\quad,
\end{eqnarray}
where $f_2(\Omega_2)$ is given by:
{\allowdisplaybreaks
\begin{align}
\nonumber f_2\left(\Omega_2\right) &= \frac{\Sigma_1}{\Sigma_2} \qquad\text{with}\\
\nonumber&\qquad \Sigma_1=\left(\Omega_1+\Omega_2\right)\Omega_5\left(-2\Omega_4+2\Omega_6\right)+\left(\Omega_4^2-2\Omega_4\Omega_6\right)\Omega_5\\
\nonumber&\qquad\qquad +\left(\Omega_1+\Omega_2\right)\Omega_4^2v_R^2\left(\frac{v_d}{v_u^3}+\frac{v_u}{v_d^3}\right)+\left(4\Omega_1^2+3\Omega_1\Omega_2\right)\Omega_6\frac{1}{v_R^2}\left(\frac{v_d^3}{v_u}+\frac{v_u^3}{v_d}\right)\\
\nonumber&\qquad\qquad -\left(\Omega_1+\Omega_2\right)\Omega_5\Omega_6\left(\frac{v_d^2}{v_u^2}+\frac{v_u^2}{v_d^2}\right)+2\left(\Omega_1+\Omega_2-\Omega_4+2\Omega_6\right)\Omega_4^2\frac{v_R^2}{v_dv_u}\\
\nonumber&\qquad\qquad -2\left(2\Omega_1+\Omega_2\right)\left(2\Omega_1+2\Omega_2-\Omega_4+2\Omega_6\right)\Omega_4\left(\frac{v_d}{v_u}+\frac{v_u}{v_d}\right)\\
\nonumber&\qquad\qquad +\left[16\Omega_1^3+8\left(4\Omega_2-\Omega_4+\Omega_6\right)\Omega_1^2+10\Omega_1\Omega_2\left(2\Omega_2-\Omega_4+\Omega_6\right)\right.\\
\nonumber&\left.\qquad\qquad\quad +\Omega_2\left(2\Omega_2-\Omega_4\right)\left(2\Omega_2-\Omega_4+2\Omega_6\right)\right]\frac{v_dv_u}{v_R^2}\\
&\qquad
\nonumber \Sigma_2=\left(\Omega_1+\Omega_2\right)\Omega_6\left(\frac{v_d^2}{v_u^2}+\frac{v_u^2}{v_d^2}\right)+2\left(\Omega_1+\Omega_2\right)\left(\Omega_4-\Omega_6\right)\\
&\qquad\qquad +2\Omega_4\Omega_6-\Omega_4^2
\label{formelfuerbeding}
\end{align}
}%
The first two conditions are in general fulfilled, but for special
values of $\tan\beta$ or $\lambda$. Putting all the above together we get
the following conditions:
\begin{equation}
f_2(\Omega_2)<\Omega_3<f_1(\Omega_2)
\end{equation}
It turns out that by taking a negative value of $\Omega_3$ ($\propto
T_\kappa$) near $f_2(\Omega_2)$ one obtains a very light singlet scalar,
whereas for a value of $\Omega_3$ near $f_1(\Omega_2)$ one gets a very
light singlet pseudoscalar. In between one finds a value of
$\Omega_3$, where both particles have the same mass. This discussion is
comparable to formula (37) in \cite{Miller:2003ay} for the
NMSSM. Moreover, a small mass of the singlet scalar and/or
pseudoscalar comes always together with a small mass of the singlet
fermion.

In the $n$ generation case similar result holds as long as
$T_{\kappa}$ and $m^2_{{\tilde \nu}^c}$ do not have off-diagonal
entries compared to $\kappa$. Inspecting Equations \eqref{SSscalar} and
\eqref{SSpseudoscalar} it is possible to show that the singlet scalars
and pseudoscalars can be heavy by appropriately chosen values for the
off-diagonal entries of $T_\kappa$ while keeping at the same time the
singlet fermions relatively light, as will be discussed later. As
pointed out in \cite{Escudero:2008jg}, the NMSSM upper bound on the
lightest doublet Higgs mass of about $\sim 150$ GeV, which also applies
in the $\mu\nu$SSM, can be relaxed to ${\cal O}(300)$ GeV, if one
does not require perturbativity up to the GUT scale.

\subsection{Neutrino masses}
\label{subsec:neutrinomass}

In the basis
\begin{equation}
\big( \psi^0 \big)^T = \big( {\tilde B}^0, {\tilde W}_3^0, {\tilde H}_d^0, {\tilde H}_u^0, \nu_s^c, \nu_i \big)
\label{eq:defbasis}
\end{equation}
the mass matrix of the neutral fermions, see Appendix 
\ref{subsec:neutralinos}, has the structure
\begin{equation}
{\cal M}_n =
\left(\begin{array}{cc}
\mathbf{M_H} & \mathbf{m} \\
\mathbf{m}^T & \mathbf{0}
\end{array} \right)\quad.
\label{eq:neutralmass}
\end{equation}
Here $\mathbf{M_H}$ is the submatrix including the heavy states, which
consists of the usual four neutralinos of the MSSM and $n$ generations of
 right-handed neutrinos. The matrix $\mathbf{m}$ mixes the heavy
 states with the left-handed neutrinos and contains the R-parity
 breaking parameters.

The matrix ${\cal M}_n$ can be diagonalized in the standard way:
\begin{equation}
\widehat{\cal M}_n = {\cal N}^* {\cal M}_n {\cal N}^{-1}
\label{eq:diagmass}
\end{equation}
As it is well known, the smallness of neutrino masses allows to 
find the effective neutrino mass matrix in a seesaw approximation
\begin{equation}\label{eq:defmnueff}
\boldsymbol{m_{\nu\nu}^{\rm eff}} = - \mathbf{m}^T \cdot \mathbf{M_H}^{-1} \mathbf{m}
 = - \xi \cdot \mathbf{m}\quad,
\end{equation}
where the matrix $\xi$ contains the small expansion parameters which
characterize the mixing between the neutrino sector and the heavy
states.

Since the superpotential explicitly breaks lepton number, at least one
mass for the left-handed neutrinos is generated at tree-level. In the case
of the $1$ $\hat\nu^c$-model the other neutrino masses are generated
at loop-level. With more than one
generation of right-handed neutrinos additional neutrino masses
are generated at tree-level, resulting in different possibilities 
to fit the neutrino oscillation data, see the discussion below.

\subsubsection{One generation of right-handed neutrinos}
\label{subsec:1genneut}

With only one generation of right-handed neutrinos the matrix $\xi$ is
given by
\begin{equation}\label{xi}
\xi_{ij} = K_{\Lambda}^j \Lambda_i - \frac{1}{\mu} \epsilon_i \delta_{j3}\quad,
\end{equation}
where the $\epsilon_i$ and $\Lambda_i$ parameters are defined as
\begin{eqnarray}
\epsilon_i & = & \frac{1}{\sqrt{2}} h_{\nu}^{i}v_{R} \\
\Lambda_i &=& \mu v_i + \epsilon_i v_d
\end{eqnarray}
and $K^j_{\Lambda}$ as 
\begin{eqnarray}\label{defK1}
K_\Lambda^1 &=& \frac{2 g' M_2 \mu}{m_\gamma}a \nonumber \\
K_\Lambda^2 &=& -\frac{2 g M_1 \mu}{m_\gamma}a \nonumber \\
K_\Lambda^3 &=& \frac{m_\gamma}{8 \mu Det(M_H)}(\lambda^2 v_d v^2 + 2 M_R \mu v_u) \nonumber\\
K_\Lambda^4 &=& -\frac{m_\gamma}{8 \mu Det(M_H)}(\lambda^2 v_u v^2 + 2 M_R \mu v_d) \nonumber \\
K_\Lambda^5 &=& \frac{\lambda m_\gamma}{4 \sqrt{2} Det(M_H)}(v_u^2 - v_d^2)
\end{eqnarray}
with
\begin{eqnarray}
m_\gamma &=& g^2M_1 +g'^2 M_2, \hskip10mm
v^2 = v_d^2 + v_u^2, \hskip10mm
M_R = \frac{1}{\sqrt{2}} \kappa v_R \\
a &=& \frac{m_\gamma}{4 \mu Det(M_H)}(v_d v_u \lambda^2 + M_R \mu)
\end{eqnarray}
and $Det(M_H)$ is the determinant of the $(5 \times 5)$ mass matrix of
the heavy states
\begin{equation}
Det(M_H) = \frac{1}{8} m_\gamma (\lambda^2 v^4 + 4 M_R \mu v_d v_u) 
- M_1 M_2 \mu(v_d v_u \lambda^2 + M_R \mu)\quad.
\end{equation}
Using these expressions the tree-level effective neutrino mass matrix
takes the form
\begin{equation}\label{eq:effone}
(\boldsymbol{m_{\nu\nu}^{\rm eff}})_{ij} = a \Lambda_i \Lambda_j\quad.
\end{equation}
The projective form of this mass matrix 
implies that only one neutrino gets a tree-level mass, while the other
two remain massless. Therefore, as in models with bilinear R-parity
violation \cite{Romao:1999up,Hirsch:2000ef,Diaz:2003as} 1-loop
corrections are needed in order to correctly explain the oscillation
data, which require at least one additional massive neutrino.
The absolute scale of neutrino mass constrains the $\vec{\Lambda}$ and 
$\vec{\epsilon}$ parameters, which have to be small. For typical SUSY masses 
order ${\cal O}(100\hskip1mm{\rm GeV})$, one finds $|\vec\Lambda|/\mu^2 \sim
10^{-7}$--$10^{-6}$ and $|\vec{\epsilon}|/\mu \sim
10^{-5}$--$10^{-4}$. This implies a ratio of
$|\vec{\epsilon}|^2/|\vec{\Lambda}|\sim 10^{-3}$--$10^{-1}$.

\begin{figure}[t]
\begin{center}
\includegraphics[width=0.5\textwidth]{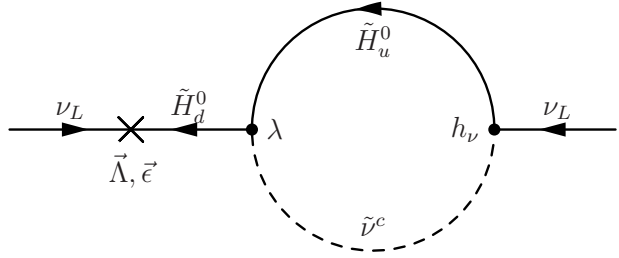}
\caption{Example of one 1-loop correction to the effective neutrino mass matrix
involving the singlet scalar/pseudoscalar.}
\label{loopdiag}
\end{center}
\end{figure}

\begin{figure}[t]
\begin{center}
\includegraphics{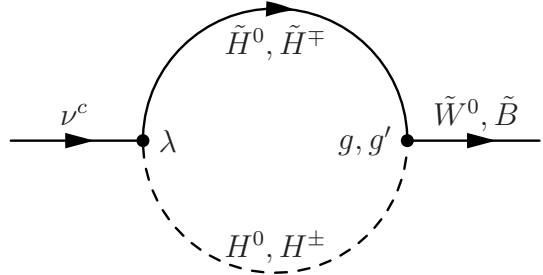}
\caption{1-loop mixing between gauginos and the right-handed neutrinos.}
\label{loopdiagMuRGaugino}
\end{center}
\end{figure}

General formulas for the 1-loop contributions can be found in
\cite{Hirsch:2000ef} and adjusted to the $\mu\nu$SSM with appropriate
changes in the index ranges for neutralinos and scalars. Important
contributions to the neutrino mass matrix are due to $b-{\tilde b}$
and $\tau-{\tilde\tau}$ loops as in the models with b-\rpv
\cite{Diaz:2003as}. In addition there are two new important 
contributions: (i) loops containing the singlet scalar and singlet
pseudoscalar shown in Figure \ref{loopdiag}.  As shown in
\cite{Hirsch:1997vz,Hirsch:1997dm,Grossman:1997is}, the sum of both
contributions is proportional to the squared mass difference
$\Delta_{12} = m_R^2 - m_I^2 \propto \kappa^2 v_R^2$ between the 
singlet scalar and pseudoscalar mass eigenstates. Note that this splitting 
can be much larger than the corresponding ones for the left sneutrinos. 
Thus the sum of both loops can be more important than $b
- \tilde b$ and $\tau - \tilde \tau$ loops in the current model. 
(ii) At loop-level a direct mixing between the
right-handed neutrinos and the gauginos is possible which is zero at
tree-level, see Figure \ref{loopdiagMuRGaugino}. 

\subsubsection{$n$ generations of right-handed neutrinos}
\label{subsec:ngenneut}

In this class of models with $n>1$ one can explain the neutrino data
using the tree-level neutrino mass matrix only. In general one finds
that the loop corrections are small if the conditions at the end of
this section are fulfilled.

For the sake of simplicity, let us consider two generations of
right-handed neutrinos which contains all relevant features.  The
matrix $\xi$ in Equation \eqref{eq:defmnueff} takes the form
\begin{equation}\label{xi2}
\xi_{ij} = K_{\Lambda}^j \Lambda_i + K_{\alpha}^j \alpha_i 
- \frac{\epsilon_i}{\mu} \delta_{j3}
\end{equation}
with
\begin{eqnarray}
\epsilon_i & = & \frac{1}{\sqrt{2}} h_{\nu}^{is}v_{Rs} \\
\Lambda_i &=& \mu v_i + \epsilon_i v_d \\
\alpha_i &=& v_u ( \lambda_2 h_\nu^{i1} -\lambda_1  h_\nu^{i2})\quad.
\label{defalpha}
\end{eqnarray}
The $K_\Lambda$ and $K_\alpha$ coefficients are:
\begin{eqnarray}\label{defK2}
K_\Lambda^1 = \frac{2 g' M_2 \mu}{m_\gamma}a, 
&\hskip1mm & K_\alpha^1 = \frac{2 g' M_2 \mu}{m_\gamma}b \nonumber \\
K_\Lambda^2 = -\frac{2 g M_1 \mu}{m_\gamma}a, 
&\hskip1mm & K_\alpha^2 = -\frac{2 g M_1 \mu}{m_\gamma}b \nonumber \\
K_\Lambda^3 = \frac{m_\gamma}{8 \mu Det(M_H)} \left[ v_d v^2 (M_{R1} \lambda_2^2 + M_{R2} \lambda_1^2) + 2 v_u M_{R1} M_{R2} \mu \right],
&\hskip1mm & K_\alpha^3 = \frac{b}{m_\gamma (v_u^2 - v_d^2)} (m_\gamma v^2 v_u - 4 M_1 M_2 \mu v_d) \nonumber \\
K_\Lambda^4 = - \frac{m_\gamma}{8 \mu Det(M_H)} \left[ v_u v^2 (M_{R1} \lambda_2^2 + M_{R2} \lambda_1^2) + 2 v_d M_{R1} M_{R2} \mu \right],
&\hskip1mm & K_\alpha^4 = \frac{b}{m_\gamma (v_u^2 - v_d^2)} (m_\gamma v^2 v_d - 4 M_1 M_2 \mu v_u) \nonumber \\
K_\Lambda^5 = \frac{M_{R2} \lambda_1 m_\gamma}{4 \sqrt{2} Det(M_H)} (v_u^2 - v_d^2), 
&\hskip1mm & K_\alpha^5 = - \sqrt{2} \lambda_2 c - \frac{4 Det_0 v_{R1}}{\mu m_\gamma (v_u^2 - v_d^2)}b \nonumber  \\ 
K_\Lambda^6 = \frac{M_{R1} \lambda_2 m_\gamma}{4 \sqrt{2} Det(M_H)} (v_u^2 - v_d^2), 
&\hskip1mm & K_\alpha^6 = \sqrt{2} \lambda_1 c - \frac{4 Det_0 v_{R2}}{\mu m_\gamma (v_u^2 - v_d^2)}b
\end{eqnarray}
The effective neutrino mass matrix reads as
\begin{equation}\label{eq:efftwo}
(\boldsymbol{m_{\nu\nu}^{\rm eff}})_{ij} = a \Lambda_i \Lambda_j + b (\Lambda_i \alpha_j + \Lambda_j \alpha_i) + c \alpha_i \alpha_j
\end{equation}
with
\begin{eqnarray}
a &=& \frac{m_\gamma}{4 \mu Det(M_H)} (M_{R1} \lambda_2^2 v_u v_d + M_{R2} \lambda_1^2 v_u v_d + M_{R1} M_{R2} \mu)  \\
b &=& \frac{m_\gamma}{8 \sqrt{2} \mu Det(M_H)} (v_u^2 - v_d^2)(M_{R1} v_{R1} \lambda_2 - M_{R2} v_{R2} \lambda_1) \\
c &=& - \frac{1}{16 \mu^2 Det(M_H)} \big[\mu^2(m_\gamma v^4 - 8 M_1 M_2 \mu v_u v_d) + 4 Det_0 (M_{R1} v_{R1}^2 + M_{R2} v_{R2}^2) \big] 
\end{eqnarray}
using $M_{Rs} = \frac{1}{\sqrt{2}} \kappa_s v_{Rs}$ and the determinant of the $(6\times6)$ mass matrix of the heavy states is
\begin{equation}
Det(M_H) = \frac{1}{8} \big[(M_{R2} \lambda_1^2 + M_{R1} \lambda_2^2)(m_\gamma v^4 - 8 M_1 M_2 \mu v_u v_d) + 8 M_{R1} M_{R2} Det_0 \big] 
\end{equation}
with $Det_0$ being the determinant of the usual MSSM neutralino mass matrix
\begin{equation}
 Det_0 = \frac{1}{2} m_\gamma \mu v_d v_u - M_1 M_2 \mu^2 \quad.
\end{equation}

The mass matrix in Equation \eqref{eq:efftwo} has two nonzero
eigenvalues and therefore the loop corrections are not needed to
explain the experimental data. Two different options arise:
\begin{itemize}
\item $\vec \Lambda$ generates the atmospheric mass scale, $\vec
\alpha$ the solar mass scale
\vspace{-1mm}
\item $\vec \alpha$ generates the atmospheric mass scale, $\vec
\Lambda$ the solar mass scale
\end{itemize}

In both cases one obtains in general a hierarchical spectrum.
A strong fine-tuning would be necessary to generate an inverted
hierarchy which is not stable against small variations of the parameters
or radiative corretions.
Moreover the absolute scale of neutrino mass requires both $|\vec
\Lambda|/\mu^2$ and $|\vec \alpha|/\mu$ to be small. For typical SUSY
masses order ${\cal O}(100\hskip1mm{\rm GeV})$ we find in the first
case $|\vec\Lambda|/\mu^2 \sim 10^{-7}$--$10^{-6}$ and
$|\vec\alpha|/\mu \sim 10^{-9}$--$10^{-8}$. In the second case we find
$|\vec\Lambda|/\mu^2 \sim 10^{-8}$--$10^{-7}$ and $|\vec\alpha|/\mu
\sim 10^{-8}$--$10^{-7}$. The ratios including $\vec{\epsilon}$ or
$\vec{\alpha}$ are much smaller than those in the $1$
$\widehat{\nu}^c$ case.  We find that 1-loop corrections 
to \eqref{eq:efftwo} are negligible if
\begin{eqnarray}
\frac{|\vec \alpha|^2}{|\vec \Lambda|}  \lesssim  10^{-3}  \qquad\,\, \mathrm{and}
\qquad \,\,
\frac{|\vec \epsilon|^2}{|\vec \Lambda|}  \lesssim  10^{-3} \label{seccond}
\end{eqnarray}
are fulfilled.
Note that the mixing of the neutrinos with the higgsinos, given by the
third column in the matrix $\xi$ in Equation \eqref{xi2}, depends not
only on $\alpha_i$ but also on $\epsilon_i$.  This leads to 1-loop
corrections to the neutrino mass matrix with pieces proportional to
the $\epsilon_i$ parameters, as it also happens in the $1$
$\widehat{\nu}^c$-model. Therefore, both conditions in
Equation \eqref{seccond} need to be fulfilled.
Finally, in models with more generations of right-handed neutrinos
there will be more freedom due to additional contributions to the
neutrino mass matrix. For example, the case of three generations is
discussed in \cite{Ghosh:2008yh}, where the additional freedom is 
also used to generate an inverted hierarchy for the neutrino masses.

\section{Choice of the parameters and experimental constraints}
\label{sec:Strate}

In the subsequent sections we work out collider signatures for various
scenarios. To facilitate the comparison with existing studies we adopt
the following strategy: We take existing study points and augment them
with the additional model parameters breaking R-parity. These points
are SPS1a' \cite{AguilarSaavedra:2005pw}, SPS3, SPS4, SPS9
\cite{Allanach:2002nj} and the ATLAS SU4 point \cite{Aad:2009wy}.
SPS1a' contains a relative light spectrum so that at LHC a high
statistic can be achieved, SPS3 has a somewhat heavier spectrum and in
addition the lightest neutralino and the lighter stau are close in
mass which affects also the R-parity violating decays of the lightest
neutralino. SPS4 is chosen because of the large $\tan\beta$ value and
SPS9 is an AMSB scenario where not only the lightest neutralino but
also the lighter chargino has dominant R-parity violating decay modes.
In all these points the lightest neutralino is so heavy that it can
decay via two-body modes, as long as it's not a light $\nu^c$.
In contrast for the SU4 point all two-body decay modes (at
tree-level) are kinematically forbidden. As the parameters of these
points are given at different scales we use the program {\tt SPheno}
\cite{Porod:2003um} to evaluate them at $Q=m_Z$ where we add the
additional model parameters. Note that we allow $\mu$ to depart from 
their standard SPS values to be consistent with the LEP bounds on 
Higgs masses, discussed below.

The additional model parameters are subject to theoretical and
experimental constraints. In \cite{Escudero:2008jg} the question
of color and charge breaking minimas, perturbativity up to the GUT
scale as well as the question of tachyonic states for the neutral scalar
and pseudoscalars have been investigated . The last issue has already
been addressed in Section \ref{sec:scalarmass} where we derived
conditions on the parameters. By choosing the coupling constants
$\lambda, \kappa<0.6$ in the $1$ $\widehat{\nu}^c$-model
and $\lambda_s,\kappa_s<0.5$ in the $2$ $\widehat{\nu}^c$-model, 
perturbativity up to the GUT scale is guaranteed \cite{Escudero:2008jg}.
Note, that choosing
somewhat larger values for $\lambda$ and/or $\kappa$ up to $1$ does not
change any of the results presented below. We also address the
question of color and charge breaking minimas by choosing
$\lambda_{s}>0$, $\kappa_{s}>0$, $T_{\lambda}^{s}>0$, $T_\kappa^{stu}
<0$, whereas the Yukawa couplings $h_\nu^{is}$ can either be positive
or negative, but those values are small $<\mathcal{O}(10^{-6})$ due to
constraints from neutrino physics. Our $T_{h_\nu}^{is}$ are negative,
so the condition (2.8) of \cite{Escudero:2008jg} is easy to fulfill.

Concerning experimental data we take the following constraints into
account:
\begin{itemize}
\item We check that the neutrino data are fulfilled within the $2$-$\sigma$
      range given in Table \ref{tab:neutrinos} taken from ref.\
      \cite{Schwetz:2008er} if not stated otherwise. These data can
      easily be fitted using the effective neutrino mass matrices
      given in Section \ref{subsec:neutrinomass}.
\item Breaking lepton number implies that flavour violating decays of
     the leptons like $\mu\to e \gamma$ are possible, where strong
     experimental bounds exist \cite{Amsler:2008zz}.  However, in the
     model under study it turns out that these bounds are automatically
     fulfilled once the constraints from neutrino physics are taken
     into account similar to the case of models with bilinear R-parity
     breaking \cite{Carvalho:2002bq}.
\item Bounds on the masses of the Higgs bosons
      \cite{Schael:2006cr,Amsler:2008zz}. For this purpose we have
      added the dominant 1-loop correction to the (2,2) entry of the
      scalar mass matrix in Appendix \ref{subsec:scalars}. Moreover,
      we have checked in the $1$ ${\hat \nu}^c$-model with the help of
      the program NMHDECAY \cite{Ellwanger:2005dv} that in the NMSSM
      limit the experimental constraints are fulfilled.
\item Constraints on the chargino and charged slepton masses given by
the PDG \cite{Amsler:2008zz}.
\item The bounds on squark and gluino masses from TEVATRON
      \cite{Amsler:2008zz} are automatically fulfilled by our choices
      of the study points.
\end{itemize}

\begin{table}[t]
\begin{center}
\begin{tabular}{|c|c|c|}
\hline
parameter & best fit & $2$-$\sigma$ \\
\hline \hline
 $\Delta m_{21}^2[10^{-5}\text{eV}^2]$ & $7.65^{+0.23}_{-0.20}$ & $7.25-8.11$\\
 $|\Delta m_{31}^2|[10^{-3}\text{eV}^2]$ & $2.40^{+0.12}_{-0.11}$ & $2.18-2.64$ \\
 $\sin^2\theta_{12}$ & $0.304^{+0.022}_{-0.016}$ & $0.27-0.35$ \\
 $\sin^2\theta_{23}$ & $0.50^{+0.07}_{-0.06}$ & $0.39-0.63$\\
 $\sin^2\theta_{13}$ & $0.01^{+0.016}_{-0.011}$ & $\leq 0.040$\\
\hline
\end{tabular}
\vspace{-2mm}
\caption{\label{tab:neutrinos} Best-fit values with $1$-$\sigma$ errors
and $2$-$\sigma$ intervals ($1$ d.o.f.) taken from \cite{Schwetz:2008er}.
In the following we will refer to these angles as $\theta_{12} = \theta_{sol}, \theta_{23}=\theta_{atm}$ and $\theta_{13}=\theta_R$.}
\end{center}
\end{table}

The smallness of the \rpv parameters guarantees that the direct
production cross sections for the SUSY particles are very similar to
the corresponding MSSM/NMSSM values. Note that for low values of
$\lambda$ the singlet states are decoupled from the rest of the particles, 
leading to low production rates.

\section{Phenomenology of the $1$ $\widehat{\nu}^c$-model}
\label{sec:phenoone}

In this section we discuss the phenomenology of the $1$
$\widehat{\nu}^c$-model, including mass hierarchies, mixings in the
scalar and fermionic sectors, decays of the scalar and fermionic
states and the correlations between certain branching
ratios and the neutrino mixing angles.

\begin{figure}[hbt]
\begin{center}
\vspace{0mm}
\includegraphics[width=80mm,height=60mm]{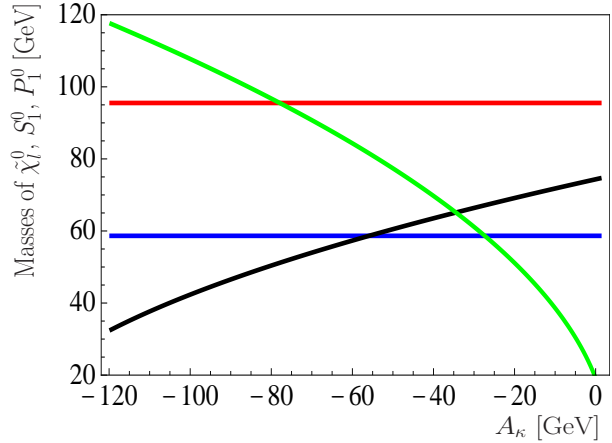} 
\end{center}
\vspace{-5mm}
\caption{Masses of the lightest neutralinos $\tilde{\chi}_l^0$ and
the lightest scalar $S_1^0=Re(\tilde{\nu}^c)$/pseudoscalar
$P_1^0=Im(\tilde{\nu}_1^c)$ as a function of $A_\kappa=T_\kappa/\kappa$ for
$\lambda=0.24$, $\kappa=0.12$, $\mu=150$GeV and $T_\lambda=360$GeV for SPS1a'.
The different colors refer to the singlino $\tilde{\chi}_1^0$ (blue),
the bino $\tilde{\chi}_2^0$ (red), the singlet scalar $S_1^0$ (black)
and the singlet pseudoscalar $P_1^0$ (green).}
\label{fig:1NuR_akappatotal}
\end{figure}

In the following discussion we call a neutralino $\tilde{\chi}_l^0$ a bino (singlino) 
if $|\mathcal{N}_{l+3,1}|^2>0.5$ ($|\mathcal{N}_{l+3,5}|^2>0.5$). 
As discussed below, light scalar $S_m^0$ or pseudoscalar states
$P_m^0$ appear, especially in case of the singlino being the lightest 
neutralino. In the following we discuss possible mass
hierarchies and mixings in more detail.

The diagonal entry of the singlet right-handed neutrino in the mass
matrix of the neutral fermions is $M_R=\tfrac{1}{\sqrt{2}}\kappa v_R$,
see Appendix \ref{subsec:neutralinos}. A singlino as lightest
neutralino is obtained by choosing small values for $\kappa$ and/or
$v_R$. Since the masses of the four MSSM neutralinos are mainly fixed
by the chosen SPS point, we can either generate a bino-like or a
singlino-like lightest neutralino by varying $\kappa$ and/or $v_R$,
where the latter case means a variation of $\lambda$ due to a fixed
$\mu$-parameter. A light singlet scalar and/or pseudoscalar can be
obtained by appropriate choices of $T_\lambda$ and $T_\kappa$. An
example spectrum is shown in Figure \ref{fig:1NuR_akappatotal}. The
MSSM parameters have been chosen according to SPS1a' except for
$\mu=150$ GeV.  The scalar state $S_2^0= h^0$ can easily get too light
to be consistent with current experimental data, although the
production rate $e^+e^-\rightarrow ZS_2^0$ is lowered, since a mixing
with the lighter singlet scalar $S_1^0= \tilde{\nu}^c$ reduces its
mass. By reducing $\mu$ the mixing can be lowered (see mass matrices)
and this problem can be solved.

\begin{figure}[ht]
\begin{center}
\vspace{2mm}
\includegraphics[width=80mm,height=60mm]{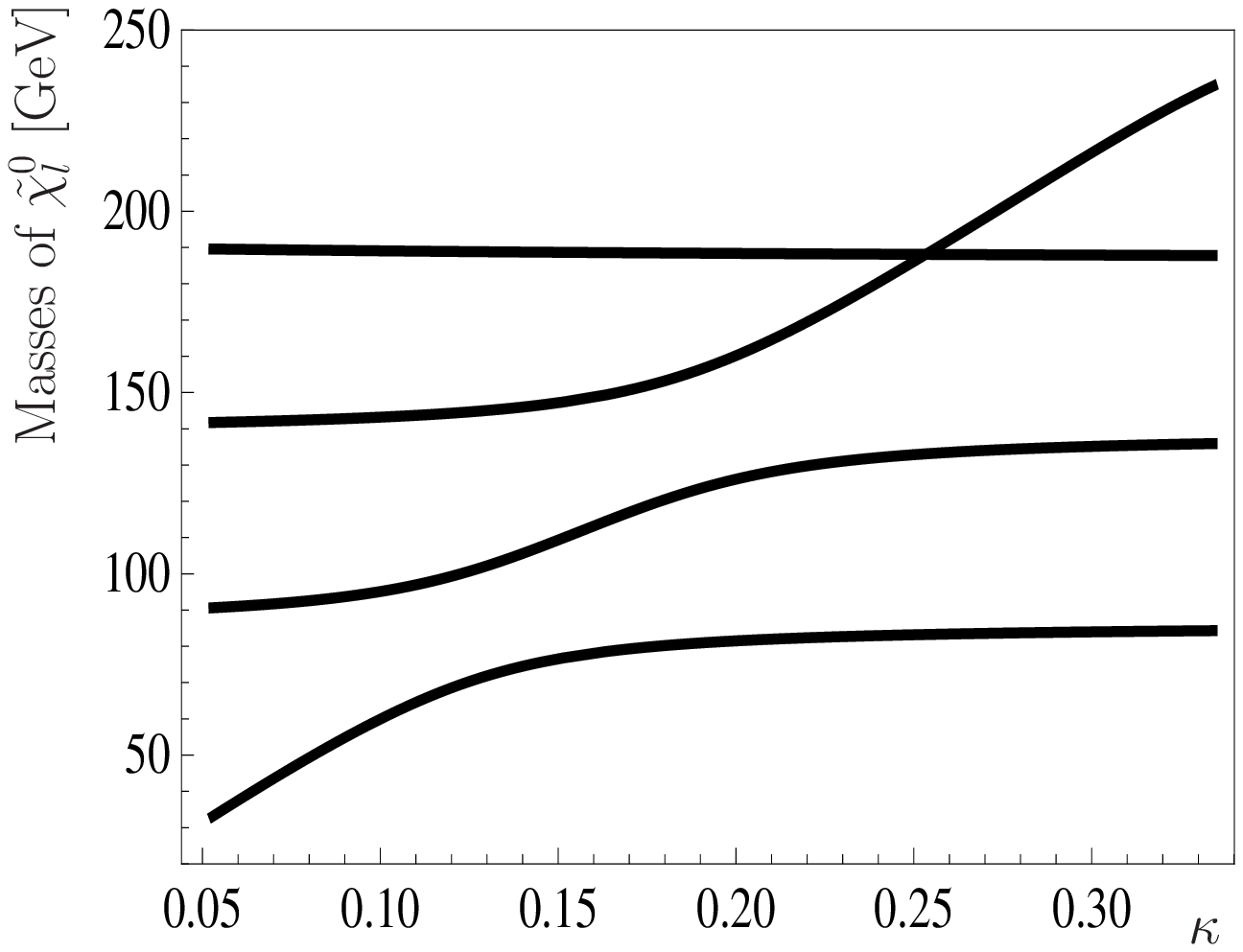} 
\hspace{2mm}
\includegraphics[width=80mm,height=60mm]{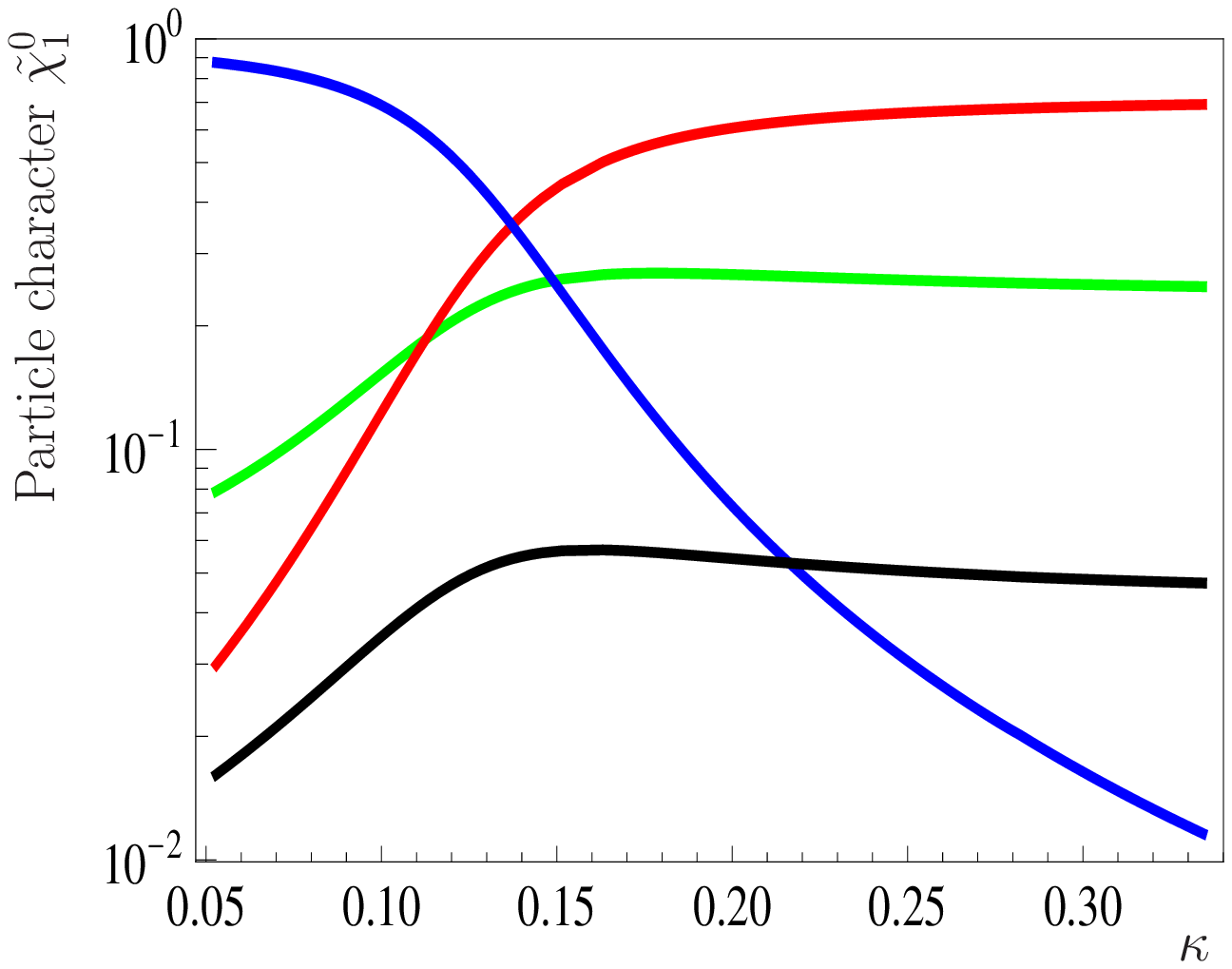} 
\vspace{2mm}
\includegraphics[width=80mm,height=60mm]{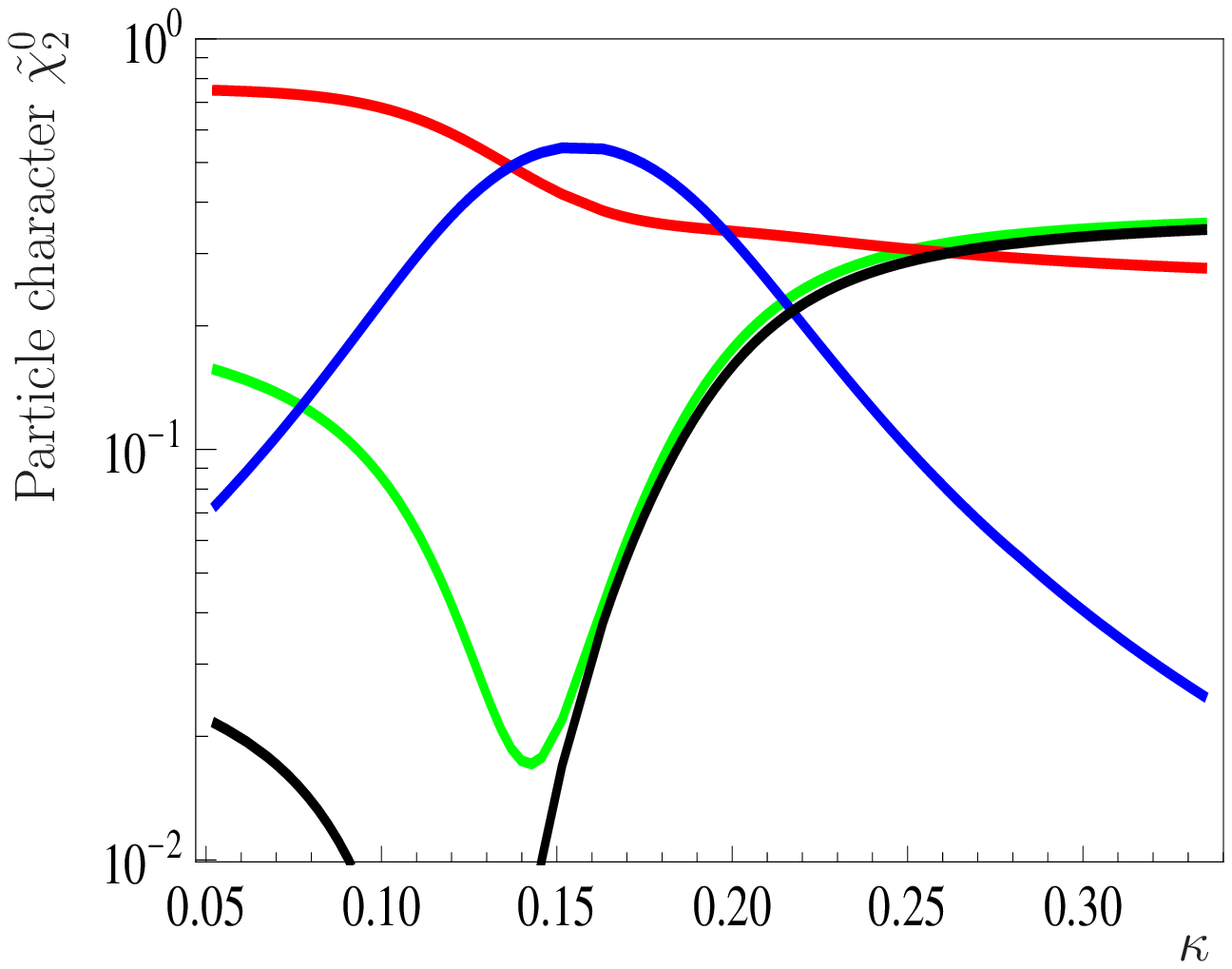} 
\hspace{2mm}
\includegraphics[width=80mm,height=60mm]{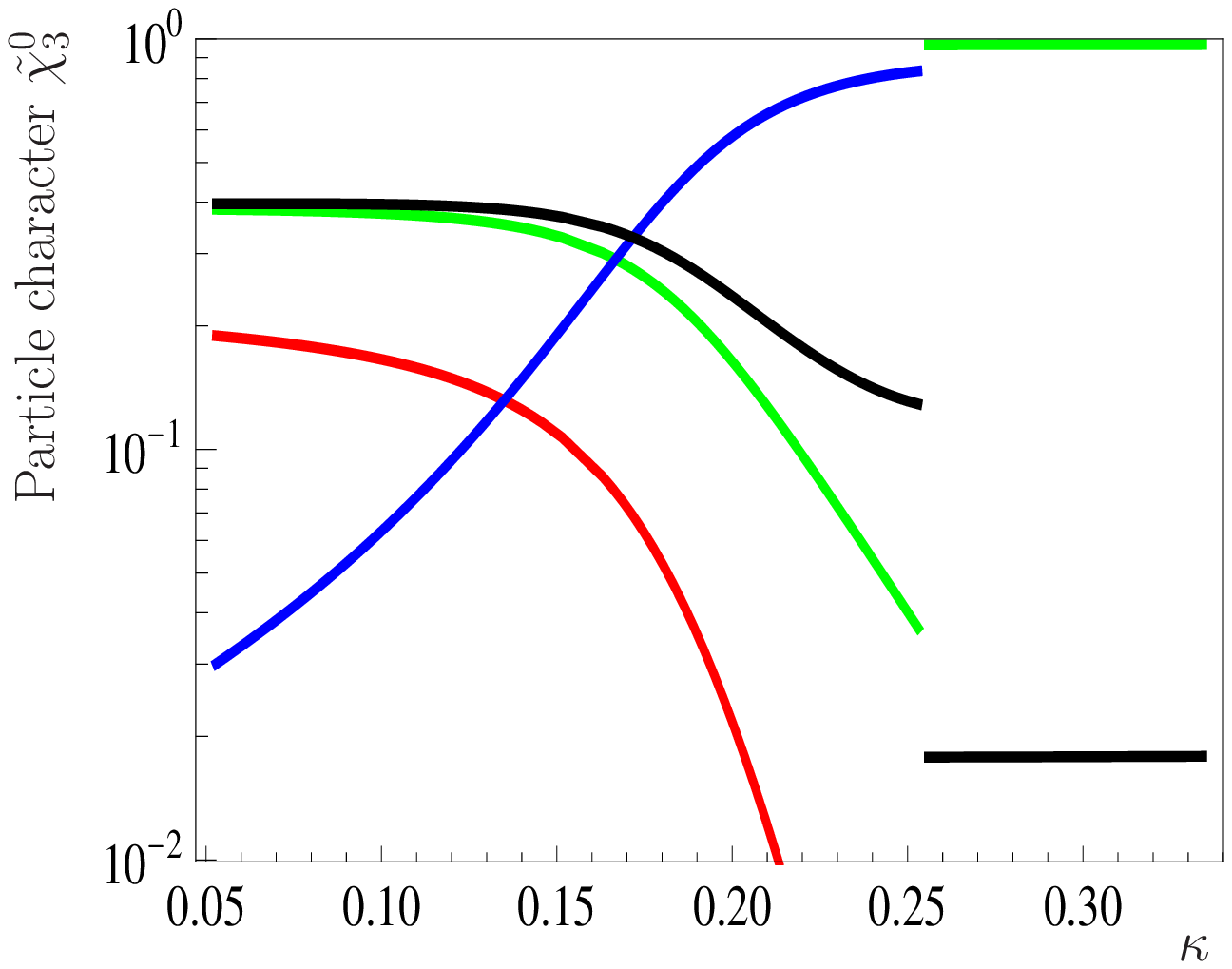} 
\end{center}
\vspace{-2mm}
\caption{Masses and particle characters of the lightest neutralinos
$\tilde{\chi}_l^0$ as a function of $\kappa$ for $\lambda=0.24$,
$\mu=170$GeV, $T_\lambda=360$GeV and $T_\kappa=-\kappa\cdot 50$GeV for SPS1a'.
The different colors refer to
singlino purity $|\mathcal{N}_{l+3,5}|^2$ (blue), bino purity
$|\mathcal{N}_{l+3,1}|^2$ (red), wino purity
$|\mathcal{N}_{l+3,2}|^2$ (black) and higgsino purity
$|\mathcal{N}_{l+3,3}|^2+|\mathcal{N}_{l+3,4}|^2$ (green).}
\label{fig:1NuR_mixingtotal}
\end{figure}

Another example spectrum for neutral fermions is shown in Figure 
\ref{fig:1NuR_mixingtotal}. Again SPS1a' parameters have been 
chosen, except $\mu=170$ GeV. As the figure demonstrates for 
this reduced value of $\mu$ the states are usually quite mixed, 
which is important for their decay properties, as discussed 
below. Note that the abrupt change in composition in $\tilde{\chi}^0_3$
is due to the level crossing in the mass eigenstates. 

\begin{figure}[htbp]
\begin{center}
\vspace{3mm}
\includegraphics[width=80mm,height=60mm]{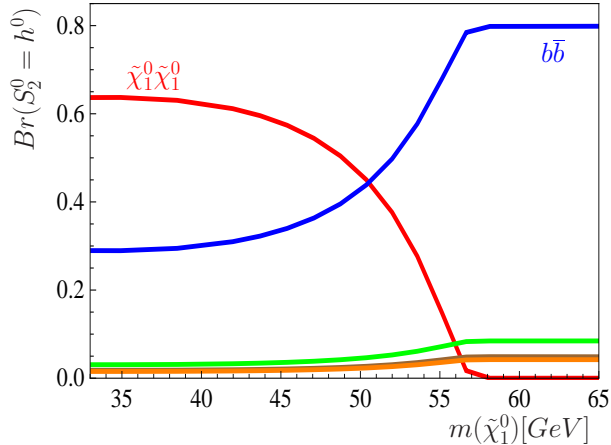}
\end{center}
\vspace{-5mm}
\caption{
Branching ratios
$Br(S_2^0 = h^0)$ as a function of $m(\tilde{\chi}_1^0)$ for
the parameter set of Figure \ref{fig:1NuR_mixingtotal} (variation of
$\kappa$). The colors indicate the different final states:
$\tilde{\chi}_1^0\tilde{\chi}_1^0$ (red), $b\overline{b}$ (blue),
$\tau^+\tau^-$ (green), $c\overline{c}$ (orange) and
$Wq\overline{q}$ (brown).}
\label{fig:1NuR_decayscalar}
\end{figure}

The decay properties of the lightest scalars/pseudoscalars are in
general quite similar to those found in the NMSSM \cite{Miller:2003ay,%
Ellwanger:2005uu}. The
lightest doublet Higgs boson similar to the $h^0$ decays mainly like
in the MSSM, apart from the possible final state $2\tilde{\chi}^0_1$,
if kinematically possible. An example is shown in Figure
\ref{fig:1NuR_decayscalar}, which display the branching ratios of
$S^0_2 = h^0$ versus $m(\tilde{\chi}_1^0)$. $\tilde{\chi}^0_1$ in this plot
is mainly a singlino (see Figure \ref{fig:1NuR_mixingtotal}),
variation of $\kappa$ varies its mass, since $v_R$ is kept fixed
here. In contrast to the NMSSM this does not lead to an invisible
Higgs, since the neutralinos themselves decay. For the range of 
parameters where the decay to $2\tilde{\chi}^0_1$ is large, 
$\tilde{\chi}^0_1$ decays mainly to $\nu b\overline{b}$, leading to 
the final state 4 $b$-jets plus missing energy. Note that the $S_1^0$
which is mainly singlet here decays dominantly to $b{\bar b}$ final
states, followed by $\tau\tau$ final states.

\subsection{Decays of a gaugino-like lightest neutralino}
\label{subsec:decaysfermions}

\begin{figure}[ht]
\begin{center}
\includegraphics[width=0.45\textwidth]{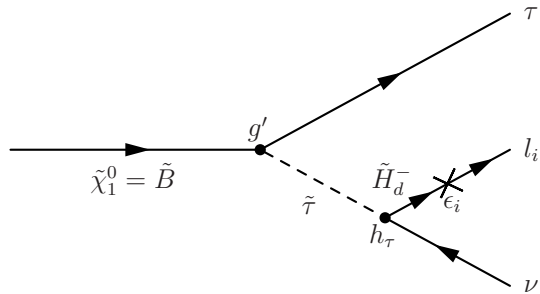}
\caption{Dominant Feynman graph for the decay $\tilde{\chi}_1^0
\rightarrow l_i\tau\nu$ with $l_i=e,\mu$.}
\label{fig:3bodydecay}
\end{center}
\end{figure}

\begin{figure}[htbp]
\begin{center}
\vspace{2mm}
\includegraphics[width=71mm,height=71mm]{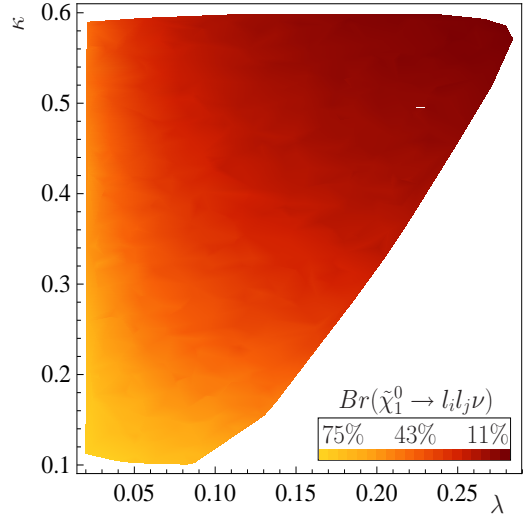}
\end{center}
\vspace{-5mm}
\caption{Dependence of allowed $\kappa(\lambda)$ for values of
$\lambda\in\left[0.02,0.5\right]$ and $\kappa\in\left[0.1,0.6\right]$
and $Br$($\tilde{\chi}_1^0\rightarrow l_il_j\nu$) as function of
$\lambda$ and $\kappa$ exemplary for SPS1a' with $\mu=390$GeV, 
$T_\lambda=\lambda\cdot 1.5$TeV and $T_\kappa=-\kappa\cdot 100$GeV.}
\label{fig:1NuR_lambdakappa}
\end{figure}

We first consider the case of a bino as lightest neutralino. Although
$m(\tilde{\chi}_1^0)>m_W$ in the SPS points we have chosen, two-body
decay modes are not necessarily dominant. The three-body decay
$\tilde{\chi}_1^0\rightarrow l_il_j\nu$ dominated by a virtual
$\tilde{\tau}$ also can have a sizeable branching ratio, see Table
\ref{table:binodecays} and Figure \ref{fig:1NuR_lambdakappa}. 
The importance of this final state can be understood from the 
Feynman graph shown in Figure \ref{fig:3bodydecay}, giving the dominant
contribution due to $\tilde{H}_d^-$-$l_i$-mixing ($l_i=e,\mu$).

In the case $l_i=\tau$ there's an additional contribution due to
$\tilde{H}_d^0$-$\nu$-mixing. As Figure \ref{fig:1NuR_lambdakappa} shows
there exist parameter 
combinations in the $\lambda$-$\kappa$-plane,
where the decay mode $\tilde{\chi}_1^0\rightarrow l_il_j\nu$ is more
important than $\tilde{\chi}_1^0\rightarrow Wl$. 
The strong variation in the branching ratios for SPS1a' is mainly due
to the strong dependence of the partial decay width of
$\tilde{\chi}_1^0\rightarrow l_il_j\nu$, where the decays with $i=j$
and $i\neq j$ both play a role. Other important final states are
$\tilde{\chi}_1^0\rightarrow Z\nu$ and in case of a light scalar with
$m(\tilde{\chi}_1^0)>m(h^0)$ the decay $\tilde{\chi}_1^0\rightarrow
h^0\nu$, as demonstrated in Table \ref{table:binodecays}.

\begin{table}[htbp]
\begin{center}
\begin{tabular}{|c|c|c|c|}
\hline
 $Br(\tilde{\chi}_1^0)$& SPS1a' & SPS3 & SPS4 \\
\hline \hline
 $Wl$ & $23-80$ & $12-55$ & $68-72$\\
 $l_il_j\nu$ & $11-75$ & $2-31$ & $2.6-3.9$ \\
 $Z\nu$ & $2.2-8.9$ & $5-28$ & $25-28$ \\
 $h^0\nu$ & $-$ & $15-53$ & $<2.0$\\
\hline
Decay length [mm] & $1.6-7.0$ & $0.1-0.5$ & $1.4-1.6$\\\hline
\end{tabular}
\caption{Branching ratios (in \%) and total decay length in mm of the
decay of the lightest bino-like neutralino for different values of
$\lambda\in\left[0.02,0.5\right]$ and $\kappa\in\left[0.1,0.6\right]$
with a dependence of allowed $\kappa(\lambda)$ similar to
\cite{Escudero:2008jg} and to Figure \ref{fig:1NuR_lambdakappa}
and $T_\lambda=\lambda\cdot 1.5$TeV and $T_\kappa=-\kappa\cdot 100$GeV.}
\label{table:binodecays}
\end{center}
\end{table}

\begin{figure}[htbp]
\begin{center}
\vspace{2mm}
\includegraphics[width=80mm,height=60mm]{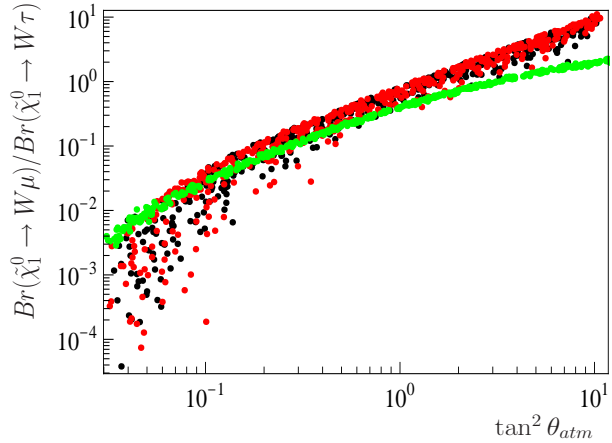} 
\end{center}
\vspace{-5mm}
\caption{Ratio $\frac{Br(\tilde{\chi}_1^0\rightarrow
W\mu)}{Br(\tilde{\chi}_1^0\rightarrow W\tau)}$ versus
$\tan^2\theta_{atm}$ for different SPS scenarios (SPS1a'
(black), SPS3 (red), SPS4 (green)) and for different values of
$\lambda\in\left[0.02,0.5\right]$ and $\kappa\in\left[0.1,0.6\right]$
with a dependence of allowed $\kappa(\lambda)$ similar to
\cite{Escudero:2008jg} and to Figure \ref{fig:1NuR_lambdakappa} 
and $T_\lambda=\lambda\cdot 1.5$TeV and $T_\kappa=-\kappa\cdot 100$GeV.}
\label{fig:1NuR_binocorr}
\end{figure}

In the $\mu\nu$SSM one finds correlation between the decays of the
lightest neutralino and the neutrino mixing angles, because neutralino
couplings depend on the same \rpv parameters as the neutrino masses.
Figure \ref{fig:1NuR_binocorr} shows the correlation between the
branching ratios of the decay $\tilde{\chi}_1^0\rightarrow Wl$ as a
function of the atmospheric angle. Although a clear correlation is 
visible it is not as pronounced as in the $n$ generation
case, see below and \cite{Ghosh:2008yh}, due to inclusion of 
1-loop effects in the neutrino masses and mixing angles. 

\begin{figure}[htbp]
\begin{center}
\vspace{2mm}
\includegraphics[width=80mm,height=60mm]{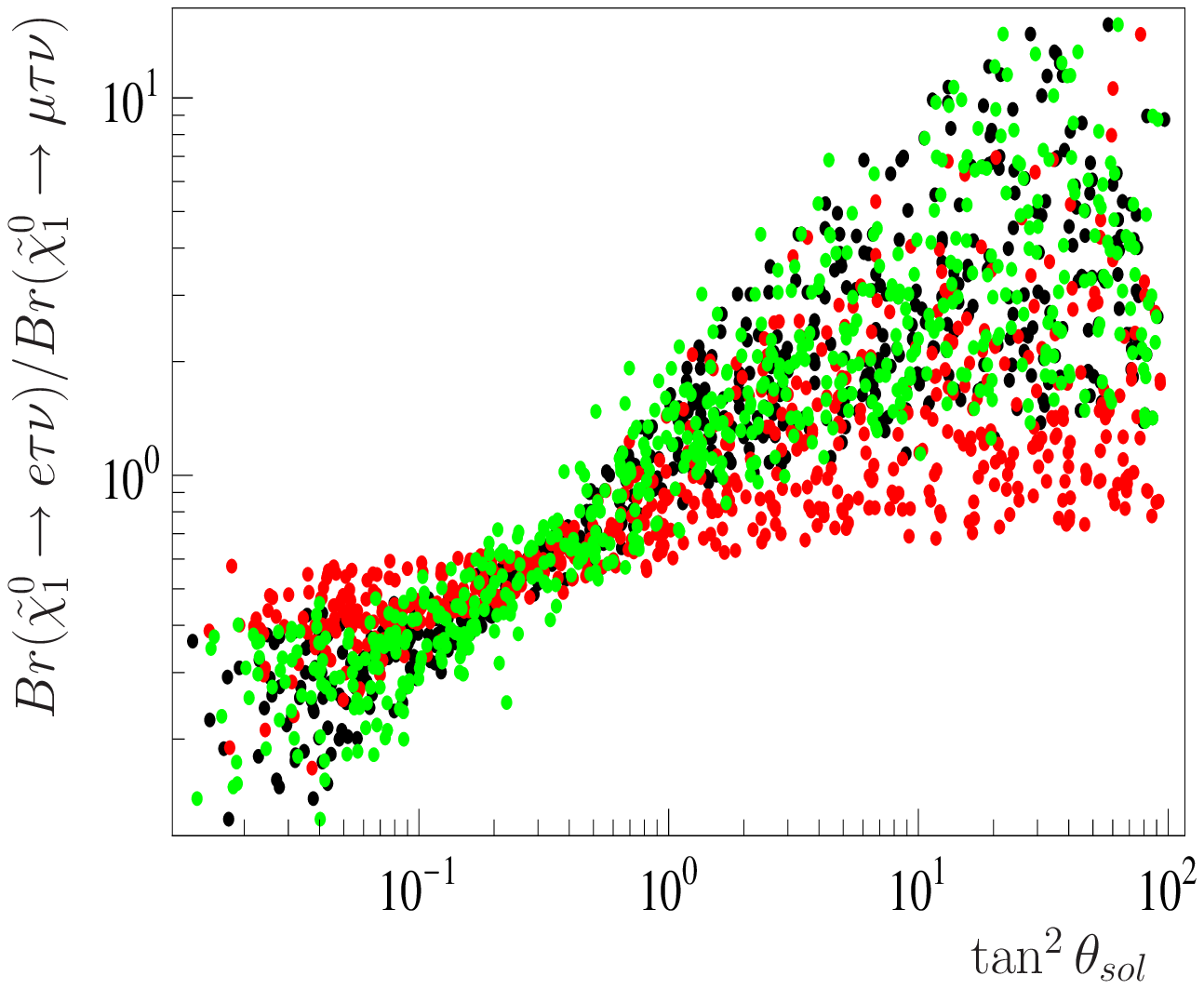}
\hspace{2mm}
\includegraphics[width=80mm,height=60mm]{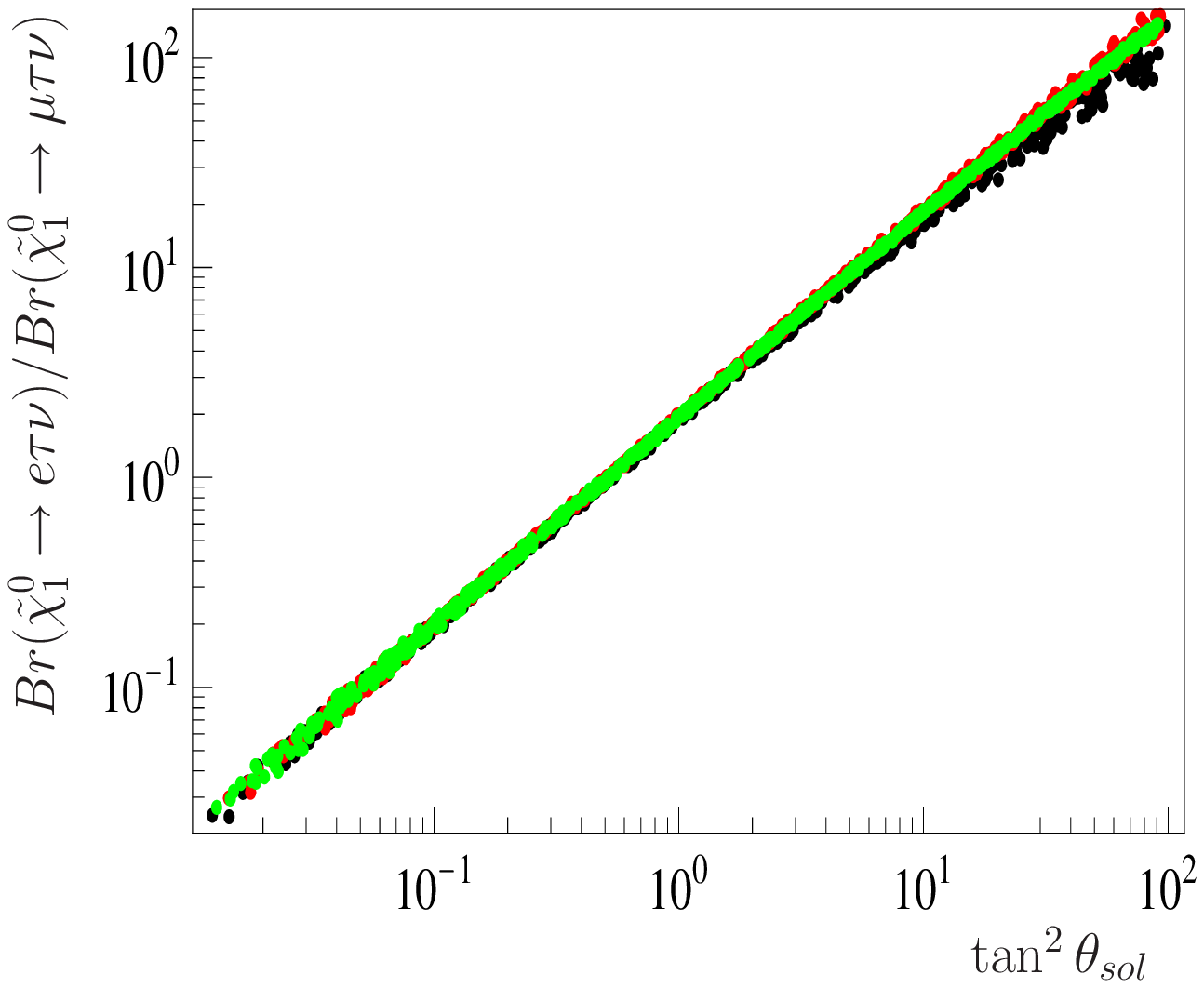} 
\end{center}
\vspace{-5mm}
\caption{Ratio $\frac{Br(\tilde{\chi}_1^0\rightarrow
e\tau\nu)}{Br(\tilde{\chi}_1^0\rightarrow \mu\tau\nu)}$ versus
$\tan^2\theta_{sol}$ with same set of parameters as Figure 
\ref{fig:1NuR_binocorr}. Bino purity $|\mathcal{N}_{41}|^2>0.97$.
To the left (a) two-body plus three-body contributions, to the right
(b) three-body contributions only. For a discussion see text.}
\label{fig:1NuR_binocorr2}
\end{figure}

Also the three-body decay $\tilde{\chi}_1^0\rightarrow l_il_j\nu$
exemplifies a correlation with neutrino physics. However, this decay 
is connected to the solar angle, see Figure \ref{fig:1NuR_binocorr2}. 
There are two main contributions to this final state: $\tilde{\chi}^0_1 
\to W l \to l_il_j\nu$ and $\tilde{\chi}^0_1 \to {\tilde\tau}^* l \to l_il_j\nu$.
While the former is mainly sensitive to $\Lambda_i$, the latter is 
dominated by $\epsilon_i$-type couplings (see Figure \ref{fig:3bodydecay}),
causing the connection to 
solar neutrino angle. In case the $W$ is on-shell as in the SPS1a' 
point, one could in principle devise kinematical cuts reducing this 
contribution. Such a cut can significantly improve the quality of 
the correlation.

\begin{figure}[htbp]
\begin{center}
\vspace{-1mm}
\includegraphics[width=80mm,height=60mm]{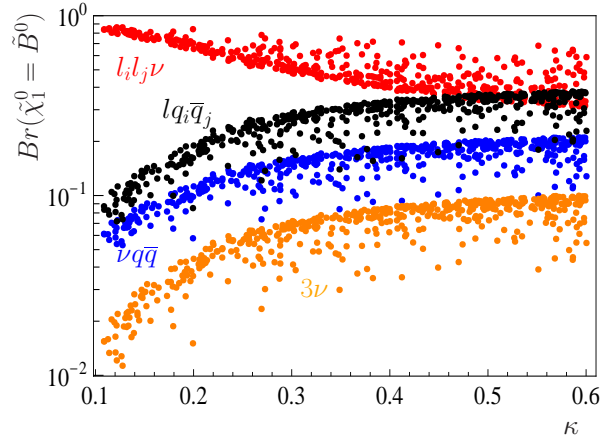}
\end{center}
\vspace{-7mm}
\caption{Decay branching ratios for bino-like lightest neutralino 
as a function of $\kappa$ for $\lambda\in\left[0.02,0.5\right]$, 
$T_\lambda=\lambda\cdot 1.5$ TeV, $T_\kappa=-\kappa \cdot 100$ GeV and for MSSM parameters 
defined by the study point SU4 of the ATLAS collaboration \cite{Aad:2009wy}. The colors
indicate the different final states: $l_il_j\nu$ (red),
$lq_i\overline{q}_j$ (black), $\nu q{\bar q}$ (blue) and $3\nu$ (orange).}
\label{fig:SU4tot}
\end{figure}

\begin{figure}[htbp]
\begin{center}
\vspace{-1mm}
\includegraphics[width=80mm,height=60mm]{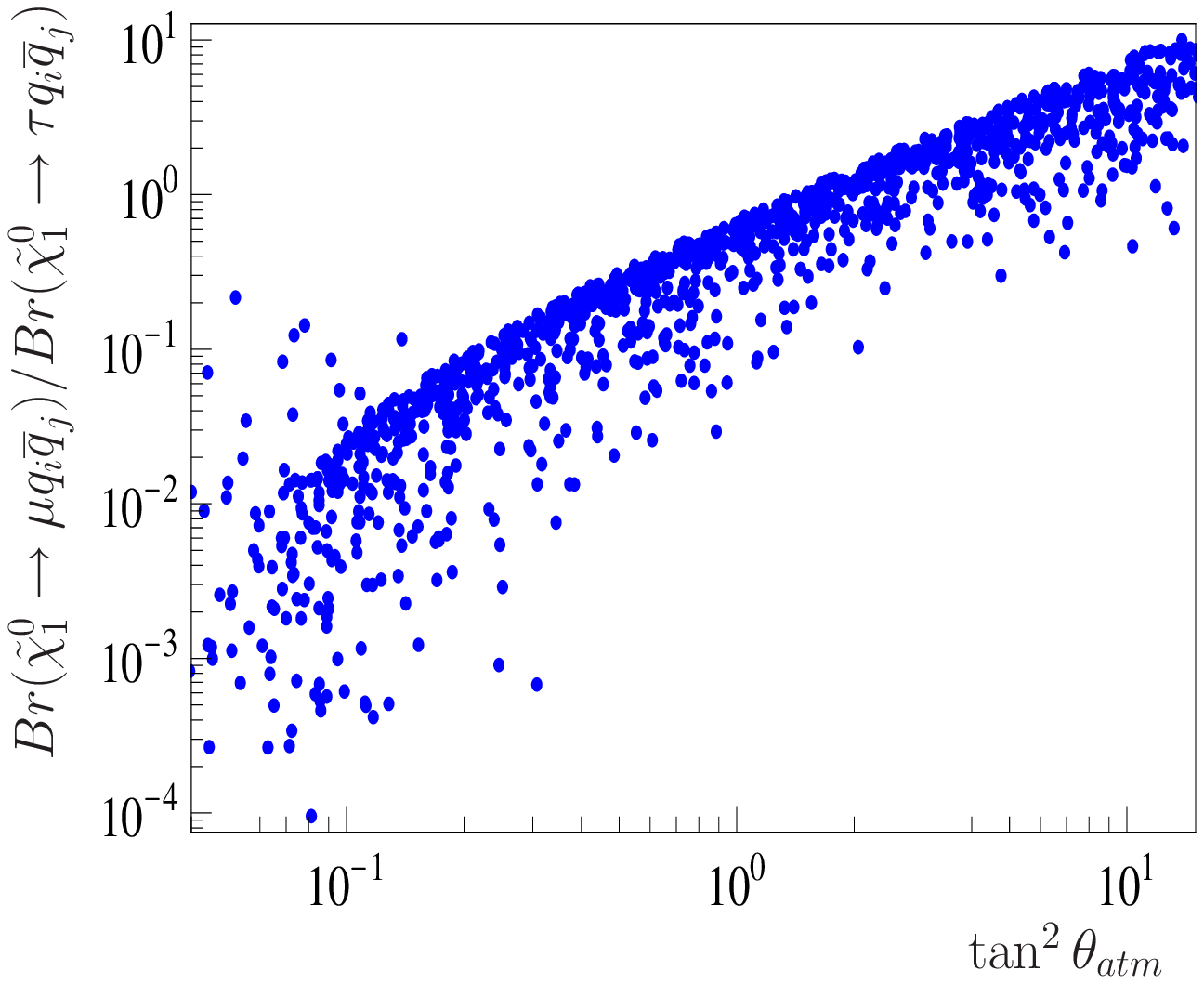} 
\hspace{2mm}
\includegraphics[width=80mm,height=60mm]{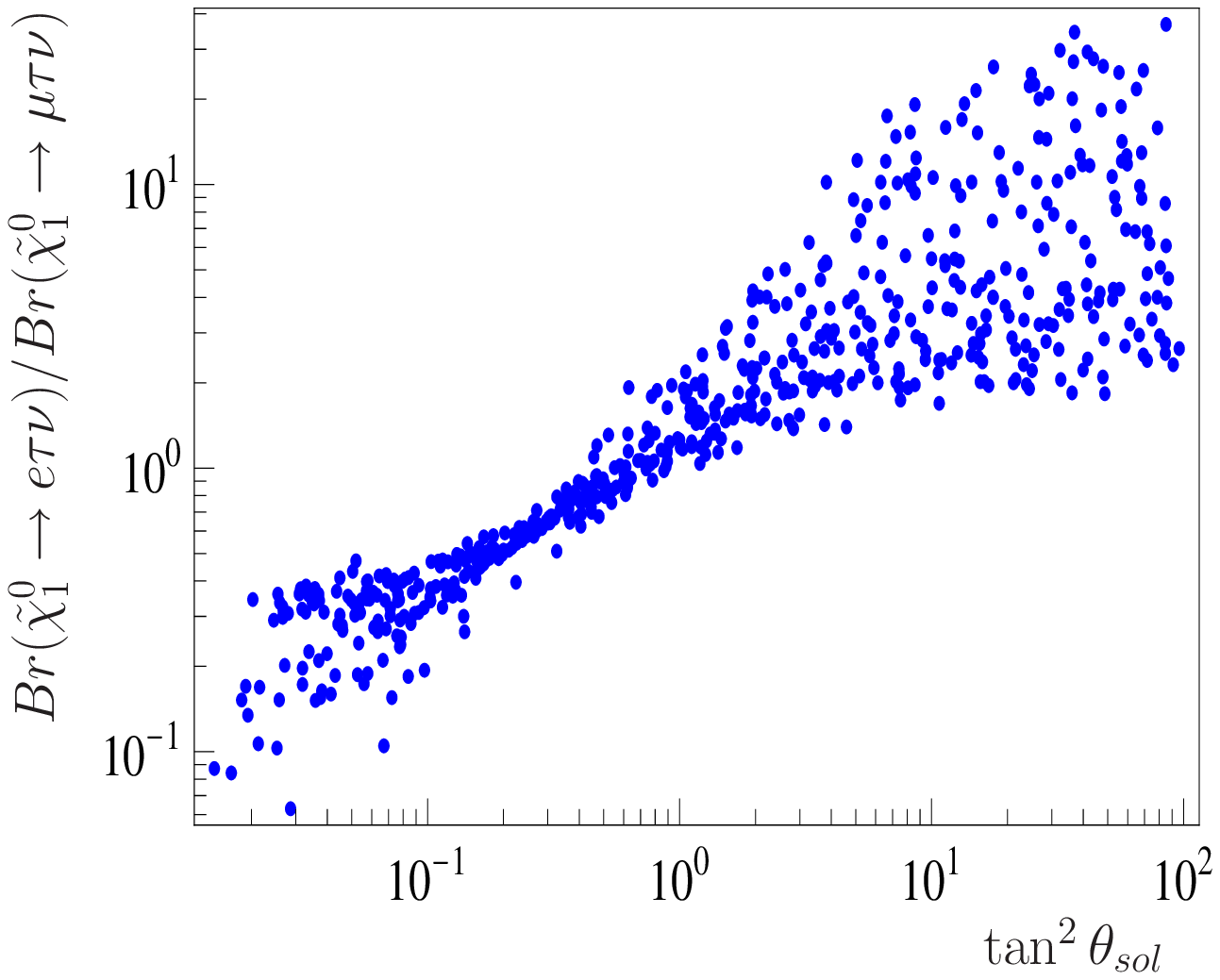} 
\end{center}
\vspace{-7mm}
\caption{To the left (a) ratio $\frac{Br(\tilde{\chi}_1^0\rightarrow
\mu q_i\overline{q}_j)}{Br(\tilde{\chi}_1^0\rightarrow
\tau q_i\overline{q}_j)}$ versus $\tan^2\theta_{atm}$ for the SU4
scenario of the ATLAS collaboration \cite{Aad:2009wy} and to the right (b) ratio
$\frac{Br(\tilde{\chi}_1^0\rightarrow
e\tau\nu)}{Br(\tilde{\chi}_1^0\rightarrow \mu\tau\nu)}$ versus
$\tan^2\theta_{sol}$ with same set of parameters as (a).
Bino purity $|\mathcal{N}_{41}|^2>0.94$.}
\label{fig:1NuR_binoSU4corr}
\end{figure}

The SU4 scenario of the ATLAS collaboration \cite{Aad:2009wy} has 
a very light SUSY spectrum close to the Tevatron bound with a bino-like
neutralino $m(\tilde{\chi}_1^0)\approx 60$ GeV. Thus, for 
SU4 the lightest neutralino has only three-body decay modes. 
Most important branching ratios are shown in Figure \ref{fig:SU4tot}.
The lightness of the bino-like neutralino $\tilde{\chi}_1^0$ in this
scenario implies a larger average decay length of $(8-90)$ cm,
depending on the parameter point in the $\lambda$-$\kappa$-plane. Note
that the decay length becomes smaller for smaller values of
$\lambda,\kappa$. In general the decay length scales as 
$L \propto m^{-4}(\tilde{\chi}^0_1)$ for $m(\tilde{\chi}^0_1) < m_W$.
Also for this point a correlation between the branching ratios 
and the neutrino mixing angles is found 
as illustrated in Figure \ref{fig:1NuR_binoSU4corr}.

\begin{figure}[hbtp]
\begin{center}
\vspace{-1mm}
\includegraphics[width=80mm,height=60mm]{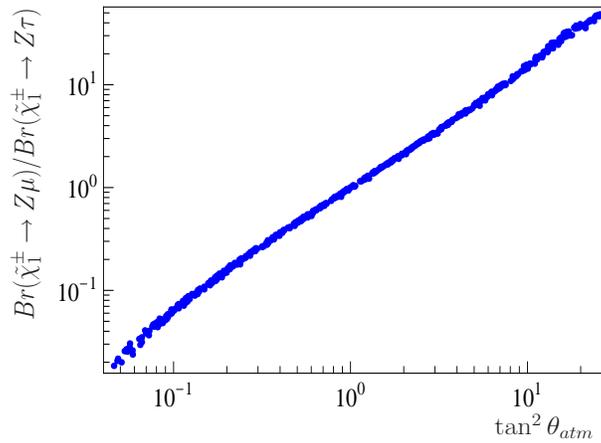}
\end{center}
\vspace{-8mm}
\caption{Ratio $\frac{Br(\tilde{\chi}_1^\pm\rightarrow
Z\mu)}{Br(\tilde{\chi}_1^\pm\rightarrow Z\tau)}$ versus
$\tan^2\theta_{atm}$ for the AMSB scenario SPS9 and for 
different values of
$\lambda\in\left[0.02,0.5\right]$, $\kappa\in\left[0.1,0.6\right]$, 
$T_\lambda=\lambda\cdot 1.5$ TeV and $T_\kappa=-\kappa \cdot 100$ GeV.}
\label{fig:1NuR_Winoatm}
\end{figure}

In addition to the SUGRA scenarios discussed up to now we have 
also studied SPS9, which is a typical AMSB point. The most 
important difference between this point and the previously discussed 
cases is the near degeneracy between lightest neutralino and 
lightest chargino. This near degeneracy is the reason that the chargino 
decay is dominated by \rpv final states. Varying 
$\lambda$ and $\kappa$ as before we find a total decay length of
$(0.12-0.16)$mm with $Br(\tilde{\chi}_1^\pm\rightarrow
W\nu)=(42-57)$\%, $Br(\tilde{\chi}_1^\pm\rightarrow Zl)=(20-26)$\% and
$Br(\tilde{\chi}_1^\pm\rightarrow h^0l)=(17-40)$\%. This is 
especially interesting since, similar to $Wl$ in case of the gaugino-like
lightest neutralino, the decay to $Zl$ of the chargino is linked 
to the atmospheric angle, see Figure \ref{fig:1NuR_Winoatm}.

\subsection{Decays of a singlino-like lightest neutralino}

\begin{figure}[htbp]
\begin{center}
\vspace{0mm}
\includegraphics[width=80mm,height=60mm]{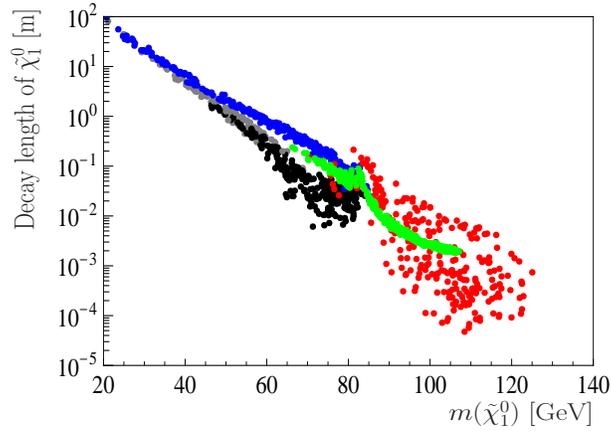}
\end{center}
\vspace{-5mm}
\caption{Decay length of the lightest neutralino $\tilde{\chi}_1^0$ in
m as a function of its mass $m(\tilde{\chi}_1^0)$ in GeV for different
values of $\lambda\in\left[0.2,0.5\right]$,
$\kappa\in\left[0.025,0.2\right]$ and $\mu\in[110,170]$GeV with a
dependence of allowed $\kappa(\lambda)$ similar to
\cite{Escudero:2008jg} and to Figure \ref{fig:1NuR_lambdakappa}
and $T_\lambda=\lambda\cdot 1.5$TeV, whereas $T_\kappa\in [-20,-0.05]$GeV is
chosen in such a way, that no lighter scalar or pseudoscalar states
with $\lbrace m(S_1^0),m(P_1^0)\rbrace <m(\tilde{\chi}_1^0)$ appear.
Note that the different colors stand for SPS1a' (real singlino,
$|\mathcal{N}_{45}|^2>0.5$) (gray), SPS1a' (mixture state) (black),
SPS3 (real singlino) (blue), SPS3 (mixture state) (red) and SPS4
(mixture state) (green).}
\label{fig:1NuR_decaylength}
\end{figure}

We now turn to the case of a singlino-like LSP. As already explained,
this scenario is connected to
a light singlet scalar and pseudoscalar. Recall, that the particles in
the fermionic sector are mixed for
$\lambda,\kappa=\mathcal{O}(10^{-1})$ due to the reduced
$\mu$-parameter as can be seen in Figure
\ref{fig:1NuR_mixingtotal}. We will first discuss the average decay
length of the lightest neutralino $\tilde{\chi}_1^0$. Figure
\ref{fig:1NuR_decaylength} shows the average decay length in meter for
different SPS scenarios as a function of the mass of the lightest
neutralino $m(\tilde{\chi}_1^0)$.  Composition of the neutralino is
indicated by colour code, as given in the caption. $\lambda$,
$\kappa$, $T_\kappa$ and $\mu$ are varied in this plot. Note that by
variation of $T_\kappa$ the parameter points in Figure
\ref{fig:1NuR_decaylength} are chosen in such a way, that all scalar
and pseudoscalar states are heavier than the lightest neutralino.
Singlino purity in this plot increases with decreasing mass and for
pure singlinos the decay length is mainly determined by its mass and
the experimentally determined neutrino masses.  For neutralino masses 
below about 50 GeV decay lengths become larger than 1 meter, implying 
that a large fraction of neutralinos will decay outside typical 
collider detectors. Note that if one
allows for lighter scalar states so that at least one of the decays
$\tilde{\chi}_1^0\rightarrow S_1^0(P_1^0)\nu$ appears, the average
decay length can be easily reduced by several orders of magnitude.

\begin{figure}[htbp]
\begin{center}
\vspace{2mm}
\includegraphics[width=80mm,height=60mm]{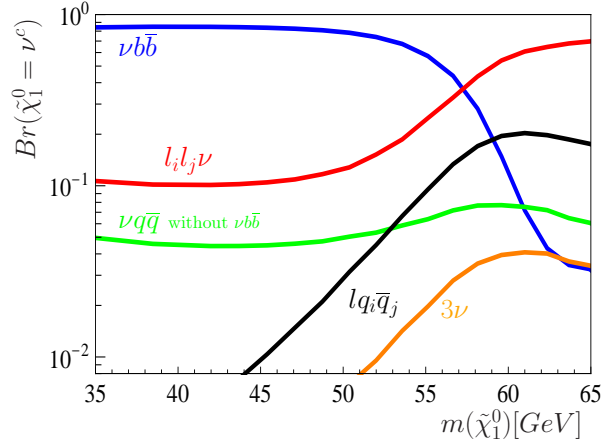}
\end{center}
\vspace{-5mm}
\caption{Singlino decay branching ratios as a function of its mass, 
for the same parameter choices as in Figure \ref{fig:1NuR_decayscalar}. 
The colors indicate the different final states: $\nu b{\bar b}$ (blue), $l_il_j\nu$ (red), 
$lq_i\overline{q}_j$ (black), $3\nu$ (orange) and $\nu q{\bar q}$ ($q \ne b$, 
green).}  
\label{fig:1NuR_singlinodecay}
\end{figure}

\begin{figure}[htbp]
\begin{center}
\vspace{2mm}
\includegraphics[width=80mm,height=60mm]{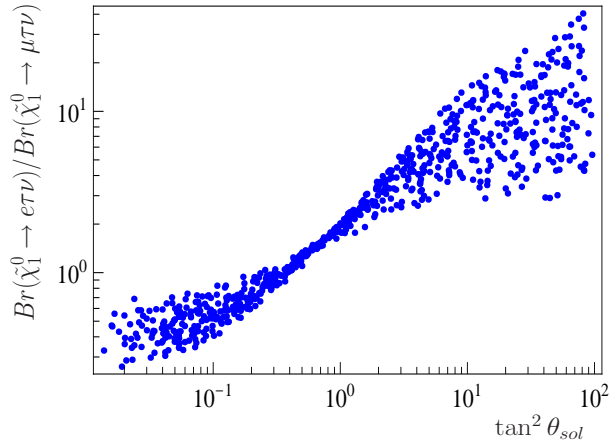} 
\end{center}
\vspace{-5mm}
\caption{Ratio $\frac{Br(\tilde{\chi}_1^0\rightarrow
e\tau\nu)}{Br(\tilde{\chi}_1^0\rightarrow \mu\tau\nu)}$ versus
$\tan^2\theta_{sol}$ for the SPS1a' scenario and
$\lambda\in\left[0.2,0.5\right]$, $\mu\in[110,170]$GeV,
$\kappa=0.035$, $T_\lambda=\lambda\cdot 1.5$ TeV and $T_\kappa=-0.7$ GeV.}
\label{fig:1NuR_singlinocorr}
\end{figure}

Again typical decays are $Wl$, $lq_i\overline{q}_j$, $Z\nu$, $\nu
q\overline{q}$, $l_il_j\nu$ and the invisible decay to $3\nu$.  For
the region of $m(\tilde{\chi}^0_1)$ below the $W$ threshold see Figure
\ref{fig:1NuR_singlinodecay}. The dominance of $\nu b\overline{b}$
for smaller values of $m(\tilde{\chi}_1^0)$ is due to the decay chain
$\tilde{\chi}_1^0\rightarrow S_1^0 \nu\rightarrow \nu b\overline{b}$,
whereas for larger values of $m(\tilde{\chi}_1^0)$ we find
$m(S_1^0)>m(\tilde{\chi}_1^0)$. Final state ratios show correlations
with neutrino physics also in this case.  As an example we show
$l_il_j\nu$ branching ratios versus the solar neutrino mixing angle in Figure
\ref{fig:1NuR_singlinocorr}. Singlino purity for this plot
$|\mathcal{N}_{45}|^2\in[0.75,0.83]$ and mass
$m(\tilde{\chi}_1^0)\in[22,53]$ GeV. The absolute values for the
branching ratios are comparable to those of the described SU4 scenario
with a bino-like lightest neutralino.
We note that for the parameters in Figure \ref{fig:1NuR_singlinocorr} 
the light Higgs $S_2^0= h^0$ decays to $\tilde{\chi}_1^0\tilde{\chi}_1^0$ 
with a branching ratio of $Br(S_2^0= h^0\rightarrow
\tilde{\chi}_1^0\tilde{\chi}_1^0)=(21-91)$\%.

Up to now we have considered values of $\lambda$ and $\kappa$ larger 
than $10^{-2}$. For very small values of these couplings, the singlet 
sector, although very light, effectively decouples.
This implies that R-parity conserving decays of $\tilde{\chi}_2^0$, e.g. 
decays to final states like $\tilde{\chi}_1^0S_1^0$, $\tilde{\chi}_1^0P_1^0$,
$\tilde{\chi}_1^0l^+l^-$ or $\tilde{\chi}_1^0 q\overline{q}$, are strongly
suppressed and the \rpv decay modes dominate, implying decays with correlations
as in the case of the explicit b-\rpv.

\section{Phenomenology of the $n$ $\widehat{\nu}^c$-model}
\label{sec:nNuC}

In the previous section the phenomenology for the one generation case
of the model has been worked out in detail. Most of the signals discussed 
so far are independent of the number of right-handed neutrinos. 
However, the $n$ generation variants also offer some additional 
phenomenology, which we discuss here for the simplified case of $n=2$. 

In a model with one right-handed neutrino superfield a light singlino
will always imply a light scalar/pseudoscalar. This connection between
the neutral fermion sector and scalar/pseudoscalar sector is a
well-known property of the NMSSM (see again 
\cite{Franke:1995xn,Miller:2003ay}). In models with more than one generation of
singlets, the off-diagonal $T_\kappa$ terms in Equation
\eqref{eq:softsing} induce mixing between the different generations of
singlet scalars and pseudoscalars.  This opens up the possibility, not
considered in previous publications
\cite{LopezFogliani:2005yw,Escudero:2008jg,Ghosh:2008yh}, to have the
singlet scalars considerably heavier than the singlet fermions.

\begin{figure}
\begin{center}
\vspace{0mm}
\includegraphics[width=80mm,height=60mm]{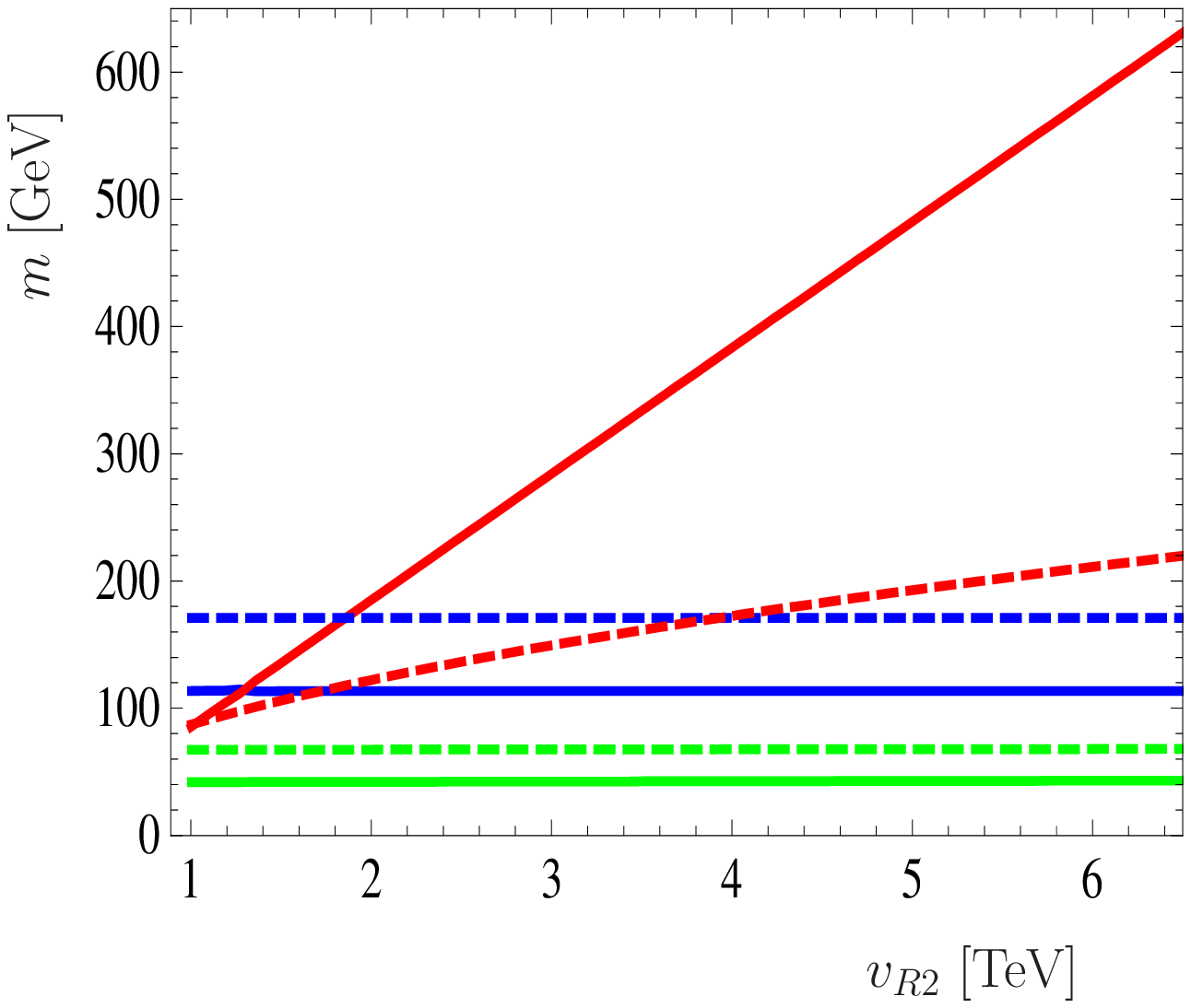}
\hspace{2mm}
\includegraphics[width=80mm,height=60mm]{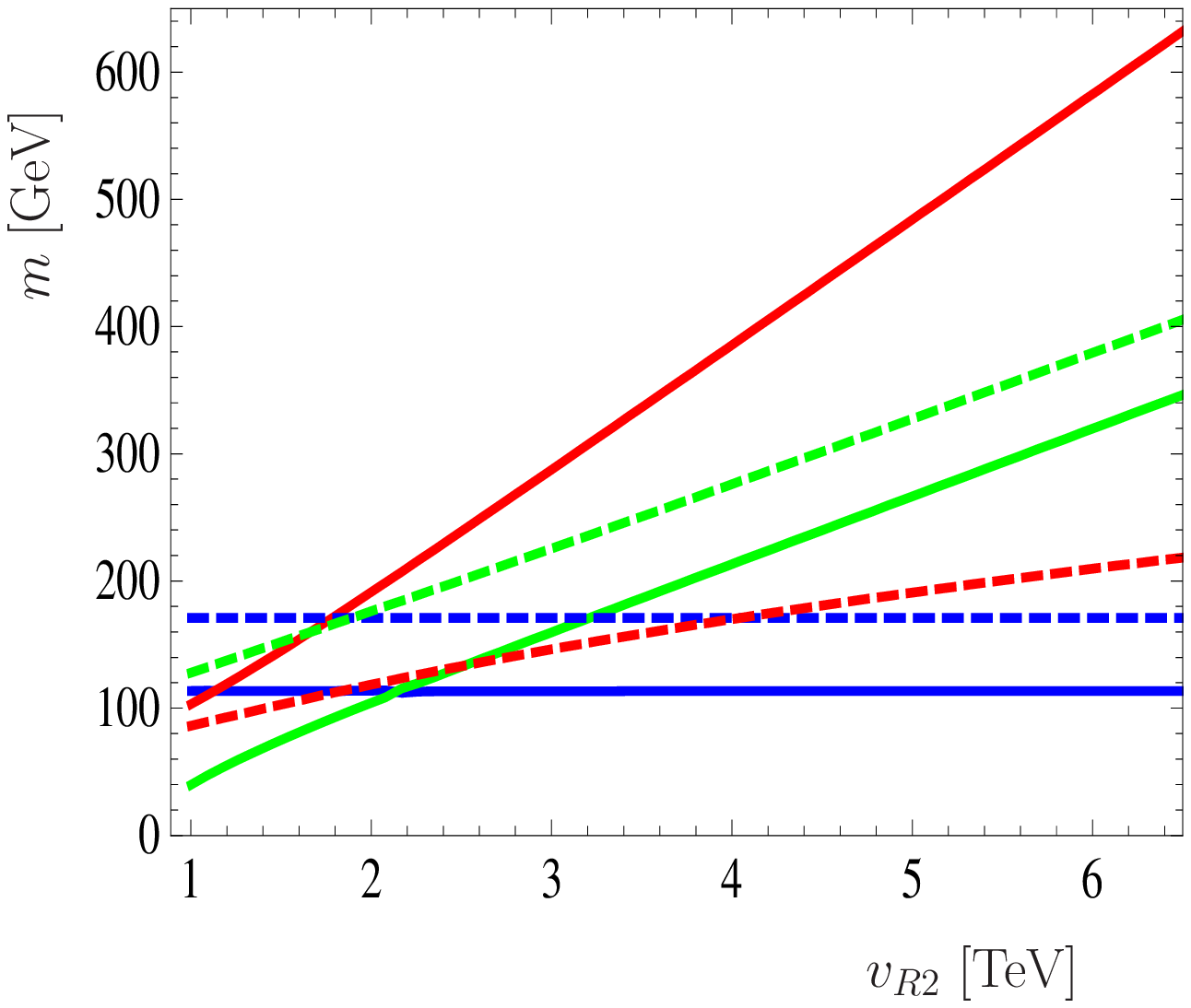}
\end{center}
\vspace{-7mm}
\caption{Masses of the scalar states $Re(\tilde{\nu}_1^c)$ (green),
$Re(\tilde{\nu}_2^c)$ (red) and $h^0$ (blue) and the pseudoscalar
states $Im(\tilde{\nu}_1^c)$ (dashed green), $Im(\tilde{\nu}_2^c)$
(dashed red) and $Im(\tilde{\nu}_1)$ (dashed blue) as a function of
$v_{R2}$ for different values of $T_\kappa^{112}=T_\kappa^{122}$. To
the left (a) $T_\kappa^{112}=T_\kappa^{122}=0$ whereas to the right (b)
$T_\kappa^{112}=T_\kappa^{122}=-2$ GeV. The MSSM parameters have been
taken such that the standard SPS1a' point is reproduced. The light
singlet parameters $\kappa_1 = 0.16$ and $v_{R1} = 500$ GeV ensure
that in all points the lightest neutralino is mostly $\nu_1^c$, with a
mass of $47-48$ GeV. In addition, $T_\lambda^1 = 300$ GeV and
$T_\lambda^2 \in [10,200]$ GeV.}
\label{fig:tkappa}
\end{figure}

Let us illustrate this feature with a simple example. Imagine a light
singlino $\nu_1^c$, and a heavy singlino $\nu_2^c$, in a model with
non-zero trilinear couplings $T_\kappa^{112}$. In that case, the
contributions to the mass of the $\tilde{\nu}_1^c$, scalar or
pseudoscalar, coming from the large value of $v_{R2}$ are proportional
to $T_\kappa^{112}$.  Without these contributions the mass of
$\tilde{\nu}_1^c$ would only depend on the small $v_{R1}$, thus making
it light like the singlino of the same generation. With non-zero
$T_\kappa^{112}$ the mass of both $\tilde{\nu}_s^c$ are dominated by the
larger of the $v_{Rs}$.  This feature is demonstrated in Figure
\ref{fig:tkappa}. In the two plots the lightest neutralino is mostly
$\nu_1^c$, with a mass of $\sim 50$ GeV. These plots show the
dependence of the masses of the singlet scalar states
$Re(\tilde{\nu}_1^c)$ and $Re(\tilde{\nu}_2^c)$ and the corresponding
pseudoscalar states $Im(\tilde{\nu}_1^c)$ and $Im(\tilde{\nu}_2^c)$
with $v_{R2}$ for different values of
$T_\kappa^{112}=T_\kappa^{122}$. The masses of the light Higgs boson
$h^0$ and the lightest left-handed sneutrino $Im(\tilde{\nu}_1)$ are
also shown for reference. Note that for
$T_\kappa^{112}=T_\kappa^{122}=0$ the mass of the state
$Re(\tilde{\nu}_1^c)$ does not depend on $v_{R2}$, whereas for
$T_\kappa^{112}=T_\kappa^{122}=-2$ GeV the lightest singlet scalar
becomes heavier for larger values of $v_{R2}$. The same feature is
present in the pseudoscalar sector, where the effect is even 
more pronounced.

\subsection{Correlations with neutrino mixing angles in the 
$n$ $\widehat{\nu}^c$-model}

The connection between decays and neutrino angles is not a
particular property of the $1$ $\widehat{\nu}^c$-model and is also
present in a general $n$ $\widehat{\nu}^c$-model. However,
since the structure of the approximate couplings
$\tilde{\chi}_1^0-W^{\pm}-l^{\mp}_i$ is different, see Appendix
\ref{sec:AppCoupXiWl}, we encounter additional features for $n=2$. 

As explained in Section \ref{subsec:ngenneut}, we have now 
two possibilities to fit neutrino
data. If the dominant contribution to the neutrino mass matrix comes
from the $\Lambda_i \Lambda_j$ term in Equation \eqref{eq:efftwo} one
can link it to the atmospheric mass scale, using the $\alpha_i\alpha_j$ term to
fit the solar mass scale. This case will be called option fit1. On the
other hand, if the dominant contribution is given by the $\alpha_i
\alpha_j$ term one has the opposite situation, where the atmospheric
scale is fitted by the $\alpha_i$ parameters and the solar scale is
fitted by the $\Lambda_i$ parameters. This case will be called option
fit2.

\begin{figure}
\begin{center}
\vspace{0mm}
\includegraphics[width=80mm,height=60mm]{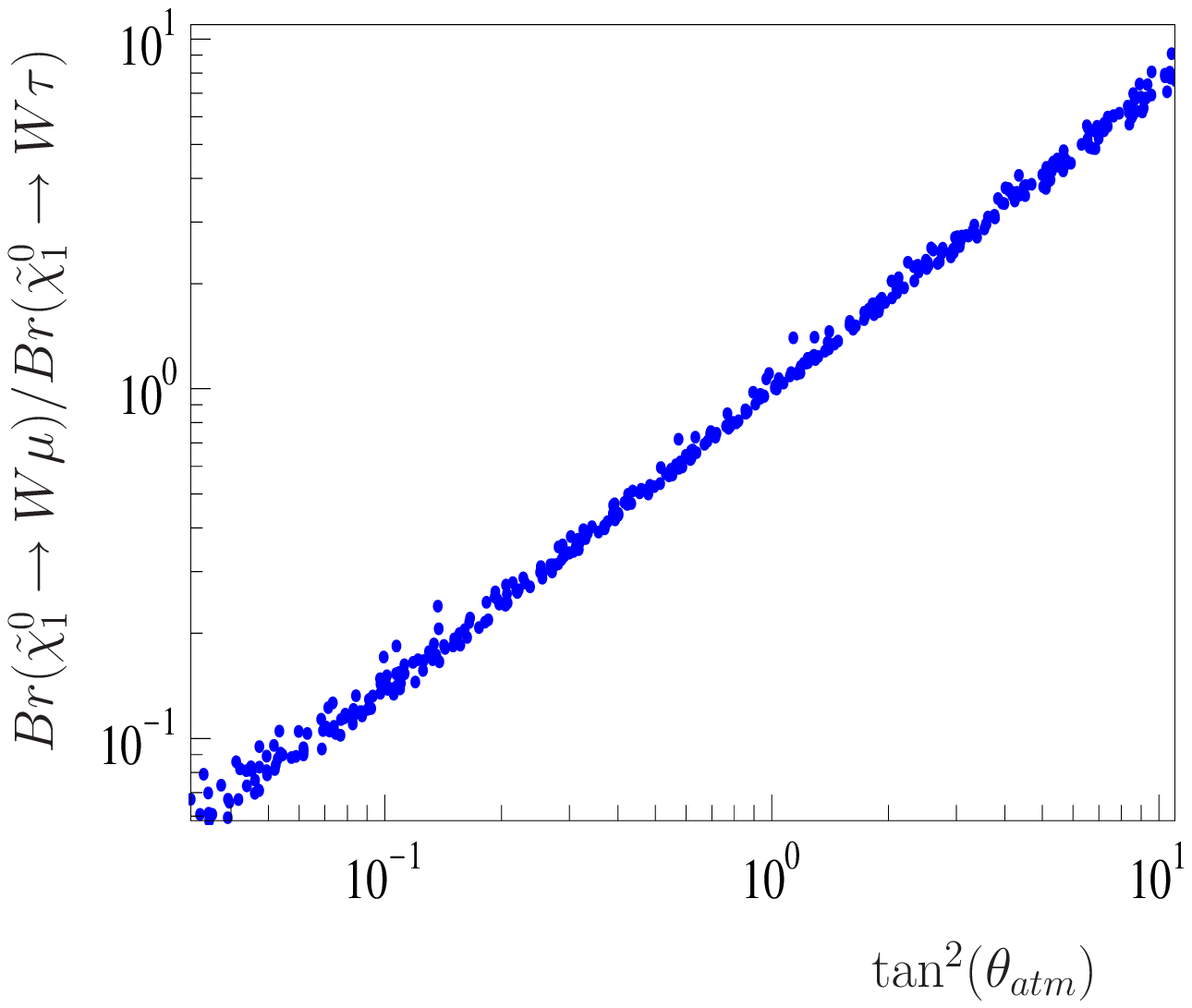}
\hspace{2mm}
\includegraphics[width=80mm,height=60mm]{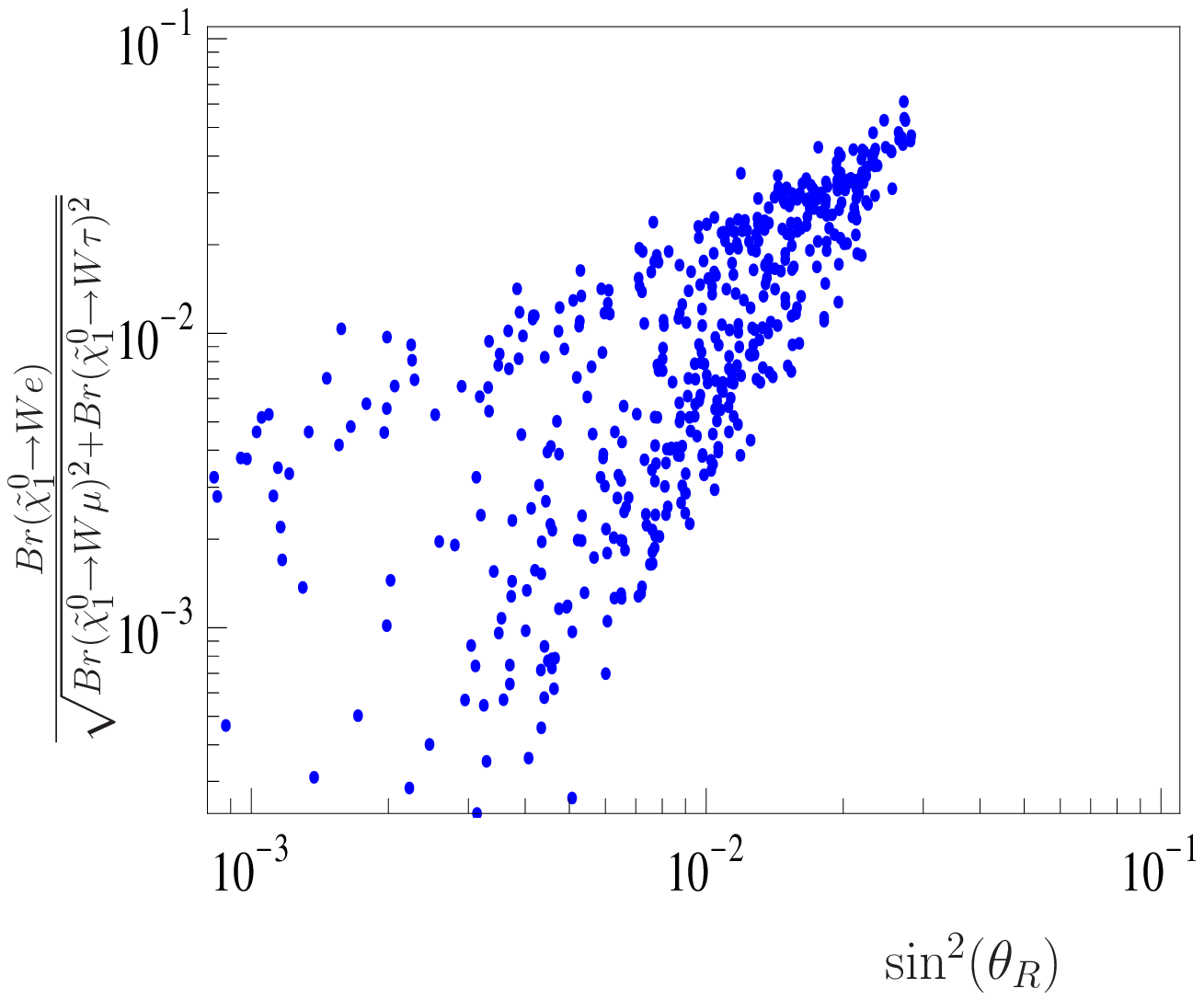}
\end{center}
\vspace{-6mm}
\caption{To the left (a) ratio $\frac{Br(\tilde{\chi}^0_1\to 
W\mu )}{Br(\tilde{\chi}^0_1\to W\tau)}$ versus $\tan^2(\theta_{atm})$
and to the right (b) ratio $\frac{Br(\tilde{\chi}^0_1\to 
We)}{\sqrt{Br(\tilde{\chi}^0_1\to W\mu )^2+Br(\tilde{\chi}^0_1\to 
W\tau )^2}}$ versus $\sin^2(\theta_R)$ for a bino LSP. Bino purity
$|\mathcal{N}_{41}|^2 > 0.9$. Neutrino data is fitted using option
fit1.}
\label{fig:binofit1}
\end{figure}

\begin{figure}
\begin{center}
\vspace{0mm}
\includegraphics[width=80mm,height=60mm]{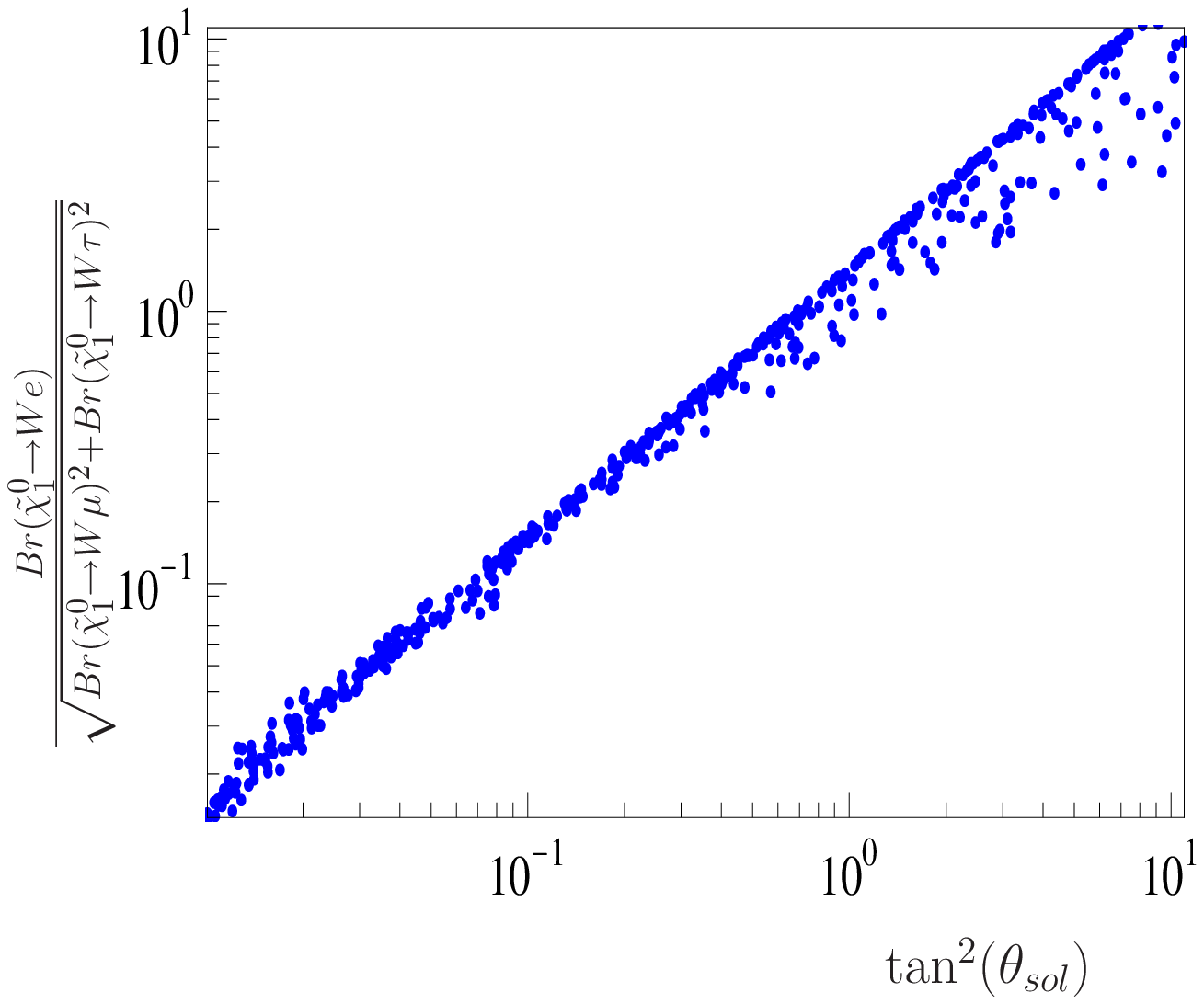}
\end{center}
\vspace{-7mm}
\caption{Ratio $\frac{Br(\tilde{\chi}^0_1\to 
We)}{\sqrt{Br(\tilde{\chi}^0_1\to W\mu )^2+Br(\tilde{\chi}^0_1\to 
W\tau )^2}}$ versus $\tan^2(\theta_{sol})$ for a bino LSP. Bino purity
$|\mathcal{N}_{41}|^2 > 0.9$. Neutrino data is fitted using option
fit2.}
\label{fig:binofit2}
\end{figure}

For the case of a bino-like lightest neutralino one can
show that the coupling is proportional to $\Lambda_i$ whereas for the
case of a singlino-like lightest neutralino the dependence is on
$\alpha_i$, as shown in Appendix \ref{sec:AppCoupXiWl}. 
Figure \ref{fig:binofit1} shows the ratio
$Br(\tilde{\chi}^0_1\to W\mu)/Br(\tilde{\chi}^0_1\to W\tau)$
versus $\tan^2(\theta_{atm})$ (left) and $Br(\tilde{\chi}^0_1\to
We)/\sqrt{Br(\tilde{\chi}^0_1\to W\mu)^2+Br(\tilde{\chi}^0_1\to
W\tau)^2}$ versus $\sin^2(\theta_R)$ (right) for a bino LSP and
option fit1. The correlation with the
atmospheric angle and the upper bound on $Br(\tilde{\chi}^0_1\to 
We)/\sqrt{Br(\tilde{\chi}^0_1\to W\mu)^2+Br(\tilde{\chi}^0_1\to 
W\tau)^2}$ from $\sin^2(\theta_R)$ is more pronounced than in the $1$ $\hat\nu^c$-model,
because we fit neutrino data with tree-level physics only.
Recall that this implies that the ratio $|\vec\epsilon|^2/|\vec\Lambda|$ is 
much smaller than in the plots shown in the previous section. 
A correlation between $Br(\tilde{\chi}^0_1\to 
We)/\sqrt{Br(\tilde{\chi}^0_1\to W\mu)^2+Br(\tilde{\chi}^0_1\to 
W\tau)^2}$ and $\tan^2(\theta_{sol})$ is found instead, if neutrino data
is fitted with option fit2, as Figure \ref{fig:binofit2} shows.

\begin{figure}
\begin{center}
\vspace{-3mm}
\includegraphics[width=80mm,height=60mm]{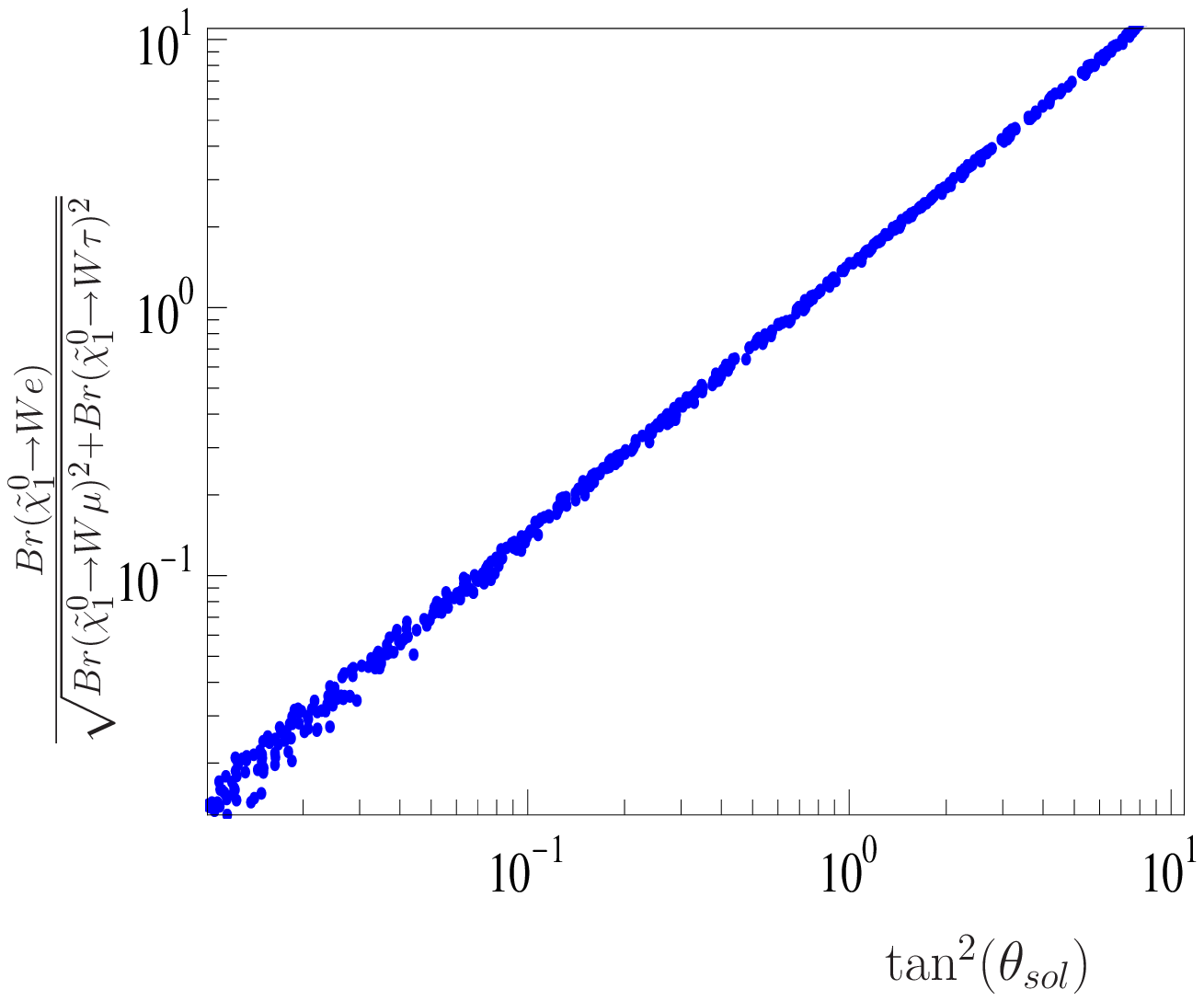}
\end{center}
\vspace{-8mm}
\caption{Ratio $\frac{Br(\tilde{\chi}^0_1\to 
We)}{\sqrt{Br(\tilde{\chi}^0_1\to W\mu )^2+Br(\tilde{\chi}^0_1\to W \tau 
)^2}}$ versus $\tan^2(\theta_{sol})$ for a singlino
LSP. Singlino purity $|\mathcal{N}_{45}|^2 > 0.9$. Neutrino data is
fitted using option fit1.}
\label{fig:singlinofit1}
\end{figure}

\begin{figure}
\begin{center}
\vspace{0mm}
\includegraphics[width=80mm,height=60mm]{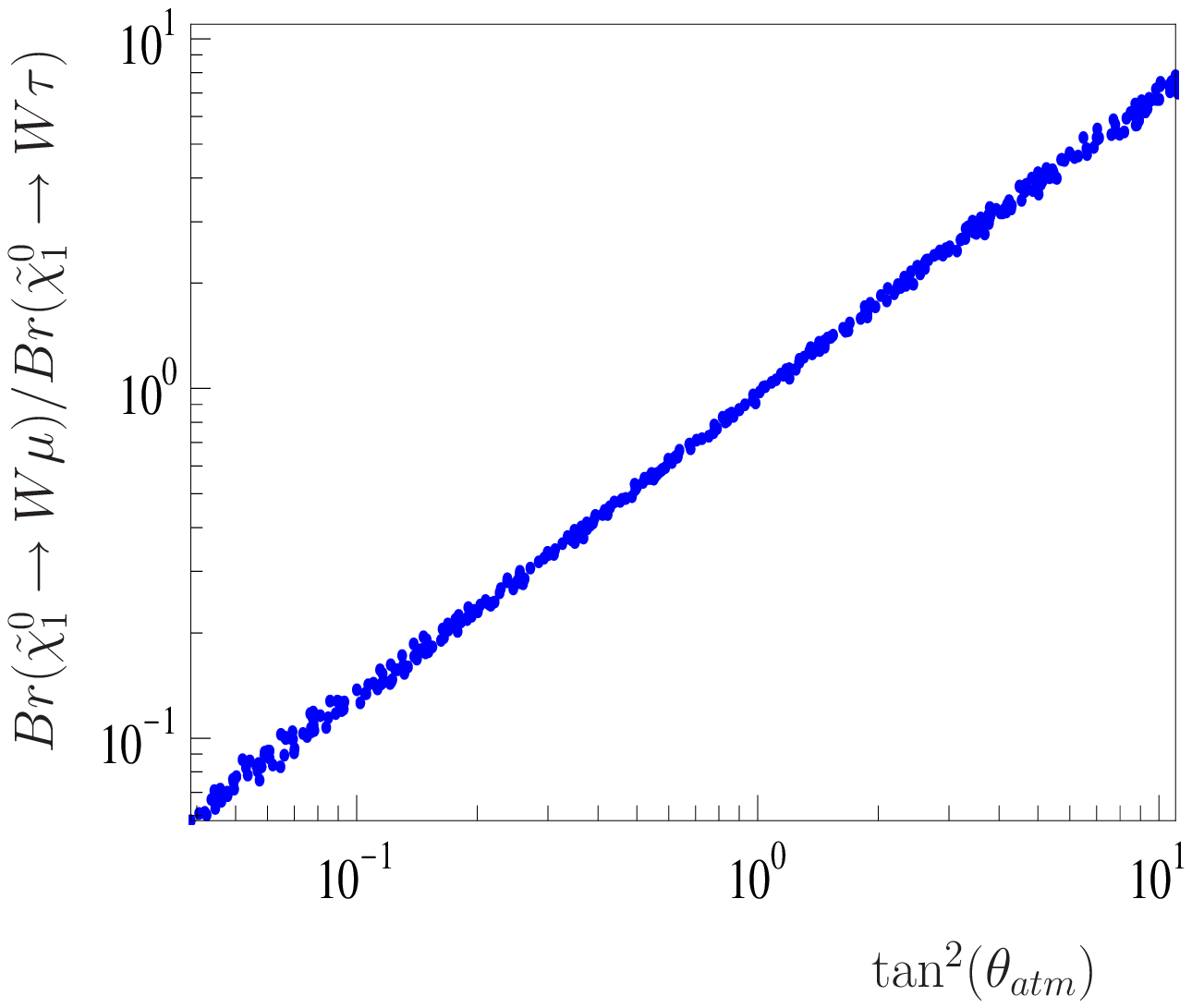}
\hspace{2mm}
\includegraphics[width=80mm,height=60mm]{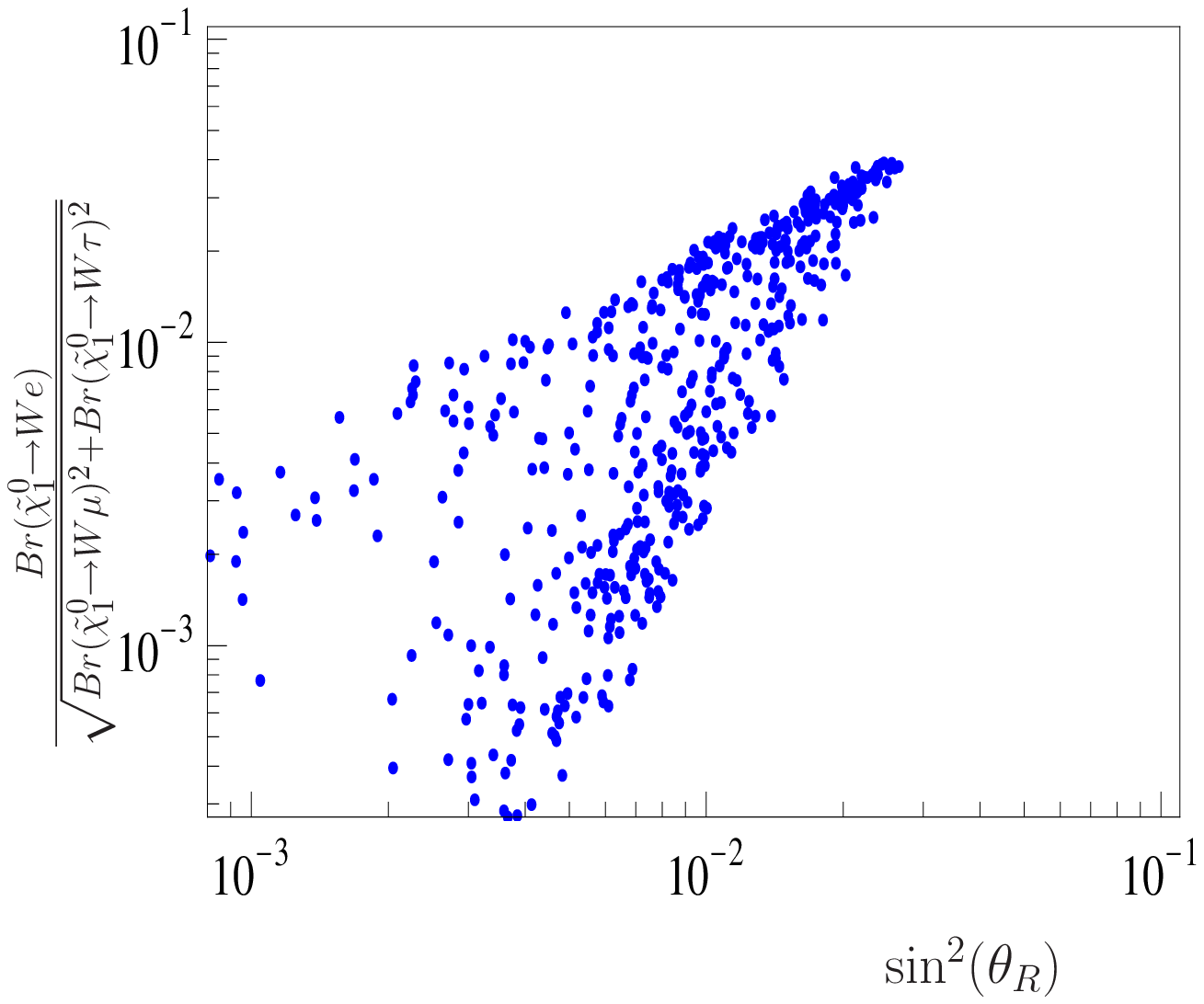}
\end{center}
\vspace{-8mm}
\caption{To the left (a) ratio $\frac{Br(\tilde{\chi}^0_1\to 
W \mu)}{Br(\tilde{\chi}^0_1\to W \tau )}$ versus $\tan^2(\theta_{atm})$
and to the right (b) ratio $\frac{Br(\tilde{\chi}^0_1\to 
We)}{\sqrt{Br(\tilde{\chi}^0_1\to W\mu )^2+Br(\tilde{\chi}^0_1\to 
W \tau)^2}}$ versus $\sin^2(\theta_R)$ for a singlino
LSP. Singlino purity $|\mathcal{N}_{45}|^2 > 0.9$. Neutrino data is
fitted using option fit2.}
\label{fig:singlinofit2}
\end{figure}

For the case of a singlino LSP the correlations and types of fit to
neutrino data are swapped with respect to the gaugino case. Since the
couplings $\tilde{\chi}_1^0-W^{\pm}-l^{\mp}_i$ are mainly proportional
to $\alpha_i$, instead of $\Lambda_i$, a scenario with a singlino LSP
and option fit1 (fit2) will be similar to bino LSP and option fit2
(fit1). This similarity is demonstrated in Figures
\ref{fig:singlinofit1} and \ref{fig:singlinofit2}.  To decide which
case is realized in nature, one would need to determine the particle
character of the lightest neutralino. This might be difficult at the
LHC, but could be determined by a cross section measurement at the
ILC. We want to note, that in the $2$ $\widehat{\nu}^c$-model
we cannot reproduce all correlations for a singlino LSP presented
for the $3$ $\widehat{\nu}^c$-model in \cite{Ghosh:2008yh}.

The results shown so far in this section were all calculated for 
the SPS1a' scenario. We have checked explicitly that for all the 
other standard points results remain unchanged. 
We have also checked that for a LSP with a mass below $m_W$ the
three-body decays $\tilde{\chi}_1^0 \to l q_i \bar{q}_j$, mediated by
virtual W bosons, show the same correlations.

A final comment is in order. In a $n$ $\widehat{\nu}^c$-model with
$n>2$, the effective neutrino mass matrix will have additional terms
with respect to \eqref{eq:efftwo}, due to the contributions coming
from the new right-handed neutrinos. For this richer structure there
is one additional contribution to $\boldsymbol{m_{\nu\nu}^{\rm eff}}$,
which could be sub-dominant. Therefore, one can imagine a scenario 
in which a third generation of singlets produces a negligible 
contribution to neutrino masses while the corresponding singlino, 
$\nu_3^c$, is the LSP. In such a scenario the correlations 
between the $\nu_3^c$ LSP decays and the neutrino mixing angles 
will be lost.

\subsection{$\tilde{\chi}_1^0$ decay length and type of fit}
\label{subsec:declengfit}

\begin{figure}
\begin{center}
\vspace{0mm}
\includegraphics[width=80mm,height=60mm]{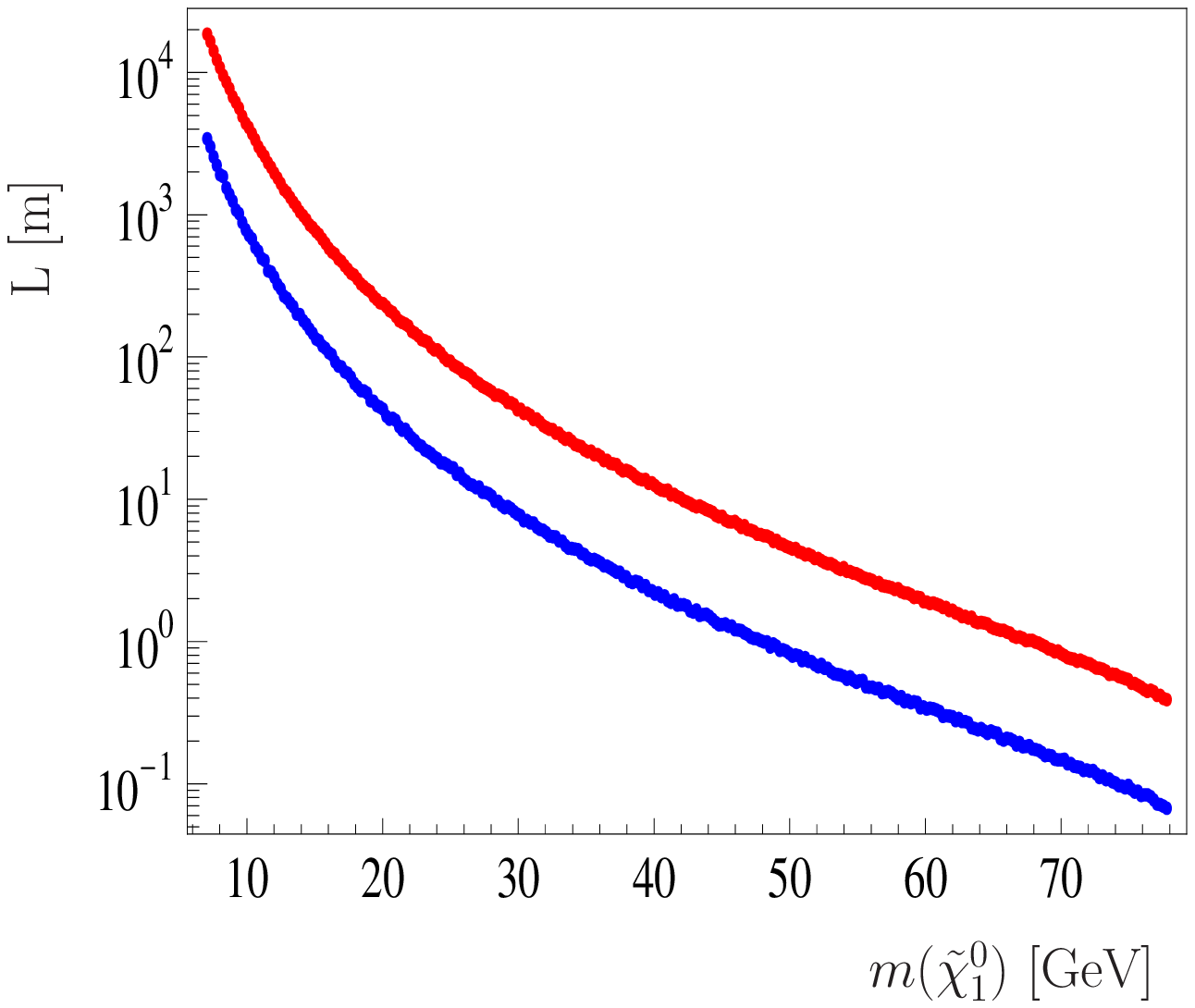}
\hspace{2mm}
\includegraphics[width=80mm,height=60mm]{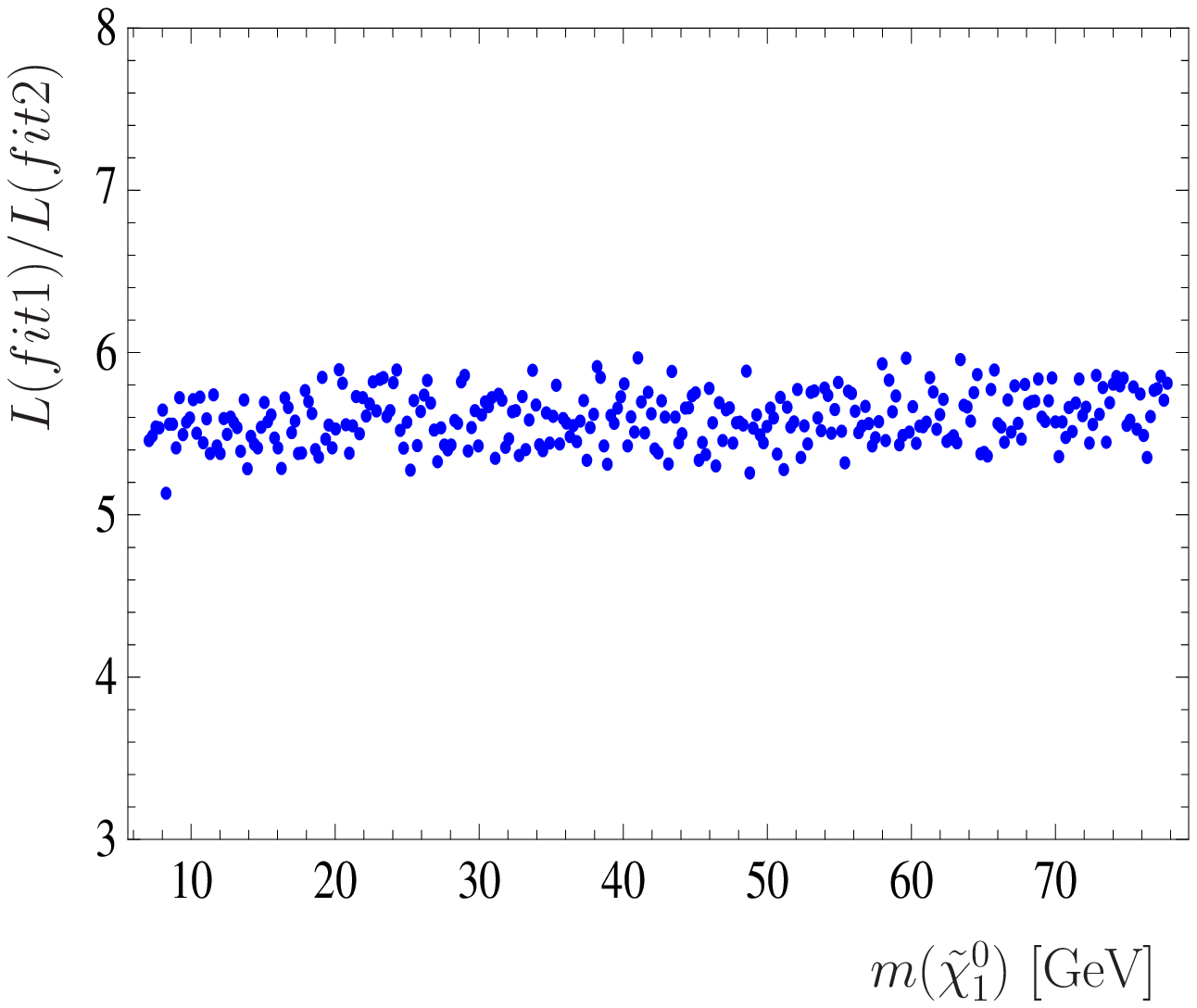}
\end{center}
\vspace{-6mm}
\caption{Decay length of the lightest neutralino and its dependence on
the type of fit to neutrino data. To the left (a) the decay length of the
lightest neutralino versus $m(\tilde{\chi}_1^0)$ for the case fit1
(red) and the case fit2 (blue). To the right (b) the ratio
$L(\text{fit1})/L(\text{fit2})$ versus $m(\tilde{\chi}_1^0)$. The MSSM
parameters have been taken such that the standard SPS1a' point is
reproduced. The light singlet parameter $\kappa$ is varied in the
range $\kappa\in\left[0.01,0.1\right]$. In all the points the lightest
neutralino has a singlino purity higher than $0.99$.}
\label{fig:nfitlength}
\end{figure}

As already discussed we have two different possiblities to fit 
neutrino data: $\vec \Lambda$ generates the atmospheric mass scale and
$\vec \alpha$ the solar mass scale (case fit1), or vice versa (case
fit2). It turns out that the decay length of the lightest neutralino
is sensitive to the type of fit, due to the proportionality between
its couplings with gauge bosons and the \rpv parameters (see Appendix
\ref{sec:AppCoupXiWl} for exact and approximated formulas of the
couplings $\tilde{\chi}_1^0-W^{\pm}-l^{\mp}_i$ and their simplified
expressions in particular limits). For example, a singlino-like
neutralino couples to the gauge bosons proportionally to the $\alpha_i$
parameters. This implies that its decay length will follow $L \propto
1/|\vec \alpha|^2$ and obeys the approximate relation
\begin{equation}
\frac{L(fit1)}{L(fit2)} \simeq \frac{m_{atm}}{m_{sol}} \simeq 6\quad.
\end{equation}
In Figure \ref{fig:nfitlength} the decay length of the lightest
neutralino and its dependence on the type of fit to neutrino data is
shown. Once mass and length are known this dependence can be used to
determine which parameters generate which mass scale.  Note that this
feature is essentially independent of the MSSM parameters. However,
this property is lost if either the lightest neutralino has a sizeable 
gaugino/higgsino component or if there are singlet scalars/pseudoscalars
lighter than the singlino.

\subsection{Several light singlets}
\label{subsec:scenariod}

In scenarios with two (or more) light singlets, the phenomenology has
additional features. The light Higgs boson $h^0$ can decay with
measurable branching ratios to pairs of right-handed neutrinos of
different generations. Similarly, the bino can decay to the different
light right-handed neutrinos. 

In the following, the case of two light singlinos and two light
scalars/pseudoscalars will be considered. For the neutral fermion sector
this implies that the mass eigenstates $\tilde{\chi}_1^0$ and
$\tilde{\chi}_2^0$ will always be the singlets $\nu_1^c$ and $\nu_2^c$
and the bino will be the $\tilde{\chi}_3^0$. In the
scalar sector one has two very light mostly singlet states $S_1^0$ 
and $S_2^0$, which are consistent with the LEP bounds. Finally, 
the state $S_3^0$ will be the light doublet Higgs boson
$h^0$. One can also have light singlet pseudoscalars.

The decays of a bino-like $\tilde{\chi}_3^0$ can be very important to
distinguish between the one generation model and models with more than
one generation of singlets. 
In principle, the most important decay channels strongly depend on the
couplings of the bino to the two generations of singlinos and the
configuration of masses of singlinos and scalars. Therefore, a general
list of signals cannot be given. Nevertheless, there are some features
which are always present:

\begin{figure}
\begin{center}
\vspace{0mm}
\includegraphics[width=80mm,height=60mm]{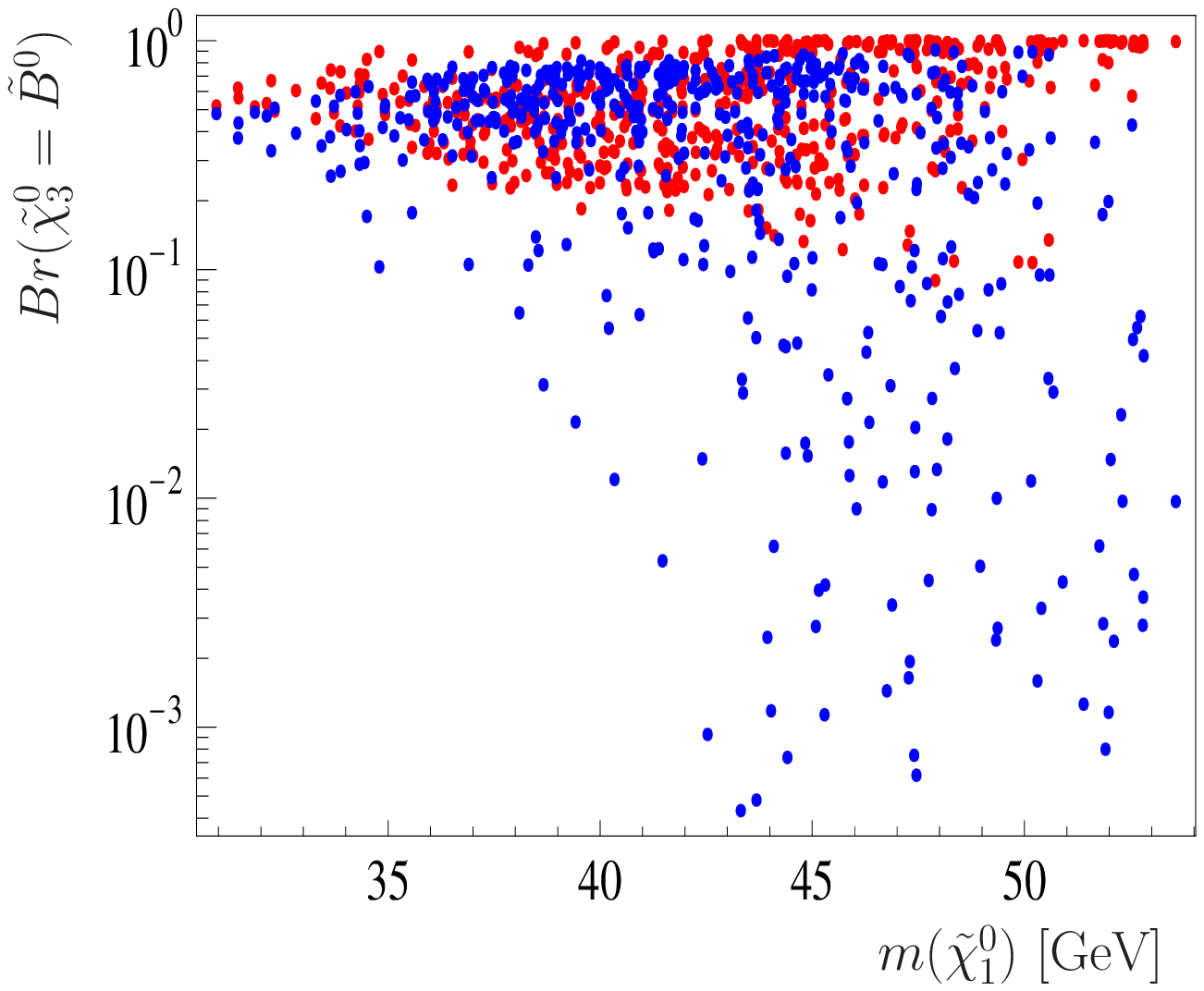}
\end{center}
\vspace{-6mm}
\caption{Branching ratios $Br(\tilde{\chi}_3^0 = \tilde{B}^0 \to
\tilde{\chi}_1^0)$ (red) and $Br(\tilde{\chi}_3^0 = \tilde{B}^0 \to
\tilde{\chi}_2^0)$ (blue) as a function of the mass of the lightest
neutralino for the scenario considered in Section
\ref{subsec:scenariod}. The MSSM parameters have been taken such that
the standard SPS1a' point is reproduced, whereas the singlet
parameters are chosen randomly in the ranges $v_{R1},v_{R2} \in
[400,600]$ GeV, $\lambda_1,\lambda_2 \in [0.0,0.4]$, $T_\kappa^{111} =
T_\kappa^{222} \in [-15,-1]$ GeV, $T_\kappa^{112} = T_\kappa^{122} \in
[-1.5,-0.005]$ GeV and $T_\lambda^1,T_\lambda^2 \in [0,600]$ GeV. $\kappa_1 =
\kappa_2 = 0.16$ is fixed to ensure the lightness of the two
singlinos.}
\label{fig:binodecD}
\end{figure}

When kinematically allowed, the decays $\tilde{\chi}_3^0 \to
\tilde{\chi}_{1,2}^0 \: S_1^0 (P_1^0)$ dominate, with the sum of the
branching ratios typically larger than 50 \%.  The relative importance
of the different channels is mainly dictated by kinematics. This
feature is illustrated in Figure \ref{fig:binodecD}, where these two
quantities are shown as a function of the mass of the lighest
neutralino. The MSSM parameters are fixed to the standard point
SPS1a', with light singlet parameters taken randomly. One can see
that the relative importance of each singlino cannot be predicted in
general, but both branching ratios are at least of order
$10^{-3}-10^{-4}$, given enough statistics. For very light singlinos
two-body decays including scalars and pseudoscalars are open, and thus
both $Br(\tilde{\chi}_3^0 \to \tilde{\chi}_1^0)$ and
$Br(\tilde{\chi}_3^0 \to \tilde{\chi}_2^0)$ are close to $50\%$,
as expected if the values of the singlet parameters are of the same
order for the two light generations. On the other hand, if the mass of
the lightest neutralino is increased some of the two-body decays are
kinematically forbidden, specially those of the $\tilde{\chi}_2^0$,
which has to be produced through three-body decays, leading to a
suppresion in $Br(\tilde{\chi}_3^0 \to \tilde{\chi}_2^0)$.  Note that
it is also possible to find points where the decay mode
$\tilde{\chi}_3^0 \to \tilde{\chi}_{1,2}^0 \: S_2^0 (P_2^0)$ has a
branching ratio about 10\%-20\%, giving additional information.

The other possible signals are the usual bino 
decays of the NMSSM. Final states with standard model particles, like
$\tilde{\chi}_{1,2}^0 l^+ l^-$ or $\tilde{\chi}_{1,2}^0 q \bar{q}$,
become very important when the decays to scalars and pseudoscalars are
kinematically forbidden.

In addition, the decays of the light Higgs boson $h^0$ can also
play a very important role in the study of the different generations,
provided it can decay to final states including
$\tilde{\chi}_1^0$ or $\tilde{\chi}_2^0$. In this case typically 
the standard Higgs boson decays are reduced to less than 40\%, 
completely changing the usual search strategies.

\begin{figure}
\begin{center}
\vspace{-2mm}
\includegraphics[width=80mm,height=60mm]{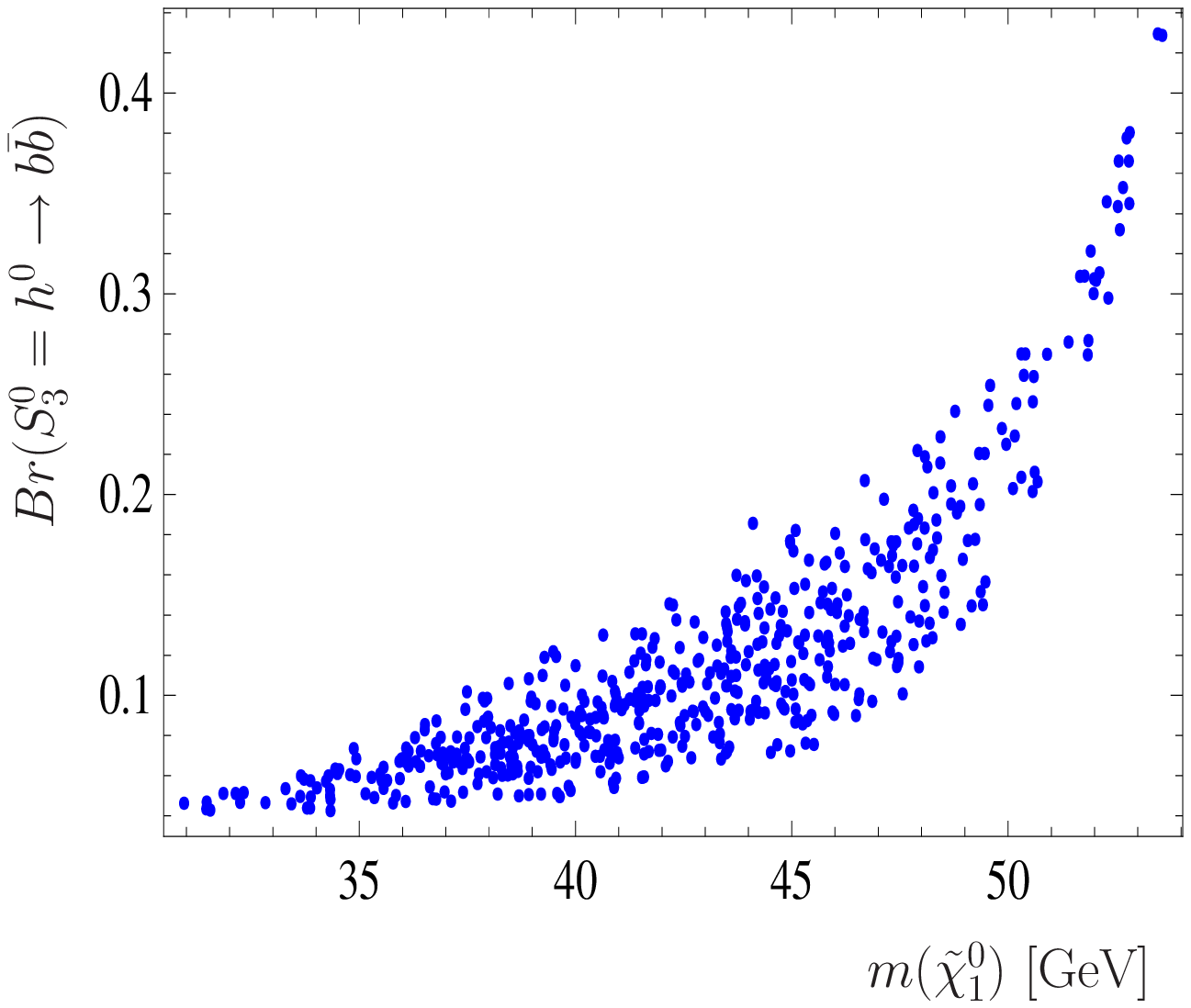}
\hspace{2mm}
\includegraphics[width=80mm,height=60mm]{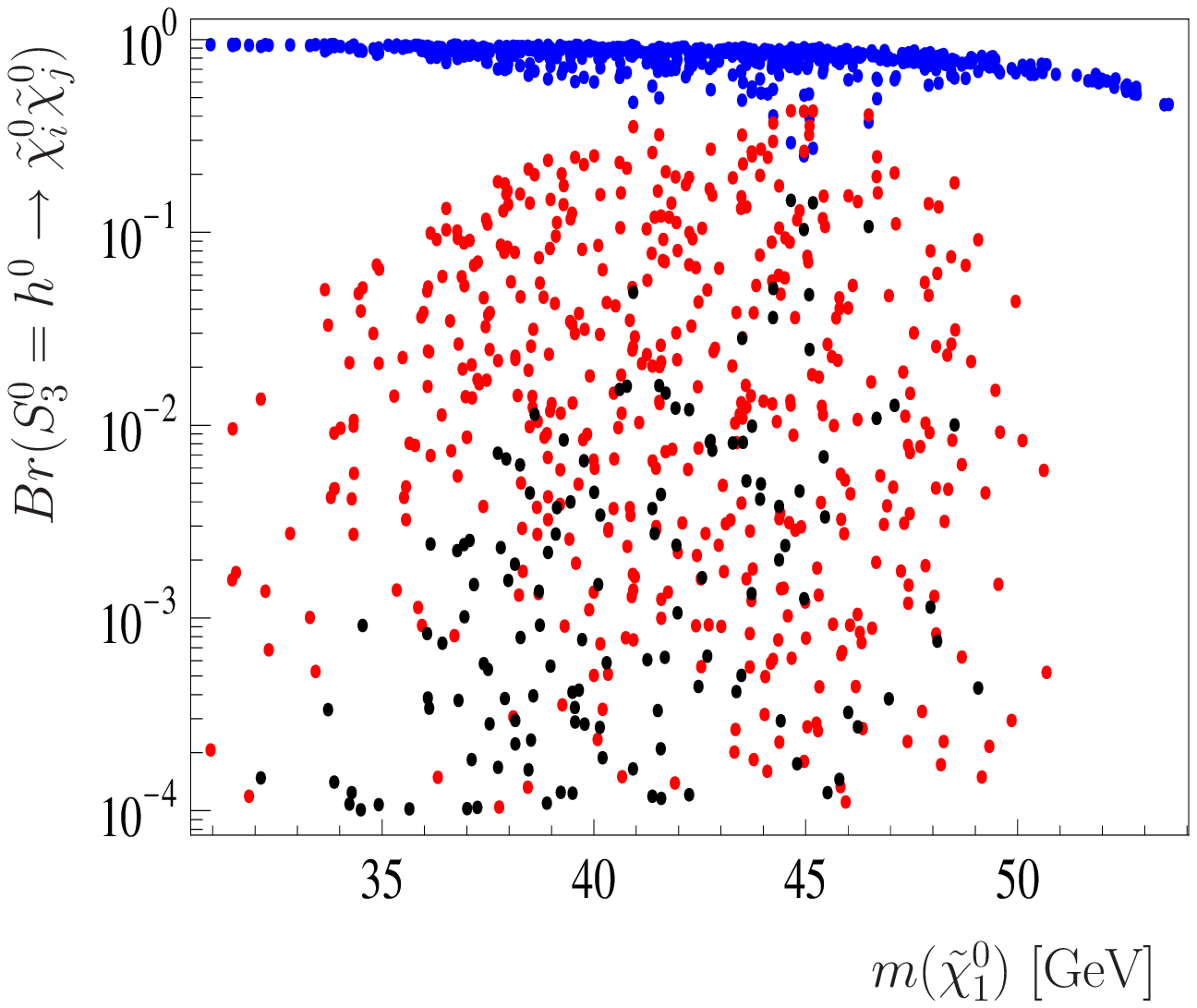}
\end{center}
\vspace{-6mm}
\caption{Higgs boson decays as a function of the mass of the lightest
neutralino for the scenario considered in Section
\ref{subsec:scenariod}. To the left (a) the standard decay channel $h^0
\to b \bar{b}$, whereas to the right (b) the exotic decays to pairs of
singlinos $h^0 \to \tilde{\chi}_1^0\tilde{\chi}_1^0$ (red), $h^0 \to
\tilde{\chi}_1^0\tilde{\chi}_2^0$ (blue) and $h^0 \to
\tilde{\chi}_2^0\tilde{\chi}_2^0$ (black). The parameters are 
chosen as in Figure \ref{fig:binodecD}.}
\label{fig:higgsdecaysD}
\end{figure}

In Figure \ref{fig:higgsdecaysD} the branching ratios of standard and
exotic Higgs boson decay channels are shown. The left plot shows the
suppressed branching ratio of the standard $b \bar{b}$ channel.  The
main decay channel is $\tilde{\chi}_1^0 \: \tilde{\chi}_1^0$, but
there is a sizeable branching ratio to $\tilde{\chi}_1^0 \:
\tilde{\chi}_2^0$. Note that $\tilde{\chi}_2^0$ decays dominantly to
$\tilde{\chi}_1^0$ plus two SM fermions. This feature allows us to
distinguish between the $1$ $\widehat{\nu}^c$-model and models with
more than one generation of singlets. Finally, the branching ratio to
$\tilde{\chi}_2^0 \: \tilde{\chi}_2^0$ is small due to kinematics, but
leads to interesting final states with up to eight b-jets plus missing
energy.

A final comment is in order. In these kind of scenarios with many
light singlets $\tilde{\chi}_1^0$ decays to $\nu b{\bar b}$ can 
be dominant. This will reduce the available statistics in the 
interesting $l_i l_j \nu$ and $l q_i \overline{q}_j$ channels. Moreover, 
the correlations are less pronounced due to mixing effects in 
the singlet sector.

\section{Discussion and conclusions}
\label{sec:Cncl}

We have studied the phenomenology of the $\mu\nu$SSM. This proposal 
solves at the same time the $\mu$-problem of the MSSM and generates 
small neutrino masses, consistent with data from neutrino oscillation 
experiments. Neutrino data put very stringent constraints on the 
parameter space of the model. Both the left-sneutrino vacuum expectation 
values and the effective bilinear parameters have to be small compared 
to MSSM soft SUSY breaking parameters. As a result all SUSY production 
cross sections and all decay chains are very similar to the NMSSM, the 
only, but phenomenologically very important, exceptions being the decay 
of the LSP and NLSP (the latter only in some parts of the parameter space)
plus the decays of the lightest Higgses.

We have discussed in some details two variants of the model. In the 
simplest version with only one generation of singlets 1-loop corrections 
to the neutralino-neutrino mass matrix need to be carefully calculated 
in order to explain neutrino data correctly. The advantage of this minimal 
scheme is that effectively it contains only six new (combinations of) 
\rpv parameters, which can be fixed to a large extend by the requirement 
that oscillation data is correctly explained. This feature of the model 
is very similar to explicit bilinear R-parity breaking, although, as we have 
discussed, the relative importance of the different 1-loop contributions 
is different in the $\mu\nu$SSM and in bilinear \rpv. Certain ratios of 
decay branching ratios depend on the same parameter combinations as 
neutrino angles and are therefore predicted from neutrino physics, to a 
large extend independent of NMSSM parameters. We have also calculated the 
decay length of the LSP, which depends mostly on the LSP mass and the 
(experimentally determined) neutrino masses. Lengths sufficiently large 
to observe displaced vertices are predicted over most parts of the parameter 
space. However, for neutralinos lighter than approximately 30 GeV, decay 
lengths become larger than 10 meter, making the observation of \rpv 
difficult for LHC experiments. However, if there is a singlet scalar or 
pseudoscalar with a mass smaller than the lightest neutralino, 
$\tilde{\chi}^0_1 \to S_m^0(P_m^0) \nu$ is the dominant decay mode and the corresponding 
decay lengths become much smaller, such that the displaced vertex 
signature of \rpv might even be lost in some points of this part of parameter 
space. On the other hand, in case the mass of the lightest scalar is larger 
than twice the singlino mass, the decay $S_m^0 \to 2 \tilde{\chi}^0_1$ becomes 
important, both for $S^0_m \sim {\tilde\nu^c}$ and $S^0_m \sim h^0$. 
If this kinematical situation is realized also the Higgs search at the 
LHC will definitely be affected. 

The more involved $n$ generation variants of the $\mu\nu$SSM can explain 
all neutrino data at tree-level and therefore are {\em calculationally} 
simpler. Depending on the nature of the neutralino, neutralino LSP decays 
show different correlations with either solar or atmospheric neutrino angles. 
This is guaranteed in the two generation version of the model and likely, 
but not always true, for $n$ generations. If the NMSSM coupling $\lambda$ 
is sufficiently small also the NLSP has decays to \rpv final states 
with potentially measurable branching ratios. In this part of parameter 
space it seems possible,  in principle, to test both solar and atmospheric 
neutrino angles. If only the singlino(s) are light, i.e. the singlet 
scalars are heavier than, say, the $h^0$, the decay length of the singlino 
is very sharply predicted as a function of its mass and either the solar or 
atmospheric neutrino mass scale. If both, singlinos and singlet scalars 
(or pseudoscalars) are light, bino NLSP and $h^0$ will decay not only 
to the lightest singlinos/singlets but also to next-to-lightest states. 
This leads to enhanced multiplicities in the final states and the 
possibility to observe multiple displaced vertices.

We now briefly discuss possible differences in collider 
phenomenology of the $\mu\nu$SSM and other R-parity breaking schemes. 
Different models of R-parity breaking appear clearly distinct at 
the Lagrangian level. However, at accelerator experiments it can 
be very hard to distinguish the different proposals. This can be 
easily understood from the fact that for a heavy singlet sector 
all \rpv models approach necessarily the MSSM with explicit R-parity 
breaking terms. It is therefore an interesting question to ask, what 
- if any - kind of signals could exist, which at least might hint at 
which model is the correct description of \rpv. Given the large variety 
of possibilities and the very limited predictive power of the most 
general cases, any discussion {\em before} the discovery of SUSY must 
be rather qualitative. 

First one should mention that not all \rpv models explaining neutrino 
data show correlations between LSP decay branching ratios and neutrino 
angles. Especially the large number of free parameters in trilinear 
models exclude the possibility to make any definite predictions. 
\rpv models which do show such correlations, on the other hand, lead 
usually to very similar predictions for the corresponding LSP decays. 
For example, fitting the atmospheric data with tree-level \rpv terms, 
a bino LSP in explicit bilinear models and in the $\mu\nu$SSM decay 
with the same ratio of branching ratios into $W l$ (or $lq_i{\bar q}_j$) 
final states. Thus, to distinguish the different proposals other signals 
are needed.

We will briefly discuss the main differences in collider phenomenology 
between the following three proposals: (i) MSSM with explicit bilinear 
terms (b-\rpv); (ii) Spontaneous \rpv (s-\rpv) model and (iii) 
$\mu\nu$SSM. Table (\ref{tab:comp}) shows a brief summary of this 
comparison. Differences occur in (a) the observability of a displaced 
vertex of the lightest neutralino decay; (b) the upper limit on the 
branching ratio of the lightest neutralino decaying completely invisible 
and (c) standard versus non-standard lightest Higgs decays.

The decay length of the lightest neutralino is fixed in both, the
b-\rpv model and the $\mu\nu$SSM, essentially by the mass of the
lightest neutralino and the experimentally determined neutrino masses.
For $m(\tilde{\chi}^0_1)$ larger than the W-mass decay lengths are
typically of the order of ${\cal O}(mm)$ and proportional to
$m^{-1}(\tilde{\chi}^0_1)$.  For lighter neutralinos, larger decay
lengths are expected, see Figures \ref{fig:1NuR_decaylength} and
\ref{fig:nfitlength}, which scale like
$m^{-4}(\tilde{\chi}^0_1)$. Shorter decay lengths are not possible in
b-\rpv and possible in the $\mu\nu$SSM only if at least one
(singlet) scalar or pseudoscalar is lighter than $\tilde{\chi}^0_1$,
when ${\tilde{\chi}^0_1} \to S_m^0(P_m^0) \nu$ dominates. Since in the
$\mu\nu$SSM the singlet scalars decay with a short decay length to
${\bar b}b$, one expects that in the $\mu\nu$SSM short
$\tilde{\chi}^0_1$ decay lengths correlate with the dominance of
${\bar b}b$ + missing energy final states. In the s-\rpv, on the other
hand, the $\tilde{\chi}^0_1$ decay length can be shorter than in the
b-\rpv, due to the new final state $\tilde{\chi}^0_1 \to J + \nu$,
where $J$ is the Majoron. Therefore, different from the $\mu\nu$SSM,
the neutralino decay length in the s-\rpv model anti-correlates with
the branching ratio for the invisible neutralino decay.

Finally, in the b-\rpv one expects that the decay properties of the 
lightest Higgs ($h^0$) are equal to the MSSM expectations, the only 
exception being the case when $h^0 \to 2 \tilde{\chi}^0_1$ is possible 
kinematically, in which the $\tilde{\chi}^0_1$ decays themselves can then 
lead to a non-standard signal in the Higgs sector. This is different 
in s-\rpv, where for a low-scale of spontaneous R-parity breaking, 
the $h^0$ can decay to two Majorons, i.e. large branching ratios 
of Higgs to invisible particles are possible. In the $\mu\nu$SSM 
the $h^0$ decays can be non-standard, if the lightest singlino is 
lighter than $m(h^0)/2$. However, since the singlinos decay, 
this will not lead to an invisible Higgs, unless the mass of the 
singlino is so small, that the decays occur outside the detector.

\begin{table}

\begin{center}
\begin{tabular}{|c|c|c|c|c|}
\hline
   & Displaced vertex & Comment & Br(invisible) & Higgs decays \\
\hline
\hline
b-\rpv & Yes & Visible & $\le 10$ \% & standard \\
\hline
s-\rpv & Yes/No & anti-correlates with invisible & any & non-standard 
(invisible) \\
\hline
$\mu \nu$SSM & Yes/No & anti-correlates with non-standard Higgs 
& $\le 10$ \% & non-standard \\
\hline
\end{tabular}
\end{center}
\caption{Comparison of displaced vertex signal, completely invisible 
final state branching ratios for LSP decays and lightest Higgs decays 
for three different R-parity violating models. For a discussion see text.}
\label{tab:comp}
\end{table}

To summarize this brief discussion, b-\rpv, s-\rpv and $\mu\nu$SSM can, 
in principle, be distinguished experimentally {\em if the singlets are 
light enough to be observed} in case of s-\rpv and $\mu\nu$SSM. We note in 
passing that we have not found any striking differences in collider 
phenomenology of the $\mu\nu$SSM and the NMSSM with explicit bilinear 
terms.

In conclusion, the $\mu\nu$SSM offers a very rich phenomenology. 
Especially scenarios with light singlets deserve further, much  
more detailed studies.

\section*{Acknowledgments}
This work was supported by Spanish grants FPA2008-00319/FPA and Accion 
Integrada NO HA-2007-0090 (MEC) and by the ``Fonds zur F\"orderung der 
wissenschaftlichen Forschung (FWF)'' of Austria, project No. P18959-N16.
The authors acknowledge support from EU under the MRTN-CT-2006-025505
and MTRN-CT-2006-503369 network programmes.
A.B. was supported by the Spanish grants SAB 2006-0072, FPA 2005-01269,
FPA 2005-25348-E and FPA 2008-00319/FPA of the Ministero de Education
y Ciencia. A.V. thanks the Generalitat Valenciana for financial support.
W.P. and S.L. are supported by the DAAD, project number D/07/13468 and
by the Initiative and Networking Fund of the Helmholtz Association, 
contract HA-101 (``Physics at the Terascale'').

\appendix

\section{Mass matrices}
\label{sec:MassMat}

In the scalar mass matrices shown below the tadpole equations 
have not yet been used to reduce the number of free parameters.

\subsection{Charged Scalars}
\label{subsec:cscalars}

In the basis
\begin{eqnarray}
\big( {S^+}' \big)^T &=& ((H_d^-)^*,H_u^+,\tilde e_L^*, \tilde
\mu_L^*, \tilde \tau_L^*, \tilde e_R, \tilde \mu_R, \tilde \tau_R)
\nonumber \\ \big( {S^-}' \big)^T &=& (H_d^-,(H_u^+)^*,\tilde e_L,
\tilde \mu_L, \tilde \tau_L, \tilde e_R^*, \tilde \mu_R^*, \tilde
\tau_R^*)
\end{eqnarray}
the scalar potential includes the term
\begin{equation}
V \supset \big( {S^-}' \big)^T M_{S^\pm}^2 {S^+}'\quad,
\end{equation}
where $M_{S^\pm}^2$ is the $(8 \times 8)$ mass matrix of the charged
scalars. In the $\xi = 0$ gauge it can be written as
\begin{equation}
M_{S^\pm}^2 = \left( \begin{array}{c c}
M_{HH}^2 & \big( M_{H \tilde l}^2 \big)^\dagger \\
M_{H \tilde l}^2 & M_{\tilde l \tilde l}^2 \end{array} \right)\quad.
\end{equation}
The $(2 \times 2)$ $M_{HH}^2$ matrix is given by:
\begin{eqnarray}
\big( M_{HH}^2 \big)_{11} &=& m_{H_d}^2 + \frac{1}{8} [ (g^2 + g'^2)
v_d^2 + (g^2 - g'^2)(v_u^2 - v_1^2 - v_2^2 - v_3^2) ] \nonumber \\ &&
+ \frac{1}{2} \lambda_s \lambda_t^* v_{Rs} v_{Rt} + \frac{1}{2} v_i
\big( h_E h_E^\dagger \big)_{ij} v_j \nonumber \\ \big( M_{HH}^2
\big)_{12} &=& \frac{1}{4}g^2 v_u v_d - \frac{1}{2} \lambda_s
\lambda_s^* v_u v_d + \frac{1}{4} \lambda_s \kappa_s^* v_{Rs}^2 +
\frac{1}{2}v_u v_i \lambda_s (h_\nu^{is})^* + \frac{1}{\sqrt{2}}
v_{Rs} T_\lambda^s \nonumber \\ \big( M_{HH}^2 \big)_{21} &=& \big(
M_{HH}^2 \big)_{12}^* \nonumber \\ \big( M_{HH}^2 \big)_{22} &=&
m_{H_u}^2 + \frac{1}{8} [ (g^2 + g'^2) v_u^2 + (g^2 - g'^2)(v_d^2 +
v_1^2 + v_2^2 + v_3^2) ] \nonumber \\ && + \frac{1}{2} \lambda_s
\lambda_t^* v_{Rs} v_{Rt} + \frac{1}{2} v_{Rs} v_{Rt} h_\nu^{is}
(h_\nu^{it})^*
\end{eqnarray}
The $(6 \times 2)$ matrix that mixes the charged Higgs bosons with the
charged sleptons is
\begin{equation}
M_{H \tilde l}^2 = \left( \begin{array}{c}
M_{HL}^2 \\
M_{HR}^2 \end{array} \right)
\end{equation}
with:
\begin{eqnarray}
\big( M_{HL}^2 \big)_{i1} &=& \frac{1}{4} g^2 v_d v_i - \frac{1}{2}
\lambda_s^* h_\nu^{it} v_{Rs} v_{Rt} - \frac{1}{2}v_d \big( h_E
h_E^\dagger \big)_{ij} v_j \nonumber \\ \big( M_{HL}^2 \big)_{i2} &=&
\frac{1}{4} g^2 v_u v_i - \frac{1}{4} \kappa_s^* v_{Rs}^2 h_\nu^{is} +
\frac{1}{2} v_u v_d \lambda_s^* h_\nu^{is} -\frac{1}{2} v_u v_j
h_\nu^{is} (h_\nu^{js})^* - \frac{1}{\sqrt{2}} v_{Rs} T_{h_\nu}^{is}
\nonumber \\ \big( M_{HR}^2 \big)_{i1} &=& -\frac{1}{2} v_u v_{Rs}
(h_E^*)_{ji} h_\nu^{js} - \frac{1}{\sqrt{2}}v_j (T_E^*)_{ji} \nonumber
\\ \big( M_{HR}^2 \big)_{i2} &=& -\frac{1}{2} \lambda_s v_{Rs} v_j
(h_E^*)_{ji} - \frac{1}{2} v_d (h_E^*)_{ji} h_\nu^{js} v_{Rs}
\end{eqnarray}
Finally, the $(6 \times 6)$ mass matrix of the charged sleptons can be
written as
\begin{equation}
M_{\tilde l \tilde l}^2 = \left( \begin{array}{c c}
M_{LL}^2 & M_{LR}^2 \\
M_{RL}^2 & M_{RR}^2 \end{array} \right)
\end{equation}
with:
\begin{eqnarray}
\big( M_{LL}^2 \big)_{ij} &=& \big( m_{\tilde L}^2 \big)_{ij} +
\frac{1}{8} (g'^2 - g^2) (v_d^2 - v_u^2 + v_1^2 + v_2^2 +
v_3^2)\delta_{ij} + \frac{1}{4}g^2 v_i v_j \nonumber \\ && +
\frac{1}{2}v_d^2 \big( h_E h_E^\dagger \big)_{ij} + \frac{1}{2} v_{Rs}
v_{Rt} h_\nu^{is} (h_\nu^{jt})^* \nonumber \\ M_{LR}^2 &=&
-\frac{1}{2} \lambda_s^* v_{Rs} v_u h_E + \frac{1}{\sqrt{2}} v_d T_E
\nonumber \\ M_{RL}^2 &=& \big( M_{LR}^2 \big)^\dagger \nonumber \\
\big( M_{RR}^2 \big)_{ij} &=& \big( m_{\tilde R}^2 \big)_{ij} +
\frac{1}{4} g'^2 (v_u^2 - v_d^2 - v_1^2 - v_2^2 - v_3^2)\delta_{ij}
\nonumber \\ && + \frac{1}{2}v_d^2 \big( h_E^\dagger h_E \big)_{ij} +
\frac{1}{2} v_k v_m \big( h_E^\dagger \big)_{ik} \big( h_E \big)_{mj}
\end{eqnarray}

\subsection{Neutral Scalars}
\label{subsec:scalars}

In the basis
\begin{equation}
\big( {S^0}' \big)^T = Re (H_d^0, H_u^0, \tilde{\nu}_s^c, \tilde{\nu}_i)
\end{equation}
the scalar potential includes the term
\begin{equation}
V \supset \big( {S^0}' \big)^T M_{S^0}^2 {S^0}'
\end{equation}
and the $((5+n) \times (5+n))$ neutral scalar mass matrix can be written as
\begin{equation}
M_{S^0}^2 = \left(
\begin{array}{c c c}
M^2_{HH} & M^2_{HS} & M^2_{H\tilde L}\\ \big( M^2_{HS} \big)^T &
M^2_{SS} & M^2_{\tilde L S}\\ \big( M^2_{H\tilde L} \big)^T & \big(
M^2_{\tilde L S} \big)^T & M^2_{\tilde L \tilde L} \end{array} \right)\quad.
\label{eq:massscalars}
\end{equation}
The matrix elements are given as follows:
\begin{eqnarray}
\big( M_{HH}^2 \big)_{11} &=& m_{H_d}^2 + \frac{1}{8} (g^2 + g'^2)
(3v_d^2 - v_u^2 + v_1^2 + v_2^2 + v_3^2) \nonumber \\ && + \frac{1}{2}
\lambda_s \lambda_t^* v_{Rs} v_{Rt} + \frac{1}{2} v_u^2 \lambda_s
\lambda_s^* \nonumber \\ \big( M_{HH}^2 \big)_{12} &=&
-\frac{1}{4}(g^2 + g'^2)v_d v_u + \lambda_s \lambda_s^* v_d v_u -
\frac{1}{8} v_{Rs}^2 (\lambda_s \kappa_s^* + h.c. ) \nonumber \\ && -
\frac{1}{2} v_u v_i (\lambda_s^* h_\nu^{is} + h.c.) - \frac{1}{2
\sqrt{2}} v_{Rs} (T_\lambda^s + h.c. ) \nonumber \\ \big( M_{HH}^2
\big)_{21} &=& \big( M_{HH}^2 \big)_{12} \nonumber \\ \big( M_{HH}^2
\big)_{22} &=& m_{H_u}^2 -\frac{1}{8} (g^2 + g'^2) (v_d^2 - 3v_u^2 +
v_1^2 + v_2^2 + v_3^2) \nonumber \\ && + \frac{1}{2} \lambda_s
\lambda_t^* v_{Rs} v_{Rt} + \frac{1}{2}v_d^2 \lambda_s \lambda_s^* +
\frac{1}{2} v_{Rs} v_{Rt} h_\nu^{is} (h_\nu^{it})^* + \frac{1}{2} v_i
v_j (h_\nu^{is})^* h_\nu^{js} \nonumber \\ && - \frac{1}{2} v_d v_i
(\lambda_s^* h_\nu^{is} + h.c.)
\end{eqnarray}
\begin{eqnarray}
\big( M^2_{H S} \big)_{1s} &=& -\frac{1}{4} v_u v_{Rs}(\lambda_s^*
\kappa_s + h.c.) + \frac{1}{2} v_d v_{Rt} ( \lambda_s \lambda_t^* +
h.c. ) \nonumber \\ && -\frac{1}{2 \sqrt{2}} v_u (T_\lambda^s + h.c.)
- \frac{1}{4} v_i v_{Rt} ( \lambda_s^* h_\nu^{it} + \lambda_t^*
h_\nu^{is} + h.c. ) \nonumber \\ \big( M^2_{H S} \big)_{2s} &=&
-\frac{1}{4} v_d v_{Rs}(\lambda_s^* \kappa_s + h.c.) + \frac{1}{2} v_u
v_{Rt} ( \lambda_s \lambda_t^* + h.c. ) \nonumber \\ && -\frac{1}{2
\sqrt{2}} v_d (T_\lambda^s + h.c.) +\frac{1}{2 \sqrt{2}} v_t
(T_{h_\nu}^{ts} + h.c.) + \frac{1}{4} v_{Rs} v_i ( \kappa_s^*
h_\nu^{is} + h.c. ) \nonumber \\ && + \frac{1}{2} v_u v_{Rt} [
h_\nu^{is} (h_\nu^{it})^* + h.c. ]
\end{eqnarray}
\begin{eqnarray}
\big( M^2_{H \tilde L} \big)_{1i} &=& \frac{1}{4}(g^2 + g'^2)v_d v_i -
\frac{1}{4} v_u^2 ( \lambda_s^* h_\nu^{is} + h.c. ) - \frac{1}{4}
v_{Rs} v_{Rt}(\lambda_s^* h_\nu^{it} + h.c.) \nonumber \\ \big( M^2_{H
\tilde L} \big)_{2i} &=& -\frac{1}{4}(g^2 + g'^2)v_u v_i + \frac{1}{8}
v_{Rs}^2 (\kappa_s^* h_\nu^{is} + h.c.) - \frac{1}{2} v_u v_d
(\lambda_s^* h_\nu^{is} + h.c.) \nonumber \\ && + \frac{1}{2} v_u v_j
[h_\nu^{js} (h_\nu^{is})^* + h.c.] + \frac{1}{2 \sqrt{2}} v_{Rs}
(T_{h_\nu}^{is} + h.c.)
\end{eqnarray}
\begin{eqnarray}\label{SSscalar}
\big( M^2_{SS} \big)_{st} &=& \frac{1}{2} [ (m_{\tilde\nu^c}^2 )_{st}
+ (m_{\tilde\nu^c}^2 )_{ts}] + \frac{1}{4} (\lambda_s \lambda_t^* +
h.c.) (v_d^2 + v_u^2) - \frac{1}{4} v_d v_u (\lambda_s^* \kappa_s +
h.c.) \delta_{st} \nonumber \\ && + \frac{3}{4}\kappa_s \kappa_s^*
v_{Rs}^2 \delta_{st} + \frac{1}{4} v_u v_i (\kappa_s^* h_\nu^{is} +
h.c.) \delta_{st} + \frac{1}{4} v_u^2 [(h_\nu^{is})^* h_\nu^{it} +
h.c.] \nonumber \\ && + \frac{1}{4} v_i v_j [(h_\nu^{is})^* h_\nu^{jt}
+ h.c.] - \frac{1}{4} v_d v_i [\lambda_s^* h_\nu^{it} + \lambda_t
(h_\nu^{is})^* + h.c.] \nonumber \\ && + \frac{1}{2 \sqrt{2}} v_{Ru}
(T_\kappa^{stu} + h.c.)
\end{eqnarray}
\begin{eqnarray}
\big( M^2_{\tilde L S} \big)_{si} &=& \frac{1}{4}v_u v_{Rs}(\kappa_s^*
h_\nu^{is} + h.c.) - \frac{1}{4} v_d v_{Rt} (\lambda_s^* h_\nu^{it} +
\lambda_t^* h_\nu^{is} + h.c.) \nonumber \\ && + \frac{1}{2 \sqrt{2}}
v_u (T_{h_\nu}^{is} + h.c.) + \frac{1}{4} v_j v_{Rt} [ h_\nu^{jt}
(h_\nu^{is})^* + h_\nu^{js} (h_\nu^{it})^* + h.c. ]
\end{eqnarray}
\begin{eqnarray}
\big( M_{\tilde L \tilde L}^2 \big)_{ij} &=& \frac{1}{2} [
(m_L^2)_{ij} + (m_L^2)_{ji}] + \frac{1}{8}(g^2 + g'^2)(v_d^2 - v_u^2 +
v_1^2 + v_2^2 + v_3^2) \delta_{ij} \nonumber \\ && + \frac{1}{4}(g^2 +
g'^2)v_i v_j + \frac{1}{4} v_u^2 [h_\nu^{is} (h_\nu^{js})^* + h.c.] +
\frac{1}{4} v_{Rs} v_{Rt} [ h_\nu^{is} (h_\nu^{jt})^* + h.c. ]
\end{eqnarray}

\subsection{Pseudoscalars}
\label{subsec:pseudoscalars}

In the basis
\begin{equation}
\big( {P^0}' \big)^T = Im (H_d^0, H_u^0, \tilde{\nu}_s^c, \tilde{\nu}_i)
\end{equation}
the scalar potential includes the term
\begin{equation}
V \supset \big( {P^0}' \big)^T M_{P^0}^2 {P^0}'
\end{equation}
and the $((5+n) \times (5+n))$ pseudoscalar mass matrix can be written as
\begin{equation}
M_{P^0}^2 = \left(
\begin{array}{c c c}
M^2_{HH} & M^2_{HS} & M^2_{H\tilde L}\\ \big( M^2_{HS} \big)^T &
M^2_{SS} & M^2_{\tilde L S}\\ \big( M^2_{H\tilde L} \big)^T & \big(
M^2_{\tilde L S} \big)^T & M^2_{\tilde L \tilde L} \end{array} \right)\quad.
\label{eq:masspseudoscalars}
\end{equation}
The matrix elements are given as follows:
\begin{eqnarray}
\big( M_{HH}^2 \big)_{11} &=& m_{H_d}^2 + \frac{1}{8} (g^2 + g'^2)
(v_d^2 - v_u^2 + v_1^2 + v_2^2 + v_3^2) \nonumber \\ && + \frac{1}{2}
\lambda_s \lambda_t^* v_{Rs} v_{Rt} + \frac{1}{2} v_u^2 \lambda_s
\lambda_s^* \nonumber \\ \big( M_{HH}^2 \big)_{12} &=& \frac{1}{8}
v_{Rs}^2 (\lambda_s \kappa_s^* + h.c. ) + \frac{1}{2 \sqrt{2}} v_{Rs}
(T_\lambda^s + h.c. ) \nonumber \\ \big( M_{HH}^2 \big)_{21} &=& \big(
M_{HH}^2 \big)_{12} \nonumber \\ \big( M_{HH}^2 \big)_{22} &=&
m_{H_u}^2 -\frac{1}{8} (g^2 + g'^2) (v_d^2 - v_u^2 + v_1^2 + v_2^2 +
v_3^2) \nonumber \\ && + \frac{1}{2} \lambda_s \lambda_t^* v_{Rs}
v_{Rt} + \frac{1}{2}v_d^2 \lambda_s \lambda_s^* + \frac{1}{2} v_{Rs}
v_{Rt} h_\nu^{is} (h_\nu^{it})^* + \frac{1}{2} v_i v_j (h_\nu^{is})^*
h_\nu^{js} \nonumber \\ && - \frac{1}{2} v_d v_i (\lambda_s^*
h_\nu^{is} + h.c.)
\end{eqnarray}
\begin{eqnarray}
\big( M^2_{H S} \big)_{1s} &=& -\frac{1}{4} v_u v_{Rs}(\lambda_s^*
\kappa_s + h.c.) + \frac{1}{4} \sum_{t \neq s} v_i v_{Rt} (
\lambda_s^* h_\nu^{it} - \lambda_t^* h_\nu^{is} + h.c. ) \nonumber \\
&& + \frac{1}{2 \sqrt{2}} v_u (T_\lambda^s + h.c.) \nonumber \\ \big(
M^2_{H S} \big)_{2s} &=& - \frac{1}{4} v_d v_{Rs}(\lambda_s^* \kappa_s
+ h.c.) + \frac{1}{4} v_{Rs} v_i ( \kappa_s^* h_\nu^{is} + h.c. )
\nonumber \\ && + \frac{1}{2 \sqrt{2}} v_d (T_\lambda^s + h.c.) -
\frac{1}{2 \sqrt{2}} v_i (T_{h_\nu}^{is} + h.c.)
\end{eqnarray}
\begin{eqnarray}
\big( M^2_{H \tilde L} \big)_{1i} &=& - \frac{1}{4} v_u^2 (
\lambda_s^* h_\nu^{is} + h.c. ) - \frac{1}{4} v_{Rs}
v_{Rt}(\lambda_s^* h_\nu^{it} + h.c.) \nonumber \\ \big( M^2_{H \tilde
L} \big)_{2i} &=& - \frac{1}{8} v_{Rs}^2 (\kappa_s^* h_\nu^{is} +
h.c.) - \frac{1}{2 \sqrt{2}} v_{Rs} (T_{h_\nu}^{is} + h.c.)
\end{eqnarray}
\begin{eqnarray}\label{SSpseudoscalar}
\big( M^2_{SS} \big)_{st} &=& \frac{1}{2} [ (m_{\tilde\nu^c}^2 )_{st}
+ (m_{\tilde\nu^c}^2 )_{ts}] + \frac{1}{4} (\lambda_s \lambda_t^* +
h.c.) (v_d^2 + v_u^2) + \frac{1}{4} v_d v_u (\lambda_s^* \kappa_s +
h.c.) \delta_{st} \nonumber \\ && + \frac{1}{4}\kappa_s \kappa_s^*
v_{Rs}^2 \delta_{st} - \frac{1}{4} v_u v_i (\kappa_s^* h_\nu^{is} +
h.c.) \delta_{st} + \frac{1}{4} v_u^2 [(h_\nu^{is})^* h_\nu^{it} +
h.c.] \nonumber \\ && + \frac{1}{4} v_i v_j [(h_\nu^{is})^* h_\nu^{jt}
+ h.c.] - \frac{1}{4} v_d v_i [\lambda_s^* h_\nu^{it} + \lambda_t
(h_\nu^{is})^* + h.c.] \nonumber \\ && - \frac{1}{2 \sqrt{2}} v_{Ru}
(T_\kappa^{stu} + h.c.)
\end{eqnarray}
\begin{eqnarray}
\big( M^2_{\tilde L S} \big)_{si} &=& \frac{1}{4} v_u
v_{Rs}(\kappa_s^* h_\nu^{is} + h.c.) + \frac{1}{4} \sum_{t \neq s} v_d
v_{Rt} (\lambda_t^* h_\nu^{is} - \lambda_s^* h_\nu^{it} + h.c.)
\nonumber \\ && + \frac{1}{4} \sum_{t \neq s} v_j v_{Rt} [ h_\nu^{js}
(h_\nu^{it})^* - h_\nu^{jt} (h_\nu^{is})^* + h.c. ] - \frac{1}{2
\sqrt{2}} v_u (T_{h_\nu}^{is} + h.c.)
\end{eqnarray}
\begin{eqnarray}
\big( M_{\tilde L \tilde L}^2 \big)_{ij} &=& \frac{1}{2} [
(m_L^2)_{ij} + (m_L^2)_{ji}] + \frac{1}{8}(g^2 + g'^2)(v_d^2 - v_u^2 +
v_1^2 + v_2^2 + v_3^2) \delta_{ij} \nonumber \\ && + \frac{1}{4} v_u^2
[h_\nu^{is} (h_\nu^{js})^* + h.c.] + \frac{1}{4} v_{Rs} v_{Rt} [
h_\nu^{is} (h_\nu^{jt})^* + h.c. ]
\end{eqnarray}

\subsection{Neutral Fermions}
\label{subsec:neutralinos}

In the basis
\begin{equation}
\big( \psi^0 \big)^T = \big( {\tilde B}^0, {\tilde W}_3^0, {\tilde
H}_d^0, {\tilde H}_u^0, \nu_s^c, \nu_i \big)
\end{equation}
the lagrangian of the model includes the term
\begin{equation}
{\cal L} \supset -\frac{1}{2} \big( \psi^0 \big)^T {\cal M}_n \psi^0 + h.c.
\end{equation}
with the $((7+n) \times (7+n))$ mass matrix of the neutral fermions, which
can be written as:
\begin{equation}
{\cal M}_n = \left( \begin{array}{c c c}
M_{\tilde{\chi}^0} & m_{\tilde{\chi}^0 \nu^c} & m_{\tilde{\chi}^0 \nu} \\
m_{\tilde{\chi}^0 \nu^c}^T & M_R & m_D \\
m_{\tilde{\chi}^0 \nu}^T & m_D^T & 0 \end{array} \right)
\end{equation}
$M_{\tilde{\chi}^0}$ is the usual mass matrix of the neutralinos in the MSSM
\begin{equation}
M_{\tilde{\chi}^0} = \left( \begin{array}{c c c c}
M_1 & 0 & -\frac{1}{2}g' v_d & \frac{1}{2}g' v_u \\
0 & M_2 & \frac{1}{2}g v_d & -\frac{1}{2}g v_u \\
-\frac{1}{2}g' v_d & \frac{1}{2}g v_d & 0 & -\mu \\
\frac{1}{2}g' v_u & -\frac{1}{2}g v_u & -\mu & 0 \end{array} \right)
\end{equation}
with
\begin{equation}
\mu = \frac{1}{\sqrt{2}} \lambda_s v_{Rs}\quad.
\end{equation}
The mixing between the neutralinos and the singlet $\nu_s^c$ is given by
\begin{equation}
(m_{\tilde{\chi}^0 \nu^c}^T)_s = \left( \begin{array}{c c c c}
0 & 0 & -\frac{1}{\sqrt{2}} \lambda_s v_u & -\frac{1}{\sqrt{2}}\lambda_s v_d + \frac{1}{\sqrt{2}} v_i h_\nu^{is} \end{array} \right)\quad.
\end{equation}
$m_{\tilde{\chi}^0 \nu}$ is the neutralino-neutrino mixing part
\begin{equation}
m_{\tilde{\chi}^0 \nu}^T = \left( \begin{array}{c c c c}
-\frac{1}{2} g' v_1 & \frac{1}{2} g v_1 & 0 & \epsilon_1  \\
-\frac{1}{2} g' v_2 & \frac{1}{2} g v_2 & 0 &  \epsilon_2  \\
-\frac{1}{2} g' v_3 & \frac{1}{2} g v_3 & 0 &  \epsilon_3 \end{array} \right)
\end{equation}
with
\begin{equation}\label{defepstot}
\epsilon_i =  \frac{1}{\sqrt{2}} \sum_{s=1}^n v_{Rs} h_\nu^{is}\quad.
\end{equation}
The neutrino Dirac term is
\begin{equation}
(m_D)_{is} = \frac{1}{\sqrt{2}} h_\nu^{is} v_u
\end{equation}
and finally $M_R$ is 
\begin{equation}
(M_R)_{st} = \frac{1}{\sqrt{2}} \kappa_s v_{Rs} \delta_{st}\quad.
\end{equation}

\subsection{Charged Fermions}
\label{subsec:charginos}

In the basis
\begin{align}
\nonumber
\left(\psi^-\right)^T&=\left(\tilde{W}^-,\tilde{H}_d^-,e,\mu,\tau\right)\\
\left(\psi^+\right)^T&=\left(\tilde{W}^+,\tilde{H}_u^+,e^c,\mu^c,\tau^c\right),
\end{align}
the $(5 \times 5)$ mass matrix of the charged fermions is given by
\begin{equation}
{\cal M}_c = \left( \begin{array}{c c c c c}
M_2 & \frac{1}{\sqrt{2}}gv_u & 0 & 0 & 0 \\
\frac{1}{\sqrt{2}}gv_d & \mu & -\frac{1}{\sqrt{2}}h_E^{i1}v_i & -\frac{1}{\sqrt{2}}h_E^{i2}v_i & -\frac{1}{\sqrt{2}}h_E^{i3}v_i \\
\frac{1}{\sqrt{2}}gv_1 & -\epsilon_1 & \frac{1}{\sqrt{2}}h_E^{11}v_d & 0 & 0 \\
\frac{1}{\sqrt{2}}gv_2 & -\epsilon_2 & 0 & \frac{1}{\sqrt{2}}h_E^{22}v_d &0 \\
\frac{1}{\sqrt{2}}gv_3 & -\epsilon_3 & 0 & 0 & \frac{1}{\sqrt{2}}h_E^{33}v_d \end{array} \right)\quad.
\end{equation}

\section{Coupling $\tilde{\chi}_1^0-W^{\pm}-l^{\mp}_i$}
\label{sec:AppCoupXiWl}

Approximate formulas for the coupling
$\tilde{\chi}_1^0-W^{\pm}-l^{\mp}_i$ can be obtained from the general
$\tilde{\chi}_i^0-W^{\pm}-\tilde{\chi}_j^{\mp}$ interaction lagrangian
\begin{equation}
\mathcal{L} \supset \overline{\tilde{\chi}_i^-} \gamma^{\mu} \big( O_{Lij}^{cnw} P_L + O_{Rij}^{cnw} P_R \big) \tilde{\chi}_j^0 W_\mu^- + \overline{\tilde{\chi}_i^0} \gamma^{\mu} \big( O_{Lij}^{ncw} P_L + O_{Rij}^{ncw} P_R \big) \tilde{\chi}_j^- W_\mu^+\quad,
\end{equation}
where
\begin{eqnarray}\label{Oexact}
O_{Li1}^{cnw} &=& g\left[-{\cal U}_{i1}{\cal N}_{12}^*-\frac{1}{\sqrt{2}}\left({\cal U}_{i2}{\cal N}_{13}^*+\sum_{k=1}^3 {\cal U}_{i,2+k}{\cal N}_{1,5+k}^*\right)\right] \nonumber \\
O_{Ri1}^{cnw} &=& g\left(-{\cal V}_{i1}^*{\cal N}_{12}+\frac{1}{\sqrt{2}}{\cal V}_{i2}^*{\cal N}_{14}\right) \nonumber \\
O_{L1j}^{ncw} &=& \left(O_{Lj1}^{cnw}\right)^* \nonumber\\
O_{R1j}^{ncw} &=& \left(O_{Rj1}^{cnw}\right)^* \quad.
\end{eqnarray}
The matrix ${\cal N}$ diagonalizes the neutral fermion mass matrix
(see Appendix \eqref{subsec:neutralinos}) while the matrices ${\cal
U}$ and ${\cal V}$ diagonalize the charged fermion mass matrix (see
Appendix \eqref{subsec:charginos}).

As  was already mentioned for the case of neutral fermions in
Section \ref{subsec:neutrinomass}, it is possible to diagonalize the
mass matrices in very good approximation due to the fact that the \rpv
parameters are small. Defining the matrices $\xi$, $\xi_L$ and
$\xi_R$, that will be taken as expansion parameters, one gets the
leading order expressions
\begin{align}
{\cal N}=\begin{pmatrix}N&N\xi^T\\-V^T\xi&V^T\end{pmatrix},\qquad {\cal U}=\begin{pmatrix}U_c&U_c\xi_L^T\\-\xi_L&I_3\end{pmatrix},\qquad {\cal V}=\begin{pmatrix}V_c&V_c\xi_R^T\\-\xi_R&I_3\end{pmatrix}\quad,
\end{align}
where $I_3$ is the $(3 \times 3)$ identity matrix. The expansion
matrices $\xi_L$ and $\xi_R$ are 
\begin{eqnarray}
\left(\xi_L\right)_{i1} &=& \frac{g\Lambda_i}{\sqrt{2} Det_+} \nonumber \\
\left(\xi_L\right)_{i2} &=& -\frac{\epsilon_i}{\mu}-\frac{g^2v_u\Lambda_i}{2\mu Det_+} \nonumber \\
\left(\xi_R\right)_{i1} &=& \frac{gv_d h_E^{ii}}{2 Det_+}\left[\frac{v_u\epsilon_i}{\mu}+\frac{\left(2\mu^2+g^2v_u^2\right)\Lambda_i}{2\mu Det_+}\right] \nonumber \\
\left(\xi_R\right)_{i2} &=& -\frac{\sqrt{2}v_d h_E^{ii}}{2 Det_+}\left[\frac{M_2\epsilon_i}{\mu}+\frac{g^2\left(v_d\mu +M_2v_u\right)\Lambda_i}{2\mu Det_+}\right]\quad,
\end{eqnarray}
where $Det_+=-\frac{1}{2}g^2v_dv_u+M_2\mu$ is the determinant of the
MSSM chargino mass matrix, $\mu = \frac{1}{\sqrt{2}}\lambda_s v_{Rs}$
and $\epsilon_i = \frac{1}{\sqrt{2}} v_{Rs} h_\nu^{is}$.
The expressions for the matrix $\xi$ depend on the number of
singlet generations in the model. Particular cases can be found in
\eqref{xi} and \eqref{xi2}.

Using the previous equations and assuming that all parameters are real
, one gets the approximate formulas
\begin{eqnarray}\label{Oapprox}
O_{Li1}^{cnw} &=& \frac{g}{\sqrt{2}} \big[ \frac{g N_{12} \Lambda_i}{Det_+}-\big( \frac{\epsilon_i}{\mu}+\frac{g^2 v_u \Lambda_i}{2 \mu Det_+} \big)N_{13}-\sum_{k=1}^{n} N_{1k} \xi_{ik} \big] \nonumber \\
O_{Ri1}^{cnw} &=& \frac{1}{2} g (h_E)_{ii} \frac{v_d}{Det_+} \big[ \frac{g v_u N_{12} - M_2 N_{14}}{\mu} \epsilon_i +\frac{g(2 \mu^2 +g^2 v_u^2)N_{12}-g^2 (v_d \mu +M_2v_u)N_{14}}{2 \mu Det_+} \Lambda_i \big] \nonumber \\
O_{Li1}^{ncw} &=& \big(O_{Li1}^{cnw}\big)^* \nonumber\\
O_{Ri1}^{ncw} &=& \big(O_{Ri1}^{cnw}\big)^*\quad.
\end{eqnarray}
It is important to emphasize that all previous formulas, and the
following simplified versions, are tree-level results. More
simplified formulas are possible if the lightest neutralino has a
large component in one of the gauge eigenstates. These particular
limits are of great interest to understand the phenomenology:

\subsection*{Bino-like $\tilde{\chi}_1^0$}

This limit is caracterized by $N_{11}^2 = 1$ and $N_{1m} = 0$ for $m
\neq 1$. One gets
\begin{eqnarray}
O_{Li1}^{cnw} &=& - \frac{g}{\sqrt{2}} \xi_{i1} \nonumber \\
O_{Ri1}^{cnw} &=& 0\quad.
\end{eqnarray}
For the $1$ $\widehat{\nu}^c$-model this implies that a bino-like
$\tilde{\chi}_1^0$ couples to $W l_i$ proportionally to $\Lambda_i$, see
Equation \eqref{xi}, without any dependence on the $\epsilon_i$
parameters.

On the other hand, for the $2$ $\widehat{\nu}^c$-model, the more complicated
structure of the $\xi$ matrix, see Equations \eqref{xi2} and
\eqref{defK2}, implies a coupling of a bino-like $\tilde{\chi}_1^0$
with $W l_i$ dependent on two pieces, one proportional to $\Lambda_i$ and
one proportional to $\alpha_i$:
\begin{equation}
\xi_{i1} = \frac{2 g' M_2 \mu}{m_\gamma} (a \Lambda_i + b \alpha_i)
\end{equation}

However, a simple estimate of the relative importance of
these two terms is possible. By assuming that all masses are at the same scale
$m_{SUSY}$, the couplings $\kappa$ and $\lambda$ are of order $0.1$,
and the \rpv terms $h_\nu^i$ and $v_i$ are of order $h_{\text{\rpv}}$ and
$m_{SUSY} h_{\text{\rpv}}$ respectively, one can show that $a \Lambda_i \sim
200 \: b \alpha_i$. Therefore, one gets a coupling which is
proportional, in very good approximation, to $\Lambda_i$, as confirmed by
the exact numerical results shown in the main part of the paper.
Similar arguments apply for models with more generations of right-handed neutrinos.

In conclusion, for a bino-like neutralino the coupling
$\tilde{\chi}_1^0-W^{\pm}-l^{\mp}_i$ is  proportional to
$\Lambda_i$ to a good approximation.

\subsection*{Higgsino-like $\tilde{\chi}_1^0$}

This limit is caracterized by $N_{13}^2 + N_{14}^2 = 1$ and $N_{1m} =
0$ for $m \neq 3,4$. If the coupling $O_{Ri1}^{cnw}$ is neglected due
to the supression given by the charged lepton Yukawa couplings, one
gets
\begin{eqnarray}\label{Ohiggs}
O_{Li1}^{cnw} &=& - \frac{g}{\sqrt{2}} \left[ \big( \frac{\epsilon_i}{\mu}+\frac{g^2 v_u \Lambda_i}{2 \mu Det_+} + \xi_{i3} \big)N_{13} + \xi_{i4} N_{i4} \right] \nonumber \\
O_{Ri1}^{cnw} &\simeq& 0\quad.
\end{eqnarray}
Equations \eqref{xi} and \eqref{xi2} show that the $\epsilon_i$ terms
cancel out in the coupling \eqref{Ohiggs}, and therefore one gets
dependence only on $\Lambda_i$ in the $1$ $\widehat{\nu}^c$-model, and
$(\Lambda_i , \alpha_i)$ in the $2$ $\widehat{\nu}^c$-model.
However, this cancellation is not perfect in $O_{Ri1}^{cnw}$ and thus
one still has some dependence on $\epsilon_i$. 

\subsection*{Singlino-like $\tilde{\chi}_1^0$}

The limit in which the right-handed neutrino $\nu_s^c$ is the lightest
neutralino is caracterized by $N_{1m}^2 = 1$ for $m \ge 5$ 
and $N_{1l} = 0$ for $l \neq m$. One gets
\begin{eqnarray}\label{Osinglino}
O_{Li1}^{cnw} &=& - \frac{g}{\sqrt{2}} \xi_{im} \nonumber \\
O_{Ri1}^{cnw} &=& 0\quad.
\end{eqnarray}
For the $1$ $\widehat{\nu}^c$-model this expression implies that a
pure singlino-like $\tilde{\chi}_1^0$ couples to $W l_i$
proportional to $\Lambda_i$, see Equation \eqref{xi}, without any
dependence on the $\epsilon_i$ parameters.
This proportionality to $\Lambda_i$ is different to what is found in
spontaneous R-parity violation, where the different structure of the
corresponding $\xi$ matrix \cite{Hirsch:2008ur} implies that the
singlino couples to $W l_i$ proportionally to $\epsilon_i$.

For the $n$ $\widehat{\nu}^c$-model one finds that the coupling
$\tilde{\chi}_1^0-W^{\pm}-l^{\mp}_i$ for a singlino-like neutralino
has little dependence on $\Lambda_i$. For example, in the $2$
$\widehat{\nu}^c$-model one finds that the element $\xi_{i5}$,
corresponding to the right-handed neutrino $\nu_1^c$, is given by
\begin{equation}\label{singcoup}
\xi_{i5} = \frac{M_{R2} \lambda_1 m_\gamma}{4 \sqrt{2} Det(M_H)} (v_u^2 - v_d^2) \Lambda_i - \left( \sqrt{2} \lambda_2 c + \frac{4 Det_0 v_{R1}}{\mu m_\gamma (v_u^2 - v_d^2)}b \right) \alpha_i\quad.
\end{equation}
The coupling has two pieces, one proportional to $\Lambda_i$ and one
proportional to $\alpha_i$. However, the $\alpha_i$ piece gives the
dominant contribution, as can be shown using an estimate completely
analogous to the one done for a bino-like $\tilde{\chi}_1^0$. In this
case, the ratio between the two terms in Equation \eqref{singcoup} is
$\alpha_i$-piece $\sim$ $8 \: \Lambda_i$-piece, sufficient to ensure a
very good proportionality to the $\alpha_i$ parameters. This estimate
has been corroborated numerically.

\bibliographystyle{h-physrev}
\bibliography{liter}

\end{document}